\documentclass[12pt]{article}
\usepackage[margin=1in]{geometry}
\usepackage[authoryear]{natbib}
\usepackage[export]{adjustbox}
\usepackage{graphicx}
\usepackage{cancel,soul}
\usepackage{longtable}
\usepackage{multicol,multirow,tabularx}
\usepackage{subfig}

\usepackage{amsmath,amssymb,bm,bbm,leftidx}
\usepackage[colorlinks]{hyperref}
\hypersetup{
	colorlinks,
	citecolor=blue,
	filecolor=black,
	linkcolor=blue,
	urlcolor=blue
}
\usepackage[nameinlink,capitalize]{cleveref}
\usepackage[title]{appendix}
\usepackage{mathbbol}
\usepackage{titletoc}
\usepackage[font={sf,small},labelfont=bf]{caption}
\usepackage{pstricks}
\usepackage{comment}
\usepackage{ltablex}
\usepackage{blkarray}
\usepackage{algorithm}
\usepackage{algpseudocode}
\usepackage{setspace}

\usepackage{titlesec}
\titleformat*{\section}{\large \bfseries}
\titleformat*{\subsection}{\bfseries}
\titleformat*{\subsubsection}{\bfseries}
\titleformat*{\paragraph}{ \bfseries}
\usepackage{stackengine}
\usepackage{scalerel,wasysym}
\usepackage{mathtools}
\usepackage{tabularx}
\usepackage{pdflscape}

\def\ten#1{
\ifcat\noexpand#1\relax 
	{\bm{#1}}
\else
	{\bf{#1}}
\fi}
\def\cten#1{
\ifcat\noexpand#1\relax 
	{\tilde{\bm{#1}}}
\else
	{\tilde{\bf{#1}}}
\fi}
\def\cbten#1{
\ifcat\noexpand#1\relax 
	{\hat{\bm{#1}}}
\else
	{\hat{\bf{#1}}}
\fi}

\def\seq{\sigma_{\rm eq}}

\def\Eeq{E_{\rm eq}}

\DeclareSymbolFontAlphabet{\mathbb}{AMSb}
\DeclareSymbolFontAlphabet{\mathbbl}{bbold}

\newcommand\blfootnote[1]{%
  \begingroup
  \renewcommand\thefootnote{}\footnote{#1}%
  \addtocounter{footnote}{-1}%
  \endgroup
}

\begin{document}
\onehalfspacing
\let\WriteBookmarks\relax
\def\floatpagepagefraction{1}
\def\textpagefraction{.001}

\title{An assessment of mechanism-based plasticity models for polycrystalline magnesium alloys}

\author{R. Vigneshwaran$^{1,*}$, Showren Datta$^2$\\
A. A. Benzerga$^{1,3,*}$, and Shailendra P. Joshi$^{2,*}$
\blfootnote{Corresponding authors: \href{mailto:vigneshr@tamu.edu}{vigneshr@tamu.edu}, \href{mailto:benzerga@tamu.edu}{benzerga@tamu.edu}, \href{mailto:shailendra@uh.edu}{shailendra@uh.edu}} \\ 
\footnotesize $^1$Department of Aerospace Engineering, Texas A\&M University  \\
\footnotesize College Station, TX 77843 \\
\footnotesize $^2$Department of Mechanical and Aerospace Engineering, University of Houston \\
\footnotesize Houston, TX 77204 \\
\footnotesize $^3$Department of Materials Science and Engineering, Texas A\&M University \\ 
\footnotesize College Station, TX 77843}
\date{}
\maketitle
\begin{abstract}
The objective of this work is to assess computationally efficient coarse-grained plasticity models
against high-fidelity crystal plasticity simulations for magnesium polycrystals over a wide
range of textures and grain sizes.
A basic requirement is that such models are able 
to capture {\it evolving} plastic anisotropy and tension-compression asymmetry.
To this end, two-surface and three-surface plasticity models are considered.
The two-surface constitutive formulation separately accounts for slip and twinning, 
while the three-surface model further apportions the contributions of basal and nonbasal slip.
Model identification is based on stress-strain responses for loading along six orientations
under both tension and compression.
The evolution of overall plastic anisotropy, as well as microscale relative activities 
of slip and twin systems, is analyzed in detail.
The prospects of using coarse-grained plasticity models 
in guiding the development of physically sound damage models for magnesium alloys are discussed.
\end{abstract}

\section{Introduction} \label{sec:Intro}
Designing strong, damage-tolerant materials is a bit like
orchestrating a play comprising a group of talented actors,
which in this case are design degrees of freedom ({\sf ddofs}). 
It is their strengths and limitations that make the space of material design
exciting and challenging. In structural materials, the {\sf ddofs} range from our ability to manipulate precise alloying compositions via frameworks such as 
Calculation of Phase Diagrams (CALPHAD), 
to introducing interfaces (e.g., nanotwinned materials),
to controlling grain sizes and material textures. 
Each operates at a different length-scale 
but they all conspire together to produce macroscopic 
emergent mechanical behaviors that are ultimately of interest to engineers
in creating structures that are strong and resilient 
under the action of forces over their expected functional lives. Of particular interest to structural metals are the yield strength,
strain hardening, strain rate sensitivity, and strain to failure.

These macroscopic engineering properties are often anisotropic \citep{hosford_msea98,wenk_rev_min_geo02}.
One of the most discriminating cases of plastic anisotropy
is found in hexagonal-closed-packed (HCP) alloys; 
e.g., magnesium (Mg), titanium, cobalt, beryllium, and many others. 
Mg alloys, which are promising candidates in automotive and biomedical applications, 
show a remarkable range of these properties \citep{bohlen2007,agnew2005plastic}, albeit with trade-offs. 
For example, the development of Mg alloys with high tensile ductility
often comes with serious reductions in their yield strengths 
\citep{Basu17,Yu_JMST18,wei_AM21,hufnagel_mecmat22,masood_CritRev22}. 
%

A question arises: How does one navigate the space of {\sf ddofs} to
quantify the links between macroscopic material properties and their
underlying microstructural ``actors" - alloy composition, grain size, texture, etc.?

Multiscale experiments and microstructural characterization provide useful correlations, but they are costly. 
An equally effective approach is to perform high-fidelity simulations. 
In this regard, well-calibrated crystal plasticity-based finite element
polycrystal modeling and simulation \citep[e.g.,][]{Lebensohn93,Agnew06,Kalidindi98,Zhang12,Chang15}
provides a promising platform to precisely
correlate microstructural mechanisms of plasticity (slip and twinning)
with the macroscopic anisotropic plasticity of yield and post-yield behaviors.
However, such an approach quickly becomes untenable for searching through the
hyperdimensional space of {\sf ddofs}. 
At the other end, an extreme coarsening typically leads to
highly compact constitutive models based on a single yield surface
with sophisticated representations of plasticity anisotropy and tension-compression (response)
asymmetry \citep[e.g.,][]{Cazacu06}. An unresolved challenge in such models is that they typically 
are opaque to the fundamental mechanisms that drive plasticity at the fine-scale.
Moreover, they often do not naturally account for the evolution of
plastic anisotropy and asymmetry.
These aspects are especially critical in HCP materials in general
and in Mg alloys in particular, which deform by myriad slip (basal $\langle a \rangle$, prismatic $\langle a \rangle$, pyramidal $\langle a \rangle$, pyramidal $\langle c{+}a \rangle$) and twinning (extension and contraction) processes. 

A single-crystal model that explicitly accounts
for all slip and twinning systems can be considered as an extreme. 
Models based on a single asymmetric yield surface with distortional hardening
constitute one coarse-graining approach \citep{li2010,SoareIJSS,kim2017}.
Later, models based on the multisurface representation of plasticity have been shown to
offer an attractive recourse.
At a minimum, in a two-surface (2S) plasticity model \citep{Steglich16,Kondori19,jedidi20,lei23},
one anisotropic but symmetric yield surface provides a coarse-grained representation
of plasticity due to all slip mechanisms, 
and the other anisotropic and asymmetric surface 
provides a coarse-grained representation
of plasticity due to twinning (typically, extension twinning).
The 2S model has been shown to capture
some key features of Mg plasticity under
uniaxial tension and compression \citep{Kondori19,Herrington19}, 
three-point bending \citep{steglich2018}, 
sheet-metal necking \citep{jedidi20},
and more recently,
in void growth and coalescence under triaxial tension \citep{Vignesh23}.
Alternatively, three-surface (3S) models \citep{Becker16,Indurkar23} further distinguish
the various glide contributions.
For example, \citet{Indurkar23} recently proposed a 3S model
that includes two anisotropic symmetric yield surfaces
- one representing soft glide (basal slip) and
the other representing hard glide (non-basal slip mechanisms);
the third yield surface representing extension twinning
is anisotropic and asymmetric. 
The 3S model has been evaluated for a 
set of boundary-value problems \citep{Indurkar23}.

The ultimate test of any mechanism-based multisurface model (MSM) 
is its ability to predict structural failure with micromechanical rigor.
In that sense, both, 2S and 3S models are amenable to embedding 
failure micromechanics. 
However, achieving that capability first requires establishing a more fundamental criterion:
{\it What is the minimal mechanistic structure that an MSM must retain to 
reliably reproduce both macroscopic plastic anisotropy 
and the underlying micromechanical activity?}

Mg alloys provide a stringent setting for this question,
given their yield strength differentials at the level of individual mechanisms, 
the role of extension twinning, 
and the strong dependence of yield asymmetry on texture and grain size.
Therefore, any coarse-graining strategy must determine how these mechanisms
should be partitioned at the constitutive level.
The 2S formulation views all glide through the lens of a single mechanism,
whereas the 3S formulation collects all non-basal glide mechanisms into a 
single anisotropic representation but separates it from the basal (soft) glide 
to reflect their distinct activation strengths and hardening characteristics.
Whether this additional glide surface constitutes a necessary element
of a coarse-grained model has not been rigorously established.

Therefore, a direct and controlled comparison of the 2S and 3S models is necessary.
By calibrating both formulations to the same high-fidelity crystal plasticity dataset
and evaluating their performance across a wide range of textures, grain sizes, and loading orientations,
one can determine which mechanistic distinctions must be preserved for robust prediction.
Such an assessment would not only clarify the relative capabilities of the 2S and 3S models,
but also identify the degree of mechanistic resolution required 
of future MSMs intended for damage and failure modeling in Mg alloys.

In the present work, we assess the capabilities of the 2S and 3S models.
The assessment is intentionally restricted to macroscopically homogeneous uniaxial
tensile and compressive states along multiple material anisotropy orientations. 
The aim is to isolate the anisotropic and asymmetric aspects of the plastic flow in Mg alloys,
as detected by the two models. 
Although these loading paths are simple in a macroscopic sense, 
they are non-trivial in the context of the MSMs. 
The challenges involved include assessing the accuracy of activation of the glide and twinning mechanisms,
their evolution, and their role in yield asymmetry and anisotropy.
A rigorous evaluation of the two MSMs under these controlled conditions
provides a basis for answering the basic question posed above.

In what follows, both models are {\it independently} calibrated to an extensive dataset of $\sim 10^3$
full-field, finite-deformation polycrystalline simulations
based on three-dimensional crystal plasticity. 
It includes 
11 textures, 4 grain sizes, 6 loading directions, and 2 loading states \citep{Baweja23}. 
The constitutive parameters of both models are calibrated using 
orientation-dependent macroscopic stress-strain data. The so-calibrated
models are assessed for their ability to predict the 
evolution of the macroscopic deformation anisotropy
as well as the microscopic deformation mechanisms.   

The paper is organized as follows. 
\Cref{sec:constitutive} introduces the constitutive equations 
of the 2S and 3S plasticity models.
\Cref{sec:results} presents a critical assessment of the model.
Finally, \Cref{sec:disc} discusses the prospectus of coarse-grained plasticity models.
The parameters of both models and the optimization procedure are presented in \cref{app:2S} and \cref{app:3S}.
\section{Constitutive equations} \label{sec:constitutive}

\subsection{Two-surface model} \label{sec:2S}
The two-surface (2S) model formulation closely follows that of \citet{Kondori19}
with the total rate of deformation being additively decomposed as:
\begin{equation}
	\ten{d}=\ten{d}^{\rm{e}}+\ten{d}^{\rm{g}}+\ten{d}^{\rm{t}}
	\label{eq:d2s}
\end{equation}
The elastic part $\ten{d}^{\rm{e}}$ is given by isotropic hypo-elasticity  
$\ten{d}^e=\mathbb{M}:\overset{\triangledown}{\ten{\sigma}}$ where
$\mathbb{M}$ is the compliance tensor and
the symbol $\tiny \triangledown$ denotes the Jaumann rate of the Cauchy stress $\ten{\sigma}$, defined as
$\overset{\triangledown}{\ten{\sigma}}=\dot{\ten{\sigma}}+\ten{\sigma} \ten{\Omega}-\ten{\Omega} \ten{\sigma}$,
with $\ten{\Omega}$ the continuum spin. 
The plastic part of the rate of deformation has two contributions:
from glide, $\ten{d}^{\rm{g}}$, and twinning, $\ten{d}^{\rm{t}}$.
They are obtained from plastic potentials $\Phi^{\rm{g}}$ and $\Phi^{\rm{t}}$, respectively, as:
\begin{align}
	\ten{d}^{\rm{g}}=\dot{p}^{\rm{g}} \dfrac{\partial \Phi^{\rm{g}}}{\partial \ten{\sigma}} \quad \text{and} \quad 
	\ten{d}^{\rm{t}}=\dot{p}^{\rm{t}} \dfrac{\partial \Phi^{\rm{t}}}{\partial \ten{\sigma}}
	\label{eq:dgdt}
\end{align}
where $\dot{p}^{\rm{g}}, \dot{p}^{\rm{t}}$ are plastic multipliers and these are nothing
but the effective strain rates of the corresponding deformation modes.
The potentials $\Phi^{\rm{g}}$, $\Phi^{\rm{t}}$ in \cref{eq:dgdt} are expressed as:
\begin{align}
	\Phi^{\rm{g}} & =\bar{\sigma}^{\rm{g}}(\ten\sigma) - R^{\rm{g}}( p^{\rm{g}}, p^{\rm{t}}) \label{eq:Phig} \\
	\Phi^{\rm{t}} & =\bar{\sigma}^{\rm{t}}(\ten\sigma) - R^{\rm{t}}( p^{\rm{g}}, p^{\rm{t}}) \label{eq:Phit}
\end{align}
where $\bar{\sigma}^{\rm{g}}, \bar{\sigma}^{\rm{t}}$ are linear homogeneous functions of the Cauchy stress and
$R^{\rm{g}}$ and $R^{\rm{t}}$ are the glide and twinning yield strengths, respectively.
The glide and twinning strengths in \cref{eq:Phig,eq:Phit} are taken as functions of both effective strains 
$p^{\rm{g}}$ and $p^{\rm{t}}$.
These rates are calculated using Norton's law:
\begin{align}
	\dot{p}^{\alpha}=\dot{\epsilon}_0^{\alpha} \left(\frac{\bar{\sigma}^{\alpha}}{R^{\alpha}}\right)^{n^{\alpha}}
	\label{eq:norton}
\end{align}
for $\alpha \in \{\rm{g}, \rm{t}\}$, where 
$\dot{\epsilon}_0^{\alpha}$ is a reference strain rate and $n^{\alpha}$ 
is a strain rate sensitivity parameter. 

\subsubsection{Slip} \label{sss:2Sglide}
Dislocation glide in HCP materials leads to strong anisotropy.
Inspired by a versatile, non-quadratic yield criterion \citep{Barlat91},
the effective stress $\bar{\sigma}^{\rm{g}}$ entering \cref{eq:Phig} is given by:
\begin{equation}
	\bar{\sigma}^{\rm{g}} = \left[ \dfrac{1}{2}\left( |s_1^{\rm{g}}-s_2^{\rm{g}}|^{a^{\rm{g}}}+ 
		|s_2^{\rm{g}}-s_3^{\rm{g}}|^{a^{\rm{g}}}+ |s_3^{\rm{g}}-s_1^{\rm{g}}|^{a^{\rm{g}}}\right)\right]^{1/a^{\rm{g}}}
	\label{eq:sbarg}
\end{equation}
with $a^{\rm{g}}$ an exponent that controls the shape of the yield surface
in the rate-independent limit.
Here, $s_i^{\rm{g}}$ denotes an eigenvalue of a transformed stress deviator 
$\ten{s}^{\rm{g}}$, defined as:
\begin{align}
	\ten{s}^{\rm{g}}=\mathbb{L}^{\rm{g}}:\ten{\sigma}
	\label{eq:sg}
\end{align}
where $\mathbb{L}^{\rm{g}}$ is a fourth-order tensor.
Accounting for minor and major symmetries, its reduced form
in Voigt's notation is given by:
\begin{equation}
[\ten{L}^{\rm{g}}]=
\small 
	\begin{pmatrix}
	\frac{1}{3}\left(l_{\rm{TT}}^{\rm{g}}+l_{\rm{SS}}^{\rm{g}}\right) & -\frac{1}{3} l_{\rm{SS}}^{\rm{g}} & -\frac{1}{3} l_{\rm{TT}}^{\rm{g}} & 0 & 0 & 0 \\[6pt]
	-\frac{1}{3} l_{\rm{SS}}^{\rm{g}} & \frac{1}{3}\left(l_{\rm{SS}}^{\rm{g}}+l_{\rm{LL}}^{\rm{g}}\right) & -\frac{1}{3} l_{\rm{LL}}^{\rm{g}} & 0 & 0 & 0 \\[6pt]
	-\frac{1}{3} l_{\rm{TT}}^{\rm{g}} & -\frac{1}{3} l_{\rm{LL}}^{\rm{g}} & \frac{1}{3}\left(l_{\rm{LL}}^{\rm{g}}+l_{\rm{TT}}^{\rm{g}}\right) & 0 & 0 & 0 \\[6pt]
	0 & 0 & 0 & l_{\rm{LT}}^{\rm{g}} & 0 & 0 \\[6pt]
	0 & 0 & 0 & 0 & l_{\rm{TS}}^{\rm{g}} & 0 \\[6pt]
	0 & 0 & 0 & 0 & 0 & l_{\rm{SL}}^{\rm{g}} \\[6pt]
	\end{pmatrix}
	\label{eq:Lg}
\end{equation}
with L, T, and S being the principal axes along the longitudinal (rolling), transverse, and short transverse directions, while LT, TS, and SL are the $45^\circ$ off-axes directions.

\subsubsection{Twinning} \label{sss:2Stwin}
Deformation by twinning introduces a strength-differential effect along with anisotropy.
To represent this, a non-quadratic yield criterion \citep{Cazacu06} is used as basis to express
the effective stress $\bar{\sigma}^{\rm{t}}$ entering \cref{eq:Phit} as:
\begin{equation}
\bar{\sigma}^{\rm{t}} = \left[ \left(|s_1^{\rm{t}}|-k s_1^{\rm{t}} \right)^{a^{\rm{t}}}
	+ \left( |s_2^{\rm{t}}|-k s_2^{\rm{t}} \right)^{a^{\rm{t}}}
	+ \left(|s_3^{\rm{t}}|-k s_3^{\rm{t}} \right)^{a^{\rm{t}}}\right]^{1/a^{\rm{t}}}
	\label{eq:sbart}
\end{equation}
where the parameter $k$ ($-1<k<1$) accounts for the \textit{initial} strength differential effect.
For $k >0$, the yield stress in tension is greater than in compression.
As in \cref{eq:sbarg}, $a^{\rm{t}}$ controls the shape of the yield surface
in the original formulation of \cite{Cazacu06},
and $s_i^{\rm{t}}$ denotes an eigenvalue of a transformed stress deviator $\ten{s}^{\rm{t}}$, defined as:
\begin{align}
    \ten{s}^{\rm{t}}=\mathbb{L}^{\rm{t}}:\mathbb{J}:\ten{\sigma}
    \label{eq:st}
\end{align}
where $\mathbb{J}$ is the deviatoric projector and $\mathbb{L}^{\rm{t}}$ is a fourth-order tensor
the reduced form of which is:
\begin{equation}
	[\ten{L}^{\rm{t}}] =  
	\begin{pmatrix}
		l_{\rm{LL}}^{\rm{t}} & \mathcal{L}_{\rm{LT}}^{\rm{t}} & \mathcal{L}_{\rm{SL}}^{\rm{t}} & 0 & 0 & 0 \\[6pt]
		\mathcal{L}_{\rm{LT}}^{\rm{t}} & l_{\rm{TT}}^{\rm{t}} & \mathcal{L}_{\rm{TS}}^{\rm{t}} & 0 & 0 & 0 \\[6pt]
		\mathcal{L}_{\rm{SL}}^{\rm{t}} & \mathcal{L}_{\rm{TS}}^{\rm{t}} & l_{\rm{SS}}^{\rm{t}} & 0 & 0 & 0 \\[6pt]
	0 & 0 & 0 & l_{\rm{LT}}^{\rm{t}} & 0 & 0 \\[6pt]
	0 & 0 & 0 & 0 & l_{\rm{TS}}^{\rm{t}} & 0 \\[6pt]
	0 & 0 & 0 & 0 & 0 & l_{\rm{SL}}^{\rm{t}} \\[6pt]
	\end{pmatrix}
	\label{eq:Lt}
\end{equation}

\subsubsection{Strain hardening} \label{sss:2Shard}
The yield strengths, $R^{\rm g}$ and $R^{\rm t}$, entering the potentials in \cref{eq:Phig,eq:Phit}, respectively,
are taken as a combination of the Voce-like laws for self-hardening and linear laws for latent hardening.
\begin{align}
	R^{\rm g} = R^{\rm g}_0+Q_1^{\rm g}(1-\exp(-b_1^{\rm g} p^{\rm{g}}))+Q_2^{\rm g}(1-\exp(-b_2^{\rm g} p^{\rm{g}}))+ \mathcal{H}^{\rm tg} p^{\rm{t}} \label{eq:Rg} \\[2ex]
	R^{\rm t} = R^{\rm t}_0+Q_1^{\rm t}(1-\exp(-b_1^{\rm t} p^{\rm{g}}))+Q^{\rm t}(\exp(b^{\rm t} p^{\rm{t}})-1) + H^{\rm t} p^{\rm{t}}+ \mathcal{H}^{\rm gt} p^{\rm{g}}
	\label{eq:Rt}
\end{align}
For the glide yield strength, five parameters ($R^{\rm g}_0, Q_1^{\rm g}, b_1^{\rm g}, Q_2^{\rm g}, b_2^{\rm g}$)
set the initial yield and self-hardening.
For the twinning yield strength, six parameters ($R^{\rm t}_0, Q_1^{\rm t}, b_1^{\rm t}, Q^{\rm t}, b^{\rm t}, H^{\rm t}$)
set the initial yield and self-hardening.
Two parameters $\mathcal{H}^{\rm gt}$, $\mathcal{H}^{\rm tg}$ account for latent hardening between the two deformation modes.
%
\subsection{Three-surface model} \label{sec:3S}
In the three-surface (3S) model, the total rate of the deformation is written as:
\begin{equation}
    \ten{d}  = \ten{d}^{\rm e} + \ten{d}^{\rm s} + \ten{d}^{\rm h} + \ten{d}^{\rm t}
    \label{eq:d3s}
\end{equation}
where the elastic part $\ten{d}^{\rm{e}}$ is as above.
Here, the plastic part of the rate-of-deformation has three contributions:
$\ten{d}^{\rm{s}}$ (soft glide $\in$ \{basal $\langle a \rangle$\}), 
$\ten{d}^{\rm{h}}$ (hard glide $\in$ \{prismatic $\langle a \rangle$, pyramidal $\langle a \rangle$, pyramidal $\langle c{+}a \rangle$\}), 
and $\ten{d}^{\rm{t}}$ (twinning).
They are obtained from plastic potentials $\Phi^{\rm{s}}$,  $\Phi^{\rm{h}}$ and $\Phi^{\rm{t}}$ as:
\begin{align}
	\ten{d}^{\rm{s}}
	=\dot{p}^{\rm{s}} \dfrac{\partial \Phi^{\rm{s}}}{\partial \ten{\sigma}} \quad 
	; \quad
	\ten{d}^{\rm{h}}
	=\dot{p}^{\rm{h}} \dfrac{\partial \Phi^{\rm{h}}}{\partial \ten{\sigma}} 
	\quad 
	; \quad 
	\ten{d}^{\rm{t}}
	=\dot{p}^{\rm{t}} \dfrac{\partial \Phi^{\rm{t}}}{\partial \ten{\sigma}}
	\label{eq:dsdhdt}
\end{align}
where $\dot{p}^{\alpha}$ is the effective plastic strain rate for
the deformation mode $\alpha \in \{{\rm s}, {\rm h}, {\rm t}\}$ that follows \cref{eq:norton}. 
The elastic rate $\ten{d}^{\rm{e}}$ in \cref{eq:d3s} is described the same as in \cref{eq:d2s}.
The potentials in \cref{eq:dsdhdt} are expressed as:
%
\begin{equation} \label{eq:3SPhi}
	\Phi^{\alpha}  =
	\bar{\sigma}^{\alpha}(\ten\sigma) - R^{\alpha} (p^{\rm{s}}, p^{\rm{h}}, p^{\rm{t}}) \qquad \forall \alpha \in \{\rm s, h, t\} 
\end{equation}
where the effective stress $\bar{\sigma}^{\alpha}$ is a linear homogeneous function of the Cauchy stress.
The current yield strength $R^{\alpha}$ 
is taken as a function of all effective strains
$p^{\rm{s}}$, $p^{\rm{h}}$, and $p^{\rm{t}}$.

In what follows, only the effective stresses of the soft glide and hard glide modes are presented;
the effective stress of the twinning mode is the same as that presented in \cref{sss:2Stwin}.
%
\subsubsection{Dislocation glide} \label{sss:3Sglide}
While the 2S model uses the \citet{Barlat91} 
model to describe the yield due to the dislocation glide,
the 3S model adapts the \cite{Hill48} criterion for the soft and hard glide modes. 
The effective stress $\bar\sigma^\alpha$ that enters \cref{eq:3SPhi} is given by:
\begin{align}
	\bar\sigma^\alpha=\left[\frac32 \ten s^\alpha: \ten s^\alpha \right]^{1/2}
	\qquad \forall \alpha \in \{\rm s, h\}
\end{align}
where $\ten s^\alpha$ is the deviatoric stress:
\begin{align}
	\ten{s}^{\alpha}=\mathbb{J}:\mathbb{H}^{\alpha}:\ten{\sigma} 
	\qquad \forall \alpha \in \{\rm s, h\}
	\label{eq:shill}
\end{align}
and $\mathbb{H}^\alpha$ is the fourth order Hill tensor.
As reformulated by \citet{Benzerga01}, $\mathbb{H}^\alpha$ has the reduced form:
\begin{equation}
	[\mathbb{H}^{\alpha}] =  
	\begin{pmatrix}
	h_{\rm{LL}}^{\alpha} & 0 & 0 & 0 & 0 & 0 \\[6pt]
	0 & h_{\rm{TT}}^{\alpha} & 0 & 0 & 0 & 0 \\[6pt]
	0 & 0 & h_{\rm{SS}}^{\alpha} & 0 & 0 & 0 \\[6pt]
	0 & 0 & 0 & h_{\rm{LT}}^{\alpha} & 0 & 0 \\[6pt]
	0 & 0 & 0 & 0 & h_{\rm{TS}}^{\alpha} & 0 \\[6pt]
	0 & 0 & 0 & 0 & 0 & h_{\rm{SL}}^{\alpha} \\[6pt]
	\end{pmatrix}
\qquad \forall \alpha \in \{\rm s, h\}
	\label{eq:H}
\end{equation}
\subsubsection{Strain hardening} \label{sss:3Shard}
The flow strengths of individual mechanisms, $R^{\rm s}$, $R^{\rm h}$ and $R^{\rm t}$ are
combinations of linear, power, and Voce-like laws for self-hardening and linear laws for latent hardening. 
\begin{align}&R^{\rm s} 
= R_0^{\rm s} \left(1 + \dfrac{\mathcal{H}^{\rm s}}{R_0^{\rm s}} p^{\rm s}\right)^n + \mathcal{H}^{\rm hs} p^{\rm h} + 
\mathcal{H}^{\rm ts} p^{\rm t} \label{eq:Rs} \\[2ex]
&R^{\rm h} 
= R_0^{\rm h} + \mathcal{H}^{\rm h} p^{\rm h}  + 
Q^{\rm h} \left( 1 - \text{exp}\left( - b^{\rm h} p^{\rm h} \right) \right)
+ \mathcal{H}^{\rm sh} p^{\rm s} + \mathcal{H}^{\rm th} p^{\rm t} \label{eq:Rh} \\[2ex]
&R^{\rm t} 
= R_0^{\rm t} + \mathcal{H}^{\rm t} p^{\rm t} + 
Q^{\rm t} \left( \text{exp}\left( b^{\rm t} (p^{\rm t} - p^{\rm c}) \right) - 1 \right)
+ \mathcal{H}^{\rm st} p^{\rm s} + \mathcal{H}^{\rm ht} p^{\rm h}
\label{eq:Rt1}
\end{align}
For the soft-glide yield strength, 
three parameters ($R^{\rm s}_0, \mathcal{H}^{\rm s}, n$) set the initial yield and self-hardening.
For the hard-glide yield strength,
four parameters ($R^{\rm h}_0, \mathcal{H}^{\rm h}, Q^{\rm h},b^{\rm h}$) set the initial yield and self-hardening.
For the twinning yield strength, 
five parameters ($R^{\rm t}_0, \mathcal{H}^{\rm t}, Q^{\rm t},b^{\rm t}, p^{\rm c}$) set the initial strength and self-hardening.
Six parameters ($\mathcal{H}^{\rm hs}, \mathcal{H}^{\rm ts}, \mathcal{H}^{\rm sh}, \mathcal{H}^{\rm th}, \mathcal{H}^{\rm st}, \mathcal{H}^{\rm ht}$)
account for latent hardening between the three deformation modes.
Note that, the expression for $R^{\rm t}$ in \cref{eq:Rt1}
for the 3S model is different from the one in \cref{eq:Rt} for the 2S model.

\subsection{Model identification} \label{ss:}

There are a total of 35 (for 2S), and 47 (for 3S) constitutive parameters 
in each model, \cref{tab:2S_3S_parameters}.
\begin{table}[t]
\centering
\caption{List of constitutive parameters in the 2S and 3S models.}
\renewcommand{\arraystretch}{1.25}
\setlength{\tabcolsep}{4pt}
\begin{tabular}{lcc}
\hline
\textbf{Parameter group} & \textbf{2S model} & \textbf{3S model} \\ 
\hline
Elasticity &
2: $E$, $\nu$ &
2: $E$, $\nu$ \\
\hline

Rate sensitivity &
4: $\dot{\epsilon}_0^{\alpha}$, $n^{\alpha}$ ($\alpha={\rm g,t}$) &
6: $\dot{\epsilon}_0^{\alpha}$, $n^{\alpha}$ ($\alpha={\rm s,h,t}$) \\
\hline

Glide anisotropy &
6 (5 ind.): $l_{\rm XY}^{\rm g}$, $a^{\rm g}$ &
6 soft glide (5 ind.): $h_{\rm XY}^{\rm s}$ \\
& & 6 hard glide (5 ind.): $h_{\rm XY}^{\rm h}$\\ 
\hline

Glide hardening &
6: $R^{\rm g}_0$, $Q_1^{\rm g}$, $b_1^{\rm g}$, $Q_2^{\rm g}$, $b_2^{\rm g}$, $\mathcal{H}^{\rm tg}$ &
5 soft glide: $R^{\rm s}_0$, $\mathcal{H}^{\rm s}$, $n$, $\mathcal{H}^{\rm hs}$, $\mathcal{H}^{\rm ts}$ \\
& & 6 hard glide: $R^{\rm h}_0$, $\mathcal{H}^{\rm h}$, $Q^{\rm h}$, $b^{\rm h}$, $\mathcal{H}^{\rm sh}$, $\mathcal{H}^{\rm th}$ \\
\hline

Twinning anisotropy &
9 (8 ind.): $l_{\rm XY}^{\rm t}$, $\mathcal{L}_{\rm XY}^{\rm t}$, $a^{\rm t}$, $k$ &
9 (8 ind.): $l_{\rm XY}^{\rm t}$, $\mathcal{L}_{\rm XY}^{\rm t}$, $a^{\rm t}$, $k$ \\
\hline


Twinning hardening &
7: $R^{\rm t}_0$, $Q_1^{\rm t}$, $b_1^{\rm t}$, $Q^{\rm t}$, $b^{\rm t}$, $H^{\rm t}$, $\mathcal{H}^{\rm gt}$ &
7: $R^{\rm t}_0$, $\mathcal{H}^{\rm t}$, $Q^{\rm t}$, $b^{\rm t}$, $p^{\rm c}$, $\mathcal{H}^{\rm st}$, $\mathcal{H}^{\rm ht}$ \\

\hline
\end{tabular}
\label{tab:2S_3S_parameters}
\end{table}
Some are held fixed during the process of parameter identification.
The values of the elasticity parameters $E=43$ GPa, $\nu=0.29$ are obtained from the literature.
The constants entering Norton's law, \cref{eq:norton}, 
are taken to be the same for all yielding modes: 
$\dot{\epsilon}^\alpha=0.001 \ {\rm s}^{-1}, n^\alpha=50$ for $\alpha \in \{ {\rm g, s, h, t} \}$.
We also fix $a^{\rm g} = a^{\rm t} = 4$ in \cref{eq:sbarg,eq:sbart}, respectively.
In addition, the hardening laws are simplified by
setting $Q_1^{\rm t}$ and $\mathcal{H}^{\rm t}=0$ in \cref{eq:Rt} for the 2S model, and 
$\mathcal{H}^{\rm ts} =0$ in \cref{eq:Rs}
and $\mathcal{H}^{\rm th} = 0$ in \cref{eq:Rh} for the 3S model.
This reduces the number of parameters to be identified to 24 and 35 for the 2S and 3S models, respectively.

The parameters for the 2S model were obtained using the calibration procedure described in \cite{Vignesh23}; 
likewise, the 3S model parameters were obtained using the procedure described in \cite{Indurkar23}.
Details aside, the two approaches can be summarized as follows.
First, a set of initial parameters is chosen for which
twelve material point simulations are performed using a single finite element:
six uniaxial tension and six compression simulations. 
These include three principal material axes (L, T, S) and
three off-axis (45$^\circ$) simulations along the LT, TS, and SL directions.
Next, the {\it distance}, in terms of an error norm, between the 2S / 3S simulations
and the reference (\cite{Baweja23}) data is evaluated using a cost function.
Based on the error, a new set of parameters is determined using
an optimization algorithm; 
see \cref{app:2S} and \cref{app:3S} for
specific details for the 2S and 3S models, respectively.
This process is repeated until the cost function reaches a minimum value or a further minimization is not possible. 

The cost function, $\mathcal{E}$, for both models is calculated based {\it only} on the stress--strain curves, defined as:
\begin{equation}
\mathcal{E}=\sum_{i=1}^{N} e^i ;
\qquad
e^i=\frac{1}{\Eeq^i} \left(\frac{1}{\max({\sigma^i_{\rm{eq,REF}}})}\right)^2
\int_0^{\Eeq^i} 
\left( \sigma_{\rm{eq,REF}}^i - \seq^i \right)^2 d\Eeq^i
\qquad
\label{eq:cost}
\end{equation}
where $N$ denotes the total number of simulations per material (here 12).
For each simulation, $e^i$ is a measure of the difference between the equivalent stress prediction of a model, $\seq^i$,
and the equivalent stress of the reference data, $\sigma_{\rm{eq,REF}}^i$.
For uniaxial loading along a global $y$ direction, as here, the expression for the
equivalent stress is simply the uniaxial stress along the $y$ direction $\seq^i=\sigma_{yy}^i$.
Also, the expression for the equivalent strain reduces to $\Eeq^i=\sqrt{2/9}\sqrt{(E_{xx}^i-E_{yy}^i)^2+(E_{yy}^i-E_{zz}^i)^2+(E_{zz}^i-E_{xx}^i)^2}$.
Note that the error $e^i$ \cref{eq:cost} is normalized by a maximum value of equivalent stress in the reference data, denoted $\max({\sigma^i_{\rm{eq,REF}}})$.
The identified model parameters are listed in \cref{app:2S} and \cref{app:3S} for the 2S and 3S models, respectively.

\begin{figure}[t!]
	\centering
        \vspace*{2.5ex} 
	\begin{picture}(0,0)
	\thicklines
	\put(-195,5){\textsf{Strong Textures}}
	\put(-50,5){\textsf{Intermediate Textures}} 
	\put(125,5){\textsf{Weak Textures}}
	\end{picture}
        \vspace*{-5ex}
	\setlength{\unitlength}{1pt} 
	\begin{picture}(0,0)
	\thicklines
	\put(105,-348){\vector(1,0){55}}
	\put(105,-348){\vector(0,1){55}}
	\put(160,-348){$\textsf{T}$}
	\put(102,-290){\rotatebox{0}{$\textsf{L}$}}	

 
	\end{picture}	
 
	\begin{multicols}{3}
	\centering
        \vspace*{-2ex}	
		\subfloat[b][\centering {\bf A ($[20^{\circ},10^{\circ},0^{\circ}]$)}]{\includegraphics[scale=0.20]{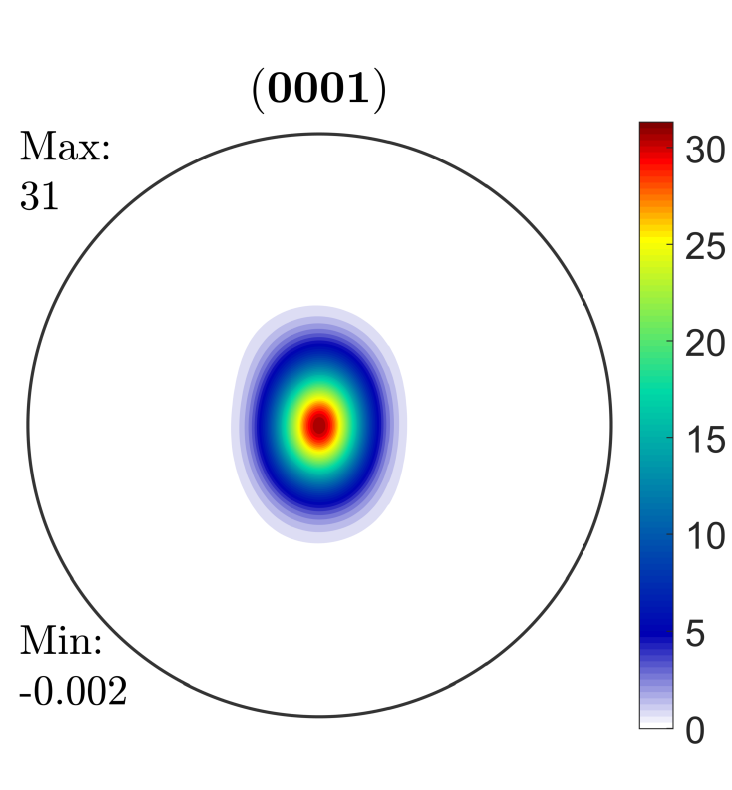}
			\includegraphics[scale=0.20]{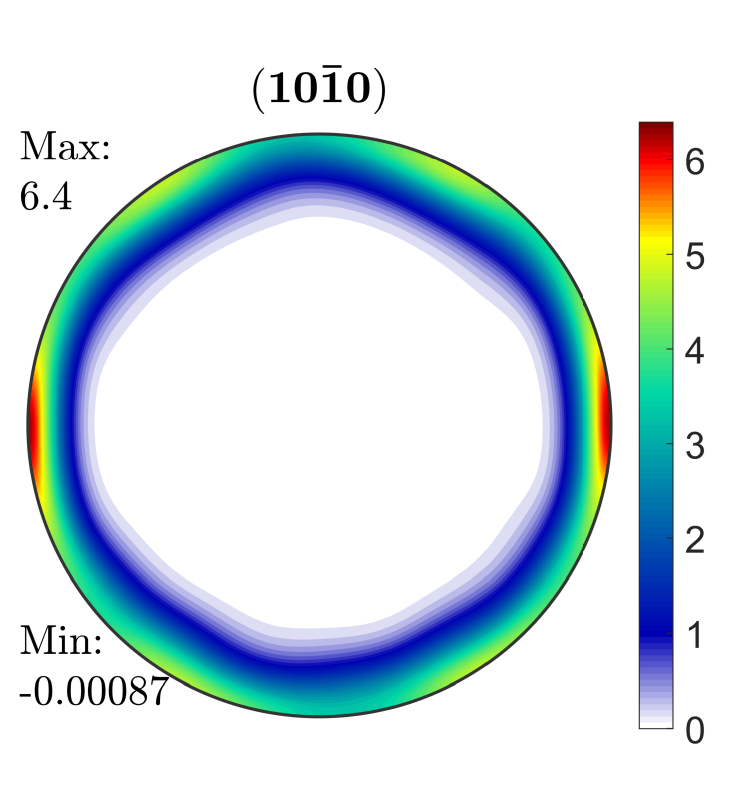}\label{f:in_TXTA2}}\par
       \vspace*{-2ex}				
		\subfloat[b][\centering {\bf B ($[10^{\circ},15^{\circ},0^{\circ}]$)}]{\includegraphics[scale=0.20]{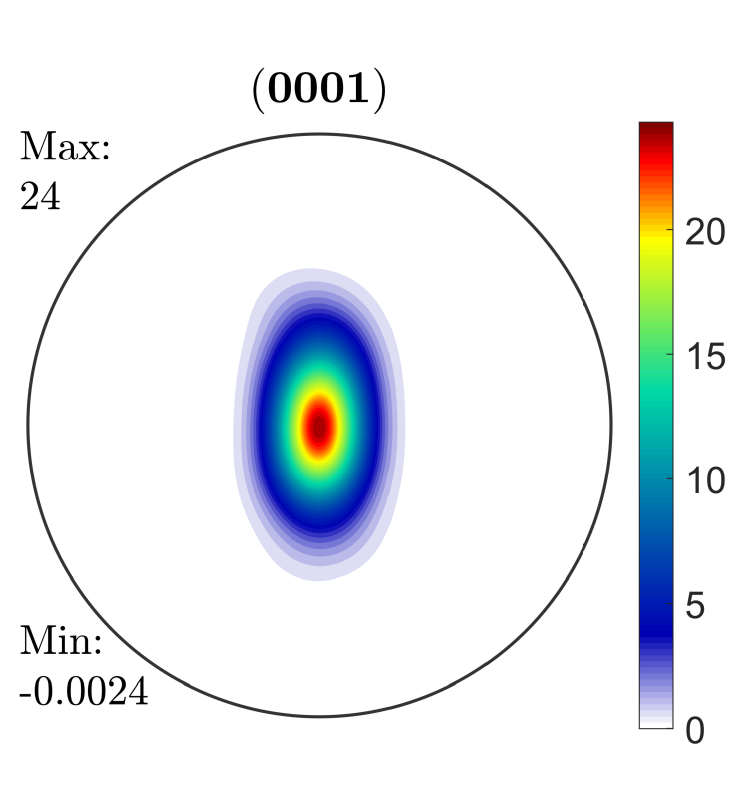}
			\includegraphics[scale=0.20]{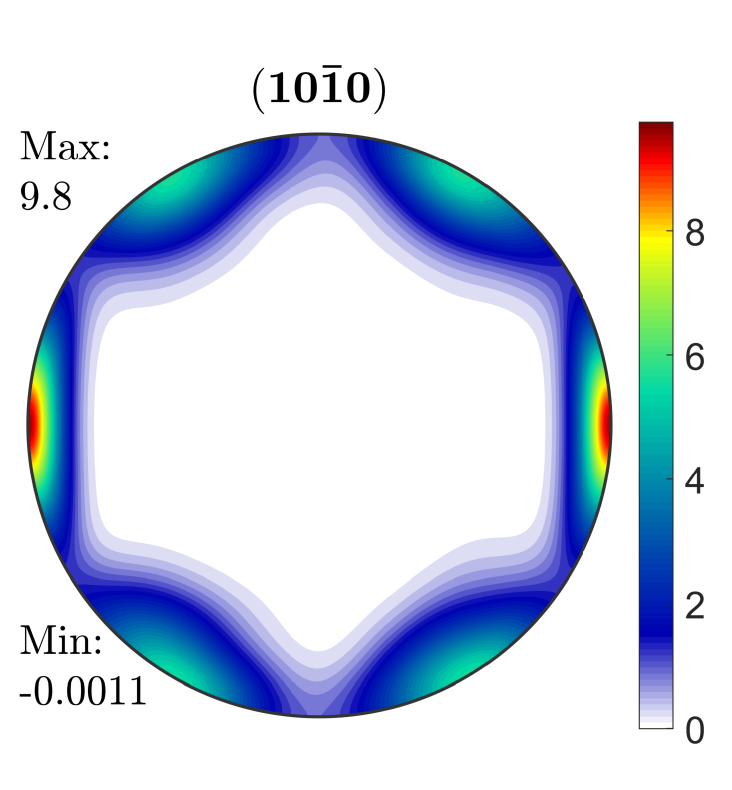}\label{f:in_TXTB2}}\par
       \vspace*{-2ex}				
		\subfloat[b][\centering {\bf C ($[20^{\circ},20^{\circ},0^{\circ}]$)}]{\includegraphics[scale=0.20]{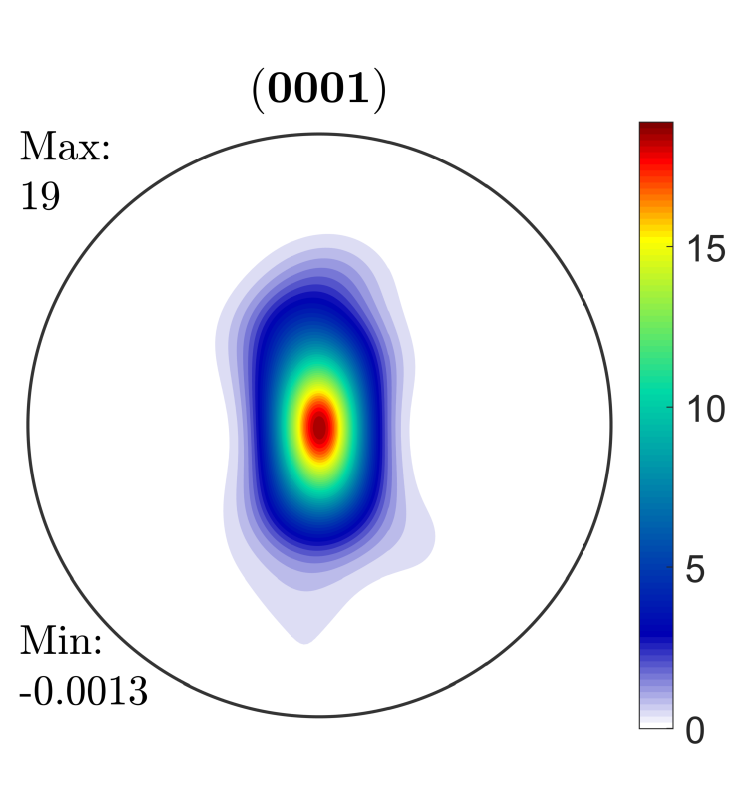}
			\includegraphics[scale=0.20]{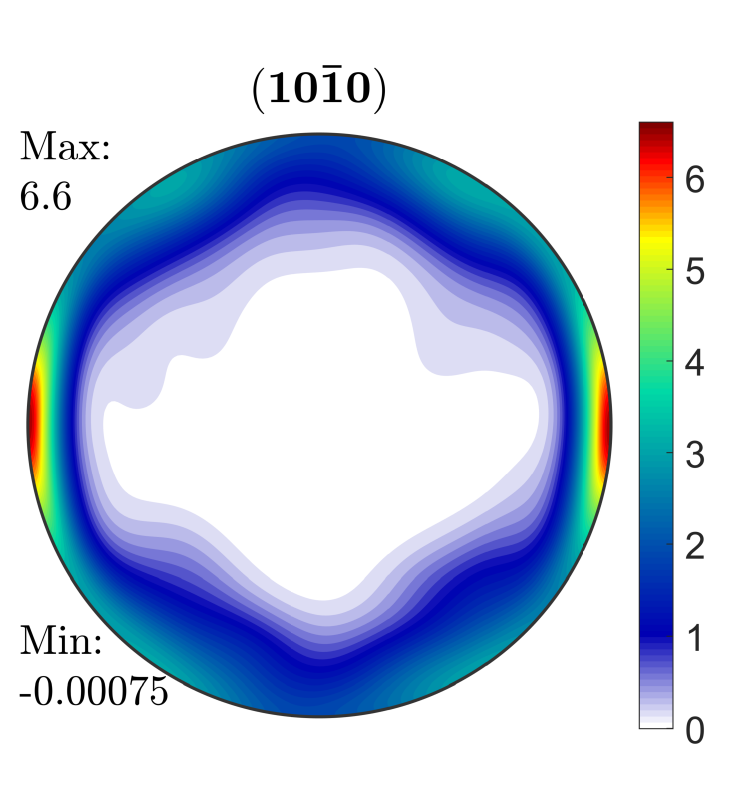}\label{f:in_TXTC2}}\par
       \vspace*{-2ex}	
		\subfloat[b][\centering {\bf D ($[30^{\circ},10^{\circ},0^{\circ}]$)}]{\includegraphics[scale=0.20]{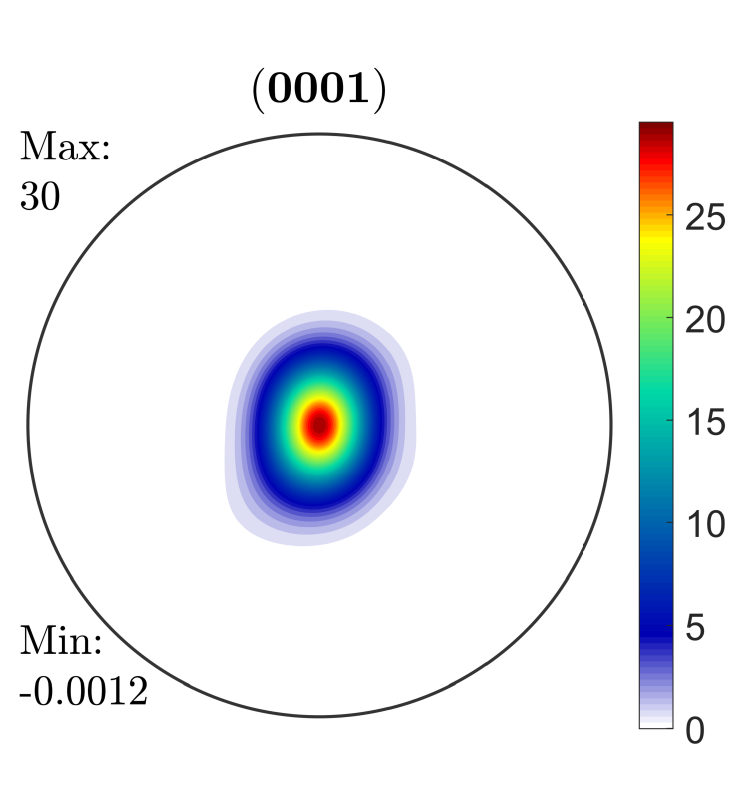}
			\includegraphics[scale=0.20]{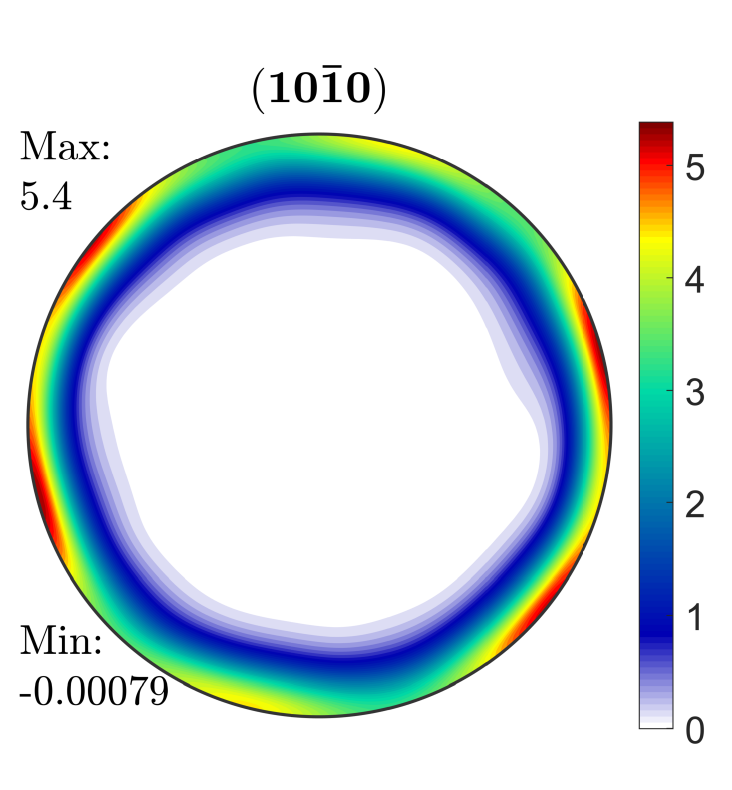}\label{f:in_TXTD2}}
       \vspace*{-2ex}	
		\subfloat[b][\centering {\bf E ($[30^{\circ},20^{\circ},0^{\circ}]$)}]{\includegraphics[scale=0.20]{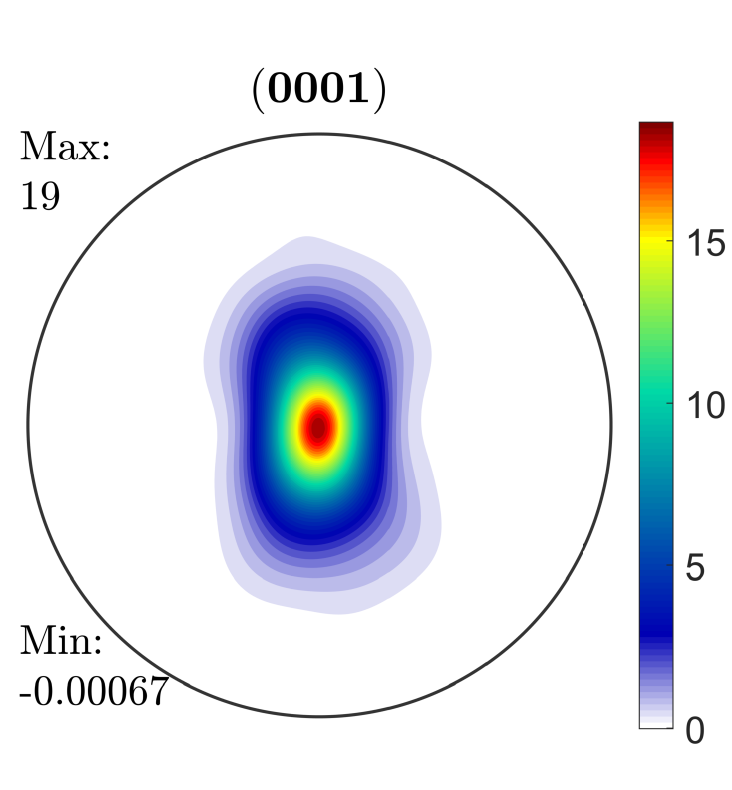}
			\includegraphics[scale=0.20]{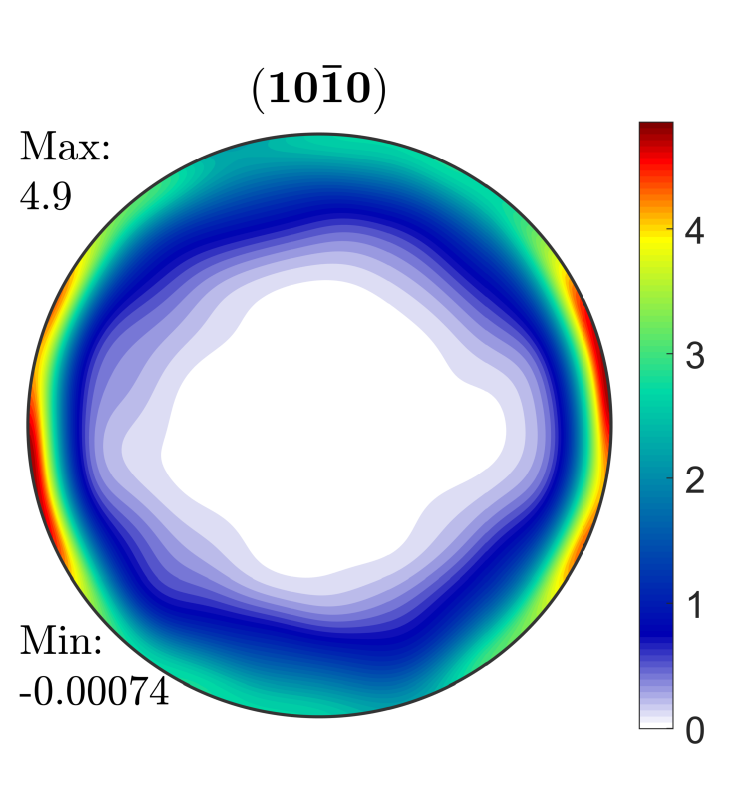}\label{f:in_TXTE2}}\par 
       \vspace*{-2ex}	
		\subfloat[b][\centering {\bf F ($[30^{\circ},20^{\circ},10^{\circ}]$)}]{\includegraphics[scale=0.20]{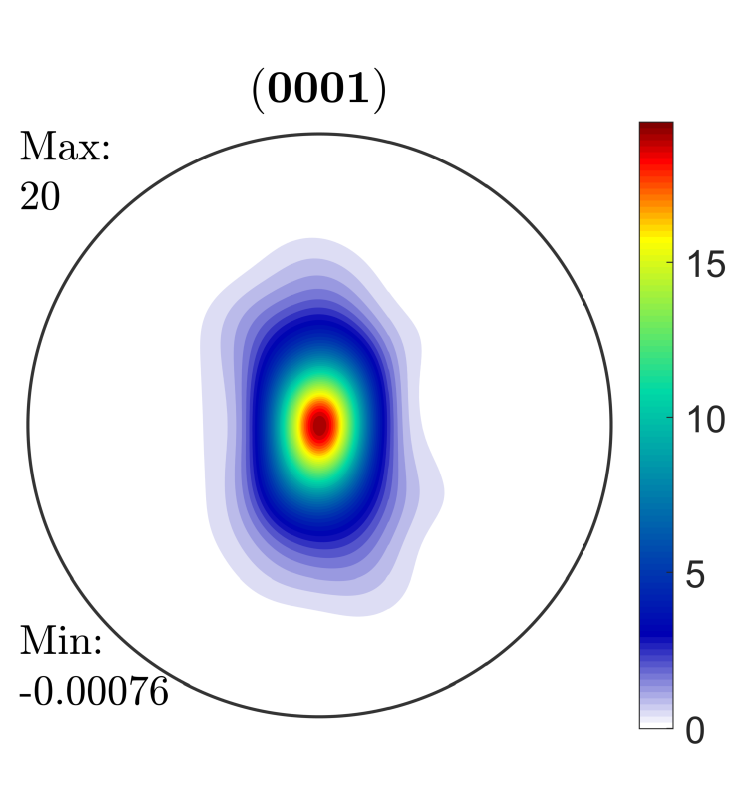}
			\includegraphics[scale=0.20]{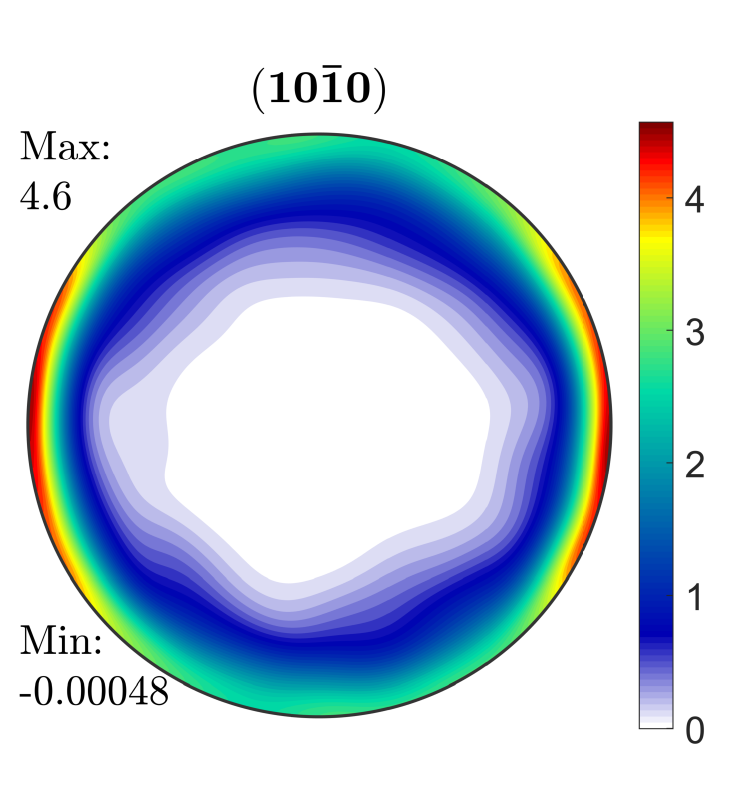}\label{f:in_TXTF2}}\par	 
       \vspace*{-2ex}	
		\subfloat[b][\centering {\bf G ($[30^{\circ},20^{\circ},20^{\circ}]$)}]{\includegraphics[scale=0.20]{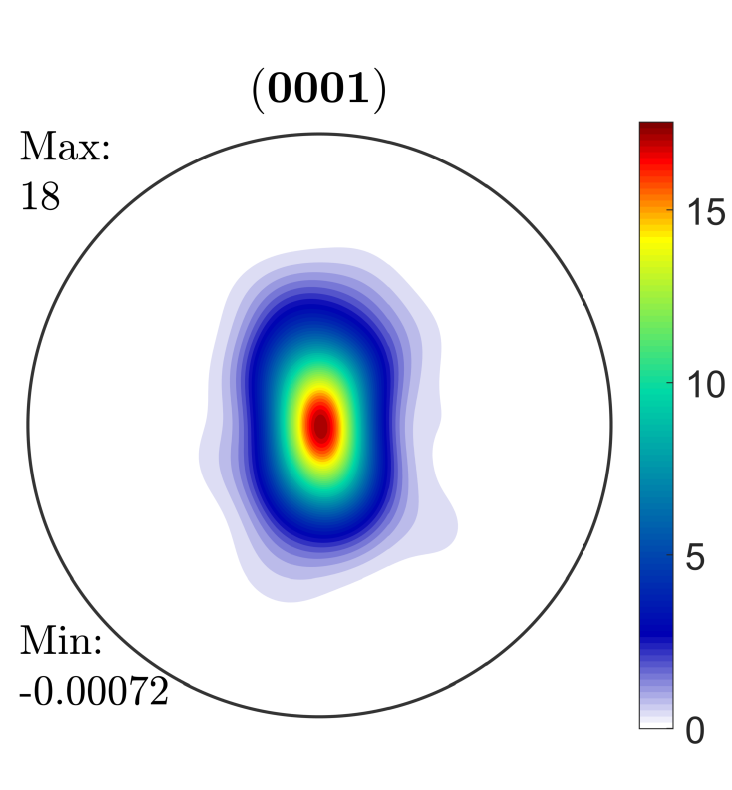}
			\includegraphics[scale=0.20]{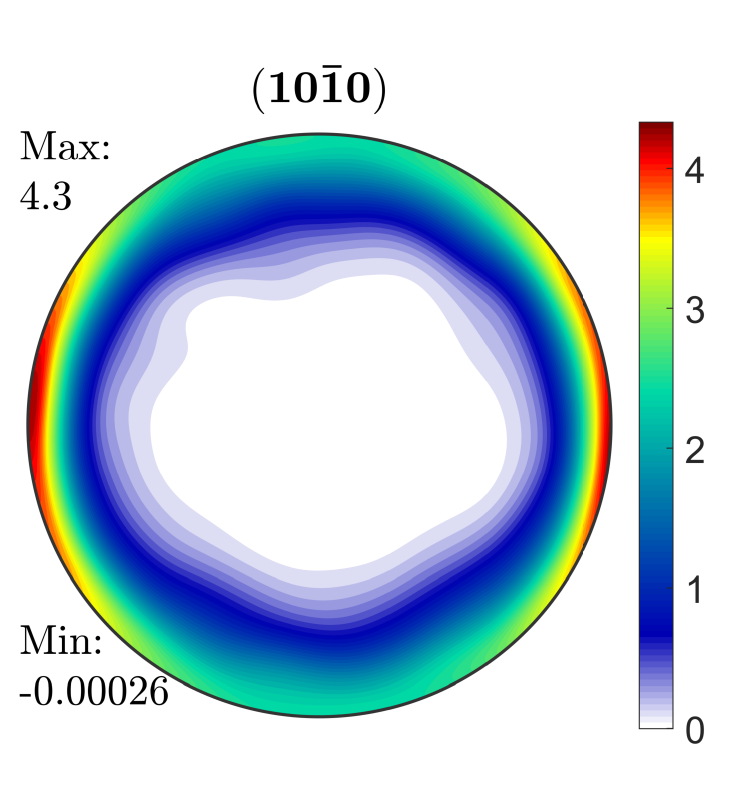}\label{f:in_TXTG2}}\par	
       \vspace*{-2ex}	
		\subfloat[b][\centering {\bf H ($[30^{\circ},30^{\circ},30^{\circ}]$)}]{\includegraphics[scale=0.20]{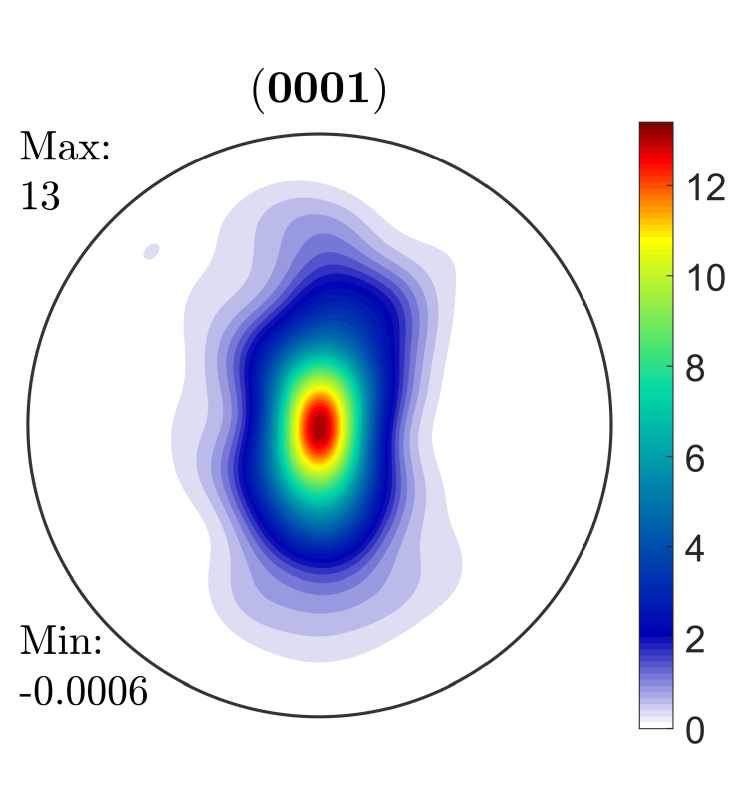}
			\includegraphics[scale=0.20]{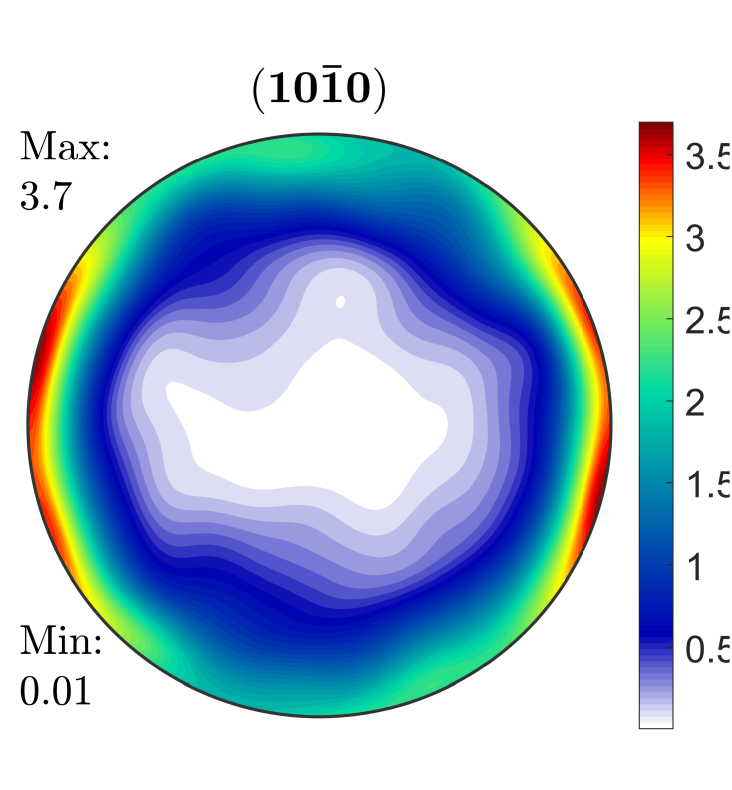}\label{f:in_TXTH2}}
       \vspace*{-2ex}	
		\subfloat[b][\centering {\bf I ($[30^{\circ},45^{\circ},30^{\circ}]$)}]{\includegraphics[scale=0.20]{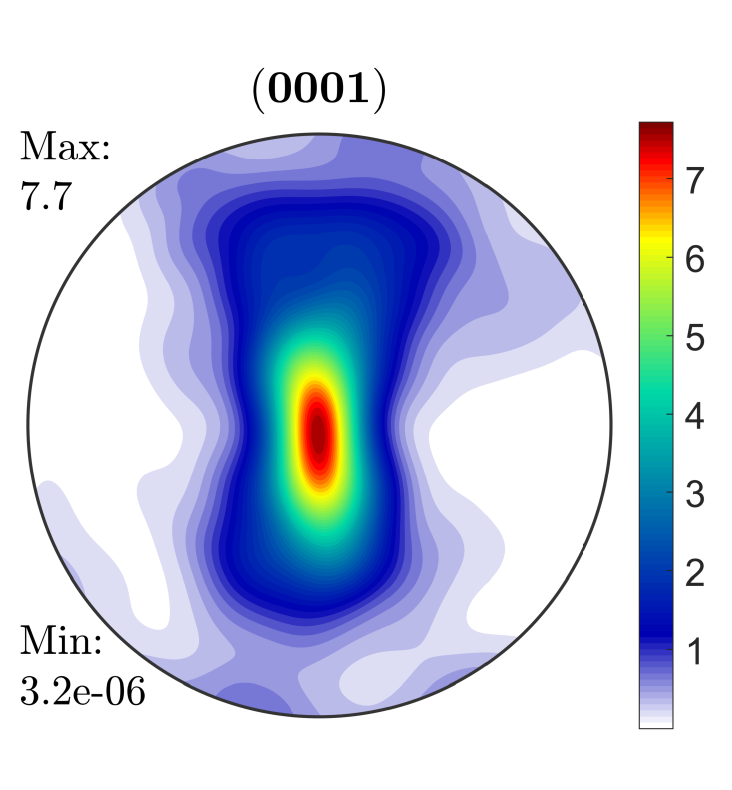}
			\includegraphics[scale=0.20]{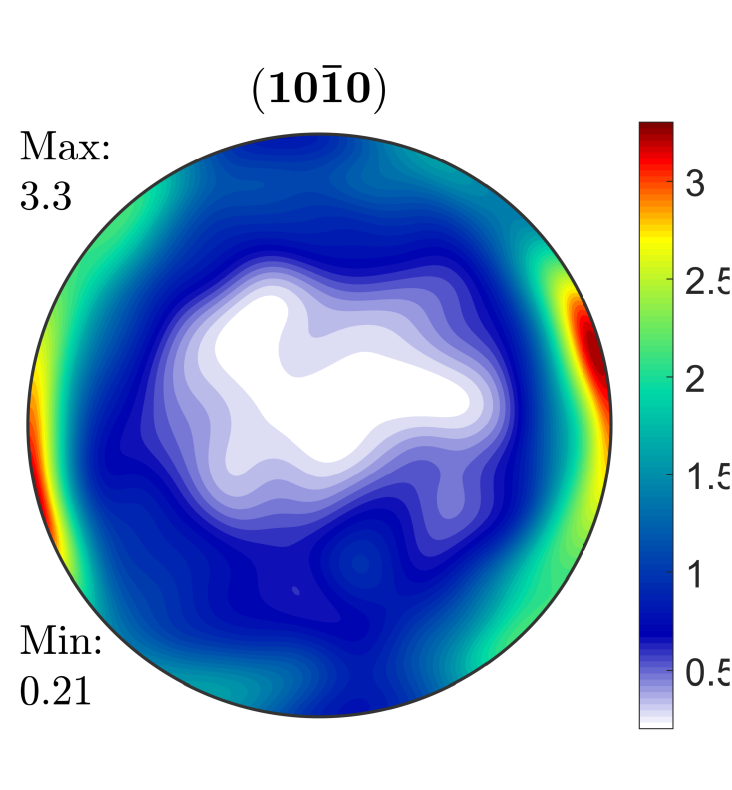}\label{f:in_TXTI2}}\par
       \vspace*{-2ex}	
		\subfloat[b][\centering {\bf J ($[45^{\circ},60^{\circ},45^{\circ}]$)}]{\includegraphics[scale=0.20]{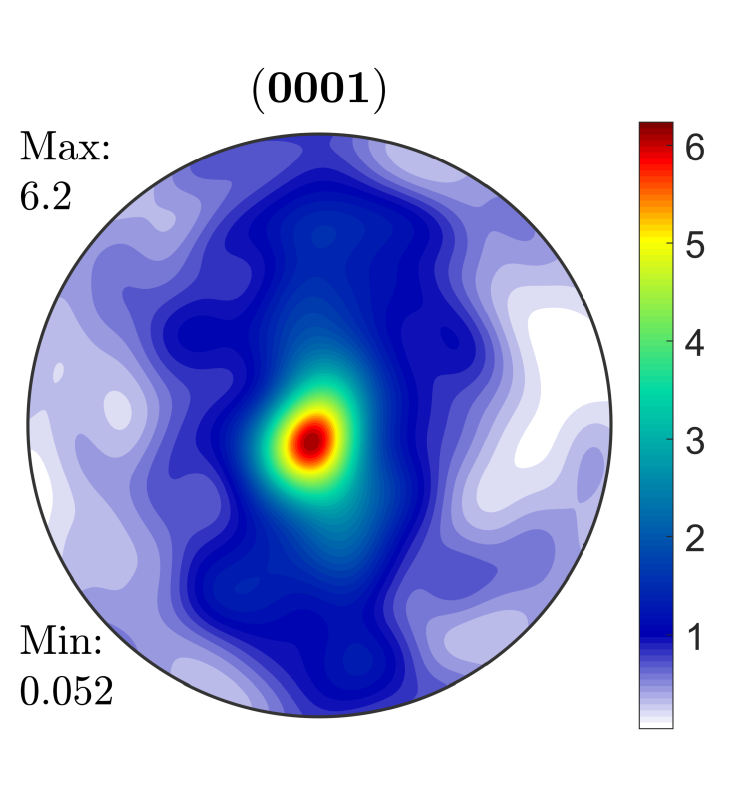}
			\includegraphics[scale=0.20]{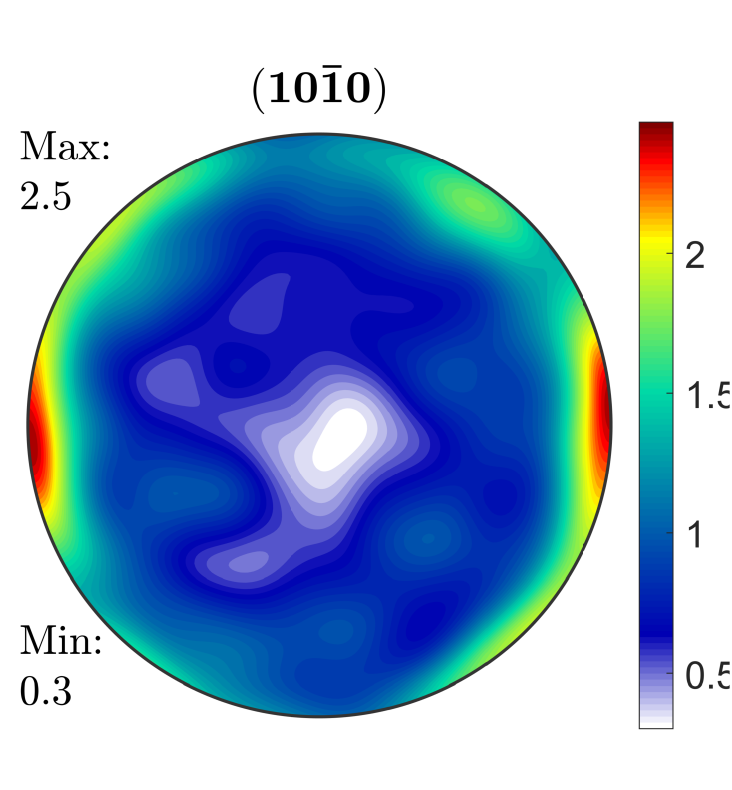}\label{f:in_TXTJ2}}\par	 
       \vspace*{-2ex}	
		\subfloat[b][\centering {\bf K ($[45^{\circ},75^{\circ},45^{\circ}]$)}]{\includegraphics[scale=0.20]{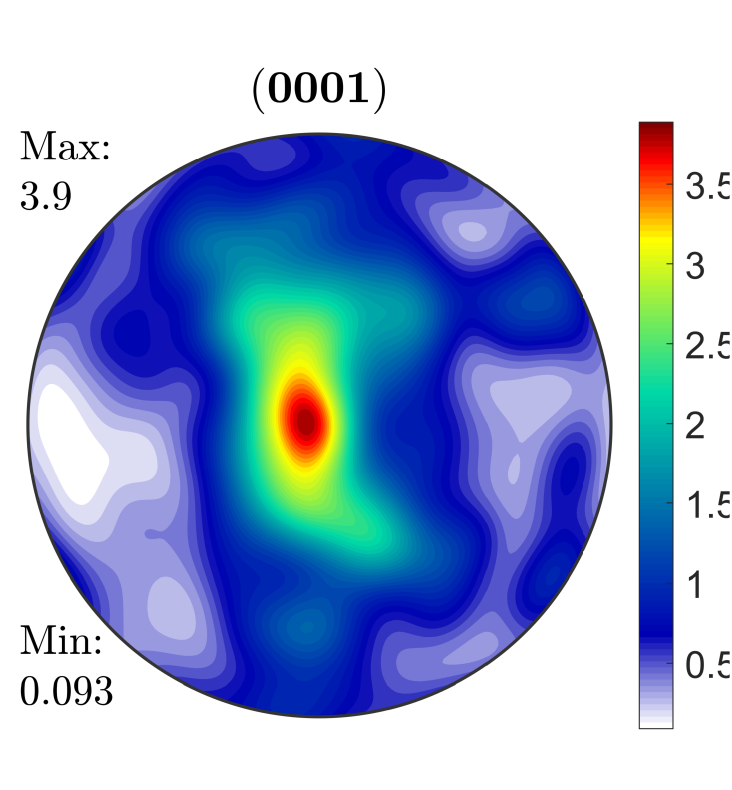}
			\includegraphics[scale=0.20]{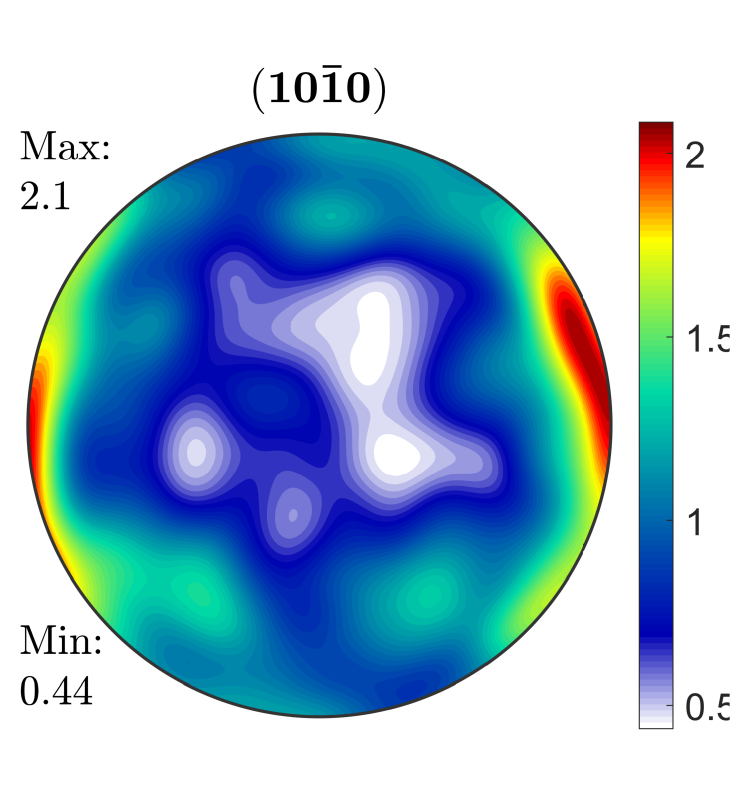}\label{f:in_TXTK2}}\par
        \vspace*{-1.2ex}  
	\end{multicols} 
	\caption{Initial $[0001]$ and $[10\bar{1}0]$ pole figures projected on the LT plane for textures A-K. For a particular texture, the angles in the brackets indicate the maximum standard deviations in the Euler angles. Adapted from \cite{Baweja23}.}\label{fig:all_init_txt}
\end{figure}
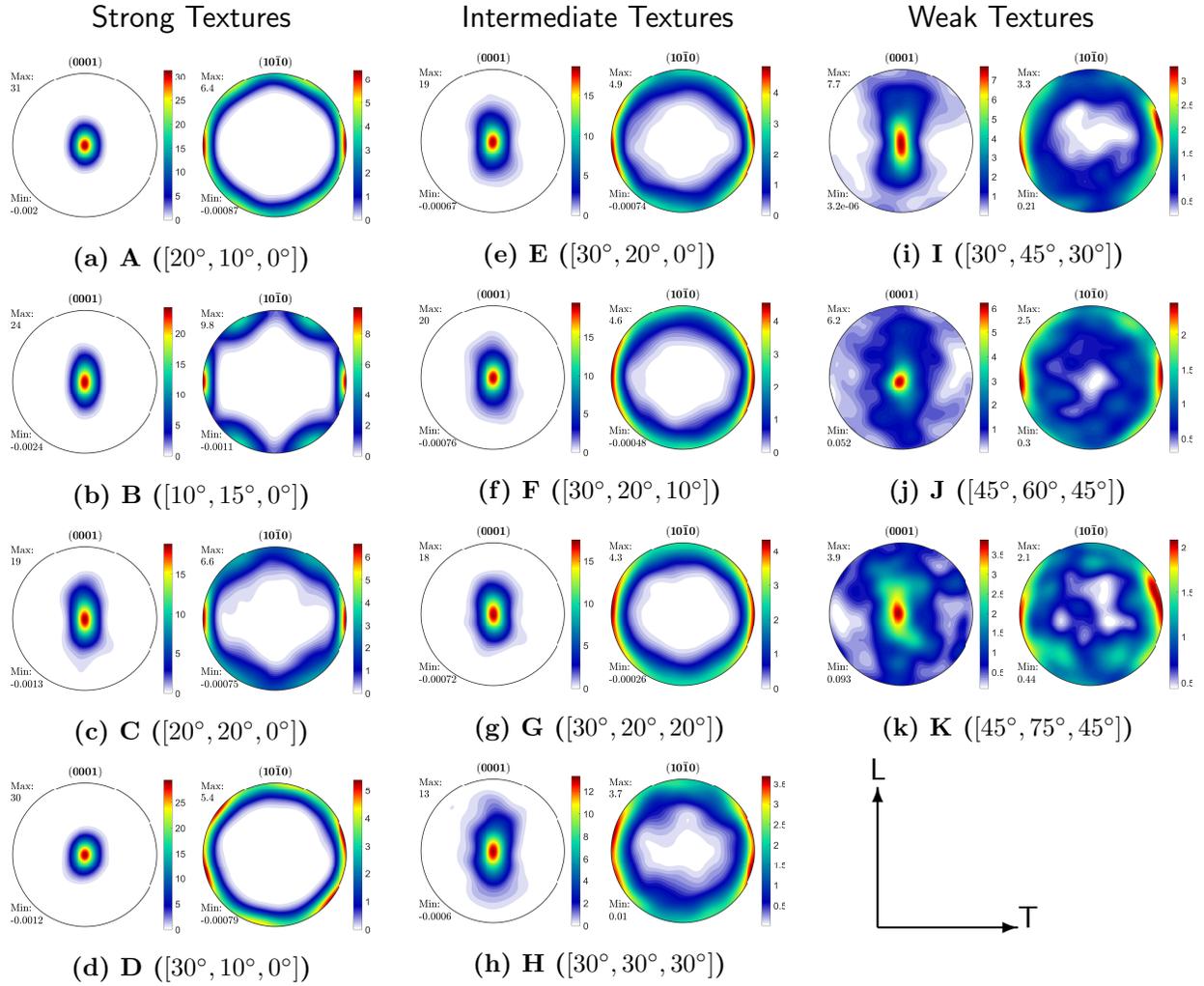
The reference data for model identification may be from experiments \citep{Basu17,Kondori19}, which are available for limited textures of Mg alloys, or from high-fidelity simulations \citep{Baweja23}, which are available for eleven synthetic textures at various grain sizes. 
Previous studies \citep{Herrington19,Kondori19} have shown the ability of the 2S model to capture experimental data for two textures. 
Here, we focus on the evaluation of both models for a wider range of textures.
Therefore, the parameters are identified using crystal plasticity (CP) data of 
magnesium polycrystals, as generated by \citet{Baweja23}.

\Cref{fig:all_init_txt} shows initial pole figures for eleven textures (A-K)
having a wide range of textural strengths \citep{Baweja23}.
Materials labeled A through D have a strong texture whereas those labeled I through K have
a weak texture.
The cases E--H are referred to as intermediate textures and are representative of 
typical rolled magnesium alloys.
The intensities of textures A and D are roughly similar 
to those of a single magnesium crystal \citep{selvarajou_AM17}, while case K resembles a random texture.  

\begin{figure}[H]
	\centering
	\subfloat[b][\centering]{\includegraphics[scale=0.2]{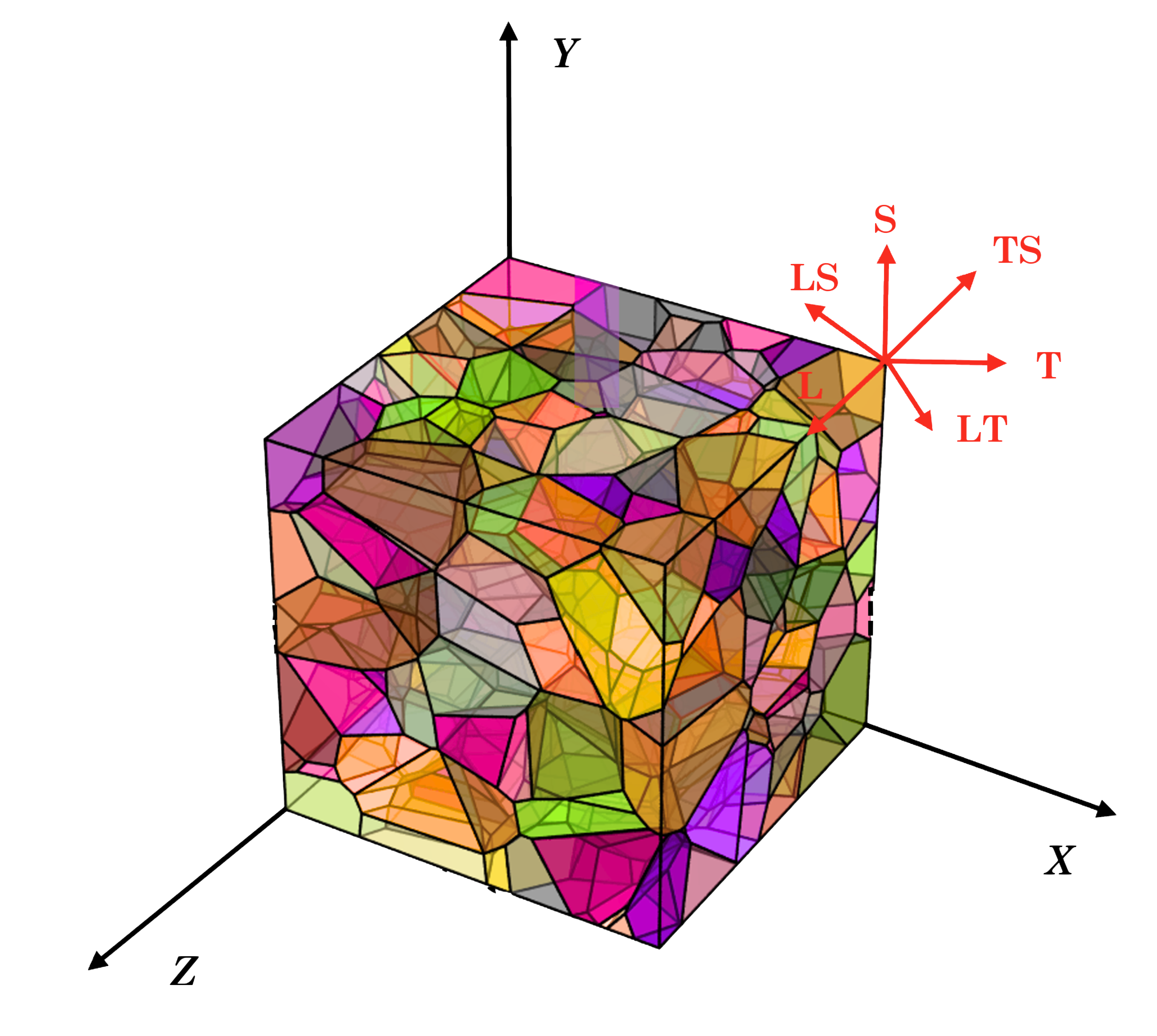}\label{f:polyX}}
	\subfloat[b][\centering] {\includegraphics[scale=0.435]{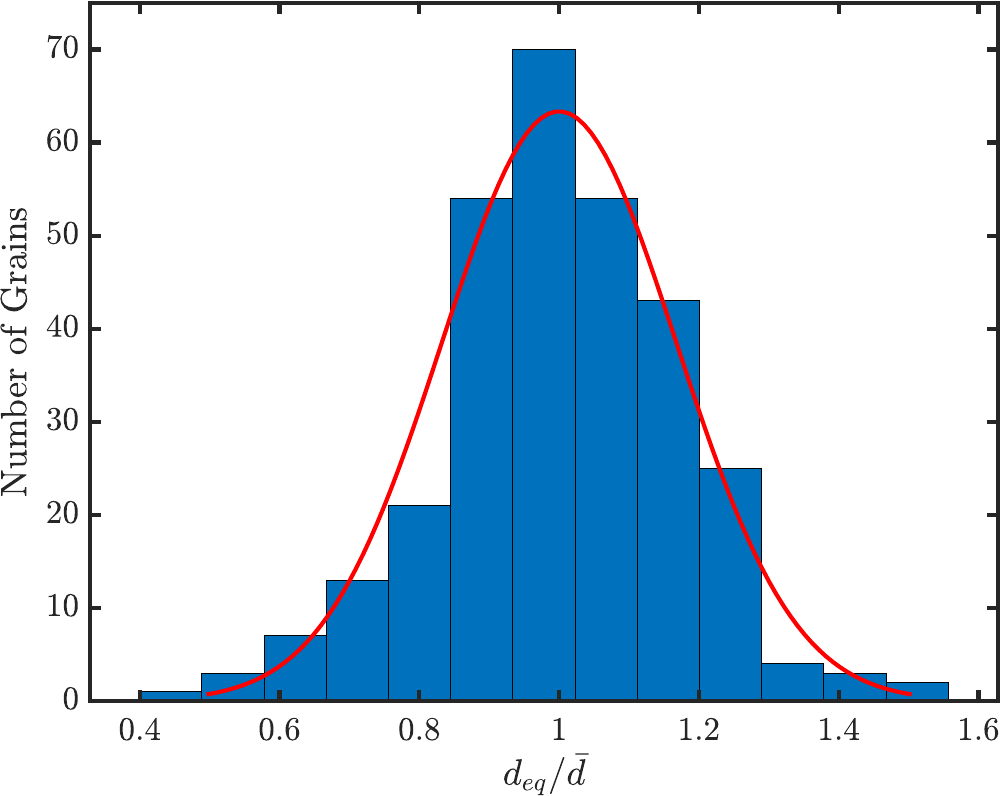}\label{f:grdist}} 
	\caption{(a) Polycrystal setup with illustrative case of a material principal direction (S) aligned with the loading (y) axis is shown. Panel (b) shows the grain size distribution. Adapted from \cite{Baweja23}.} \label{fig:polyX_model}
\end{figure}

In addition, for each texture, four average grain sizes are considered:
$\bar d = 10^4 \mu$m (base), 100$\mu$m, 10$\mu$m and 1$\mu$m.
\Cref{f:polyX} shows the three-dimensional polycrystal microstructure with a normal grain size distribution \cref{f:grdist}.
For all four cases, the microstructure (i.e., grain topology, grain  orientation allocation, and grain size distribution) remains the same, with an unchanged finite element mesh; see \cite{Baweja23} for details. The grain size dependency is through a size-dependent finite-deformation single crystal plasticity framework of \citet{Ravaji21}.

\section{Results} \label{sec:results}
A total of 44 materials were considered (eleven textures and four average grain sizes).
For brevity, detailed comparisons in terms of stress-strain response, lateral strain
evolution and relative activities are presented only for textures B and K using
the base grain size. Then, the effects of texture intensity and grain size
are demonstrated in synthetic plots. A more thorough database of comparisons
is provided under Supplementary Material.

\subsection{A strong anisotropic material}
\subsubsection{Stress-strain response}
\begin{figure}[h!]
	\centering
	\includegraphics[width=0.95\textwidth,page=7]{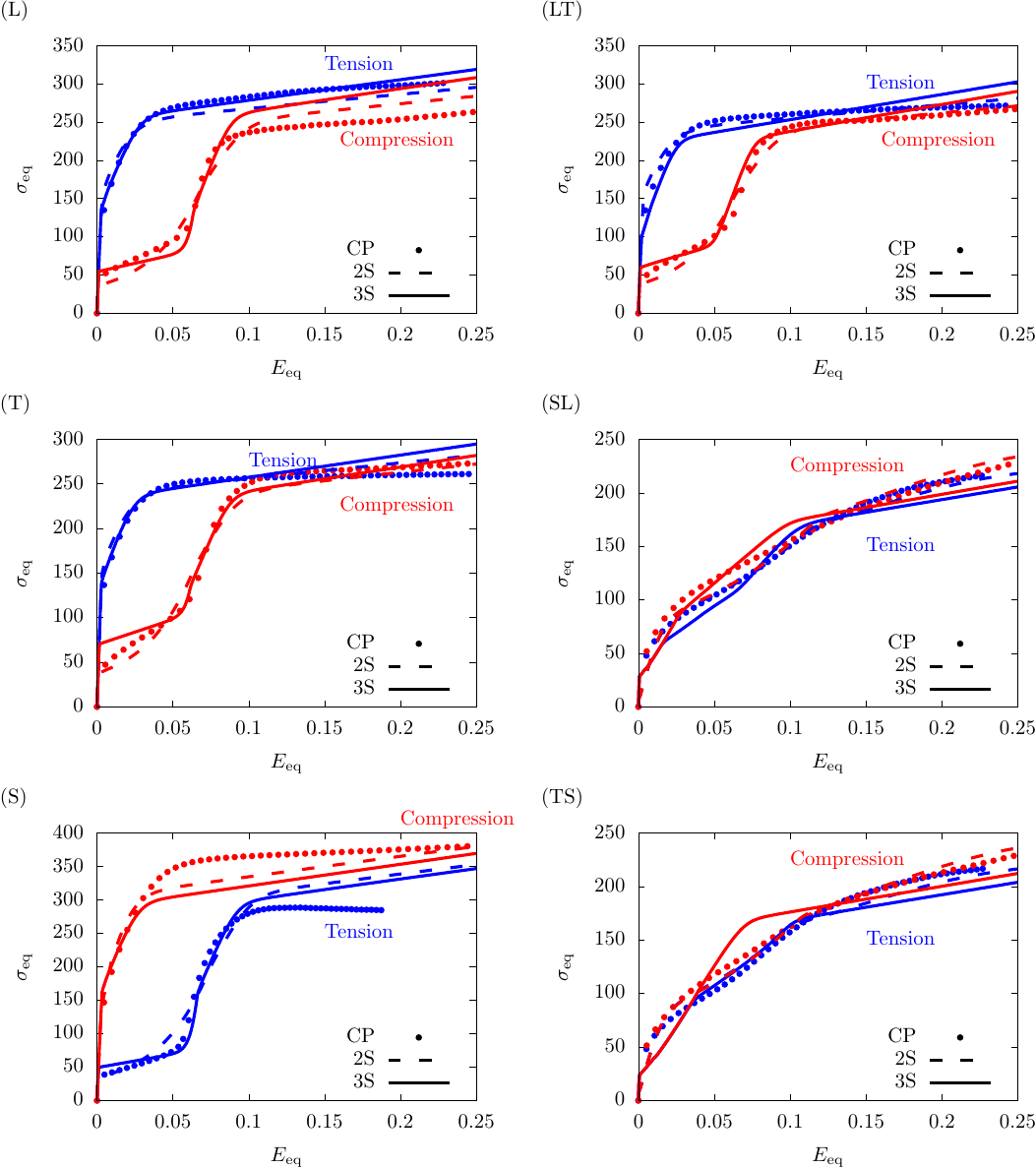}
	\caption{Calibrated stress-strain responses for material B with $\bar d = 10^4 \mu$m 
    under uniaxial loading along principal material (left column) and off-axis (right column) directions. 
    {\bf Symbols}: CP data \citep{Baweja23},
	{\bf Dashed lines}: 2S model, {\bf Solid lines}: 3S model. 
    {\red Red}: Compressive responses, {\blue Blue}: Tensile responses.
    }
	\label{fig:B_stress}
\end{figure}

\cref{fig:B_stress} compares the stress--strain responses predicted by the two coarse‐grained models
with the crystal plasticity (CP) simulations for the strongly textured material~B with base grain size ($\bar d \sim 10^4 \mu$m).
Unless otherwise noted, the dots correspond to the CP data, the dashed lines to the 2S model,
and the solid lines to the 3S model.
Also, uniaxial tension results are shown in blue whereas uniaxial compression
results are shown in red.
All three responses display the signature tension--compression asymmetry of magnesium, 
yet the origin and magnitude of this asymmetry differ subtly among them.

Focusing first on the principal axes, 
the CP data show a power-law-like hardening in the tensile responses
(blue) along the L and T directions (\cref{fig:B_stress}L,T). 
Both 2S and 3S models capture this behavior reasonably well, although they
exhibit somewhat higher strain hardening at large strains. 
In compression (red), a pronounced sigmoidal (S-shaped) response emerges, 
marked by a moderate hardening, twin-dominated regime for $E_{\rm eq} \lesssim 0.06$, 
followed by rapid stress increase due to the activation of non‐basal glide
resulting from twin-induced crystallographic reorientation. 
Subsequently, mild strain hardening results
from saturation of non-basal glide. 
The models do not explicitly track twinning reorientation 
but instead embed the sigmoidal characteristic within the hardening functions. 
Neither model reproduces the exact landscape of the sigmoidal region.  
The 3S model appears to reproduce the width of the initial plateau better
than the 2S model, although it undercaptures its slope. 
The 2S, on the other hand, does a better job of capturing the
transition from a rapid stress increase to a saturation-type behavior. 
These departures reveal how each model translates the underlying microstructural sequence. 
Recall that the 2S model merges basal and non‐basal slip into a single yielding mode,
which accelerates the transfer of load from twinning back to glide, 
while the 3S model allows that transfer but distributes the strain between
its soft and hard glide modes, slightly softening the macroscopic response.

Along the S-direction, the CP simulation reverses the asymmetry.
Both models capture this reversal qualitatively, 
though the 2S formulation again shows a stronger stiffening in the
twinning-dominated and the late‐stage responses, 
whereas the 3S formulation reproduces the twinning plateau
but overshoots the post-twinning response.  
The deviations are tied to how the two models represent latent hardening:
the coupling between glide and twinning in the 2S model causes
a more rapid build-up of latent hardening once both modes are active;
in the 3S model the additional soft glide mode appears to partially screen this interaction.

Interestingly, the responses along the LT direction 
are similar to the responses along the L direction, \cref{fig:B_stress}.
The strain values where the transition in hardening 
occurs are also similar to the values along the principal directions.
The models capture the trend quite well.
For the SL, and TS off‐axis directions, the CP data show that 
the magnitude of anisotropy diminishes and the tension--compression asymmetry fades.  
Both models follow this trend and show reasonably good quantitative agreement 
with the CP results.  

%
\subsubsection{Lateral strains} \label{sss:defaniso}
\begin{figure}[h!]
	\centering
	\includegraphics[width=0.95\textwidth,page=8]{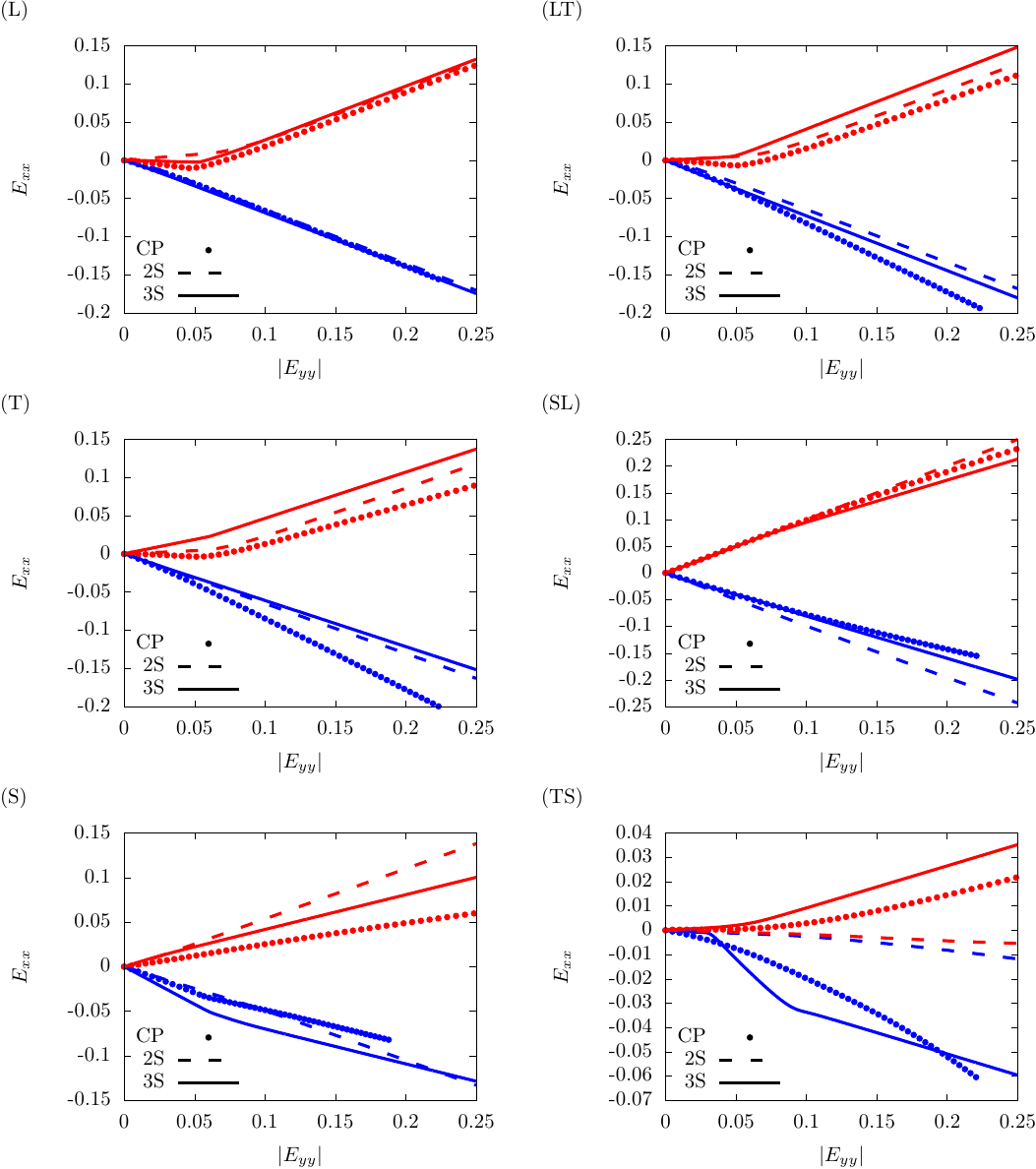}
	\caption{Predicted lateral strain, $E_{xx}$, for material B with $\bar d = 10^4 \mu$m 
    under uniaxial loading along principal material (left column) and off-axis (right column) directions. 
    {\bf Symbols}: CP data \citep{Baweja23},
	{\bf Dashed lines}: 2S model, {\bf Solid lines}: 3S model. {\red Red}: Compressive responses, {\blue Blue}: Tensile responses}
	\label{fig:B_Exx}
\end{figure}

\begin{figure}[b!]
	\centering
	\includegraphics[width=0.95\textwidth,page=9]{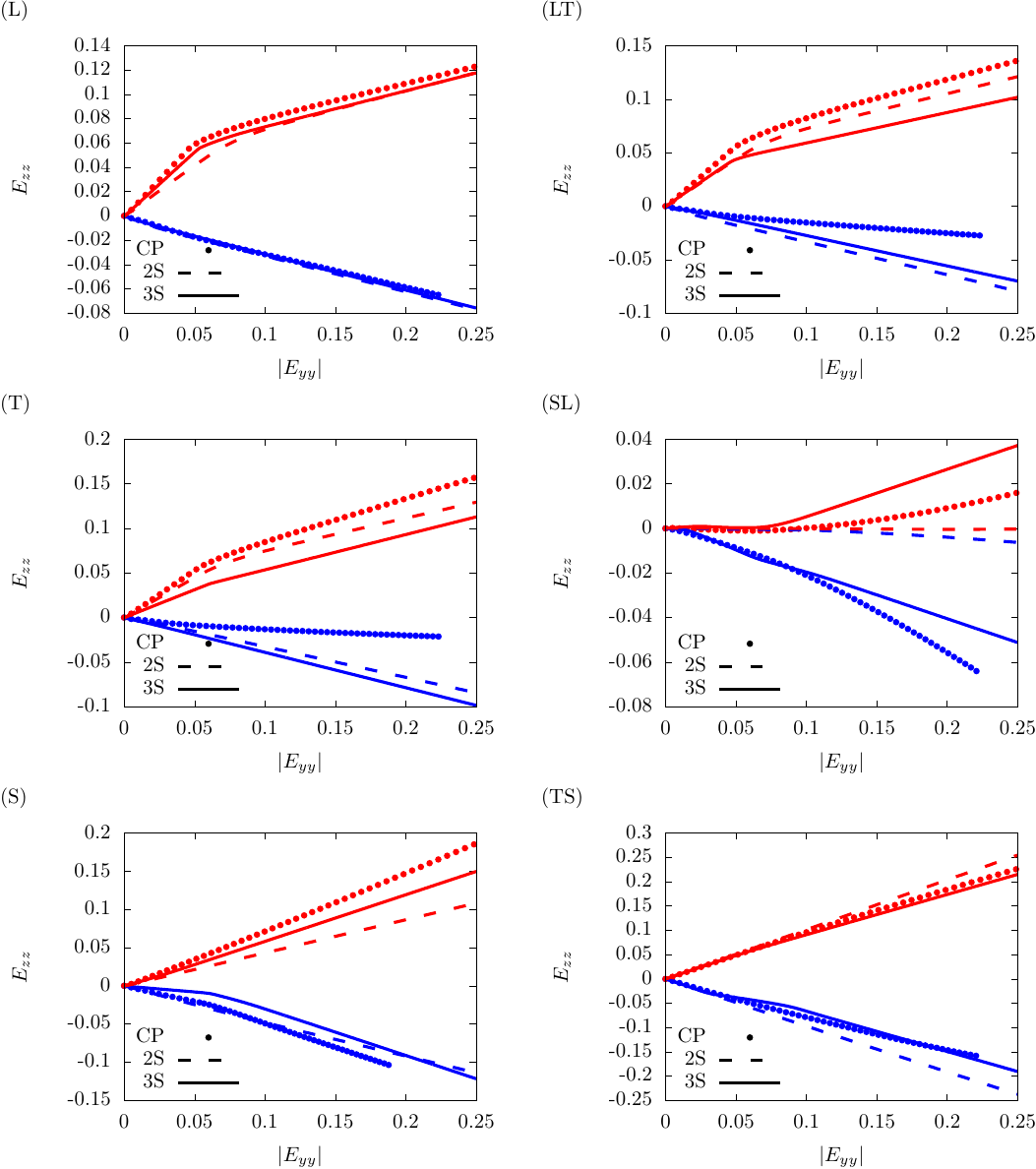}
	\caption{Predicted lateral strain, $E_{zz}$, for material B with $\bar d = 10^4 \mu$m 
    under uniaxial loading along principal material (left column) and off-axis (right column) directions. 
    {\bf Symbols}: CP data \citep{Baweja23},
	{\bf Dashed lines}: 2S model, {\bf Solid lines}: 3S model. {\red Red}: Compressive responses, {\blue Blue}: Tensile responses}
	\label{fig:B_Ezz}
\end{figure}

Figures~\ref{fig:B_Exx} and~\ref{fig:B_Ezz} present the predicted lateral strain components,
$E_{xx}$ and $E_{zz}$, for material B.  
These quantities were \emph{not} part of the calibration,
and therefore serve as a predictive power of each model
internal anisotropy and deformation kinematics.  

Both models capture the general trends observed in the CP simulations: 
contraction in the transverse directions during tension and expansion during compression,
consistent with plastic incompressibility.  
For the principal directions (L, T, S), the predicted slopes of $E_{xx}$–$|E_{yy}|$
and $E_{zz}$–$|E_{yy}|$ curves reproduce the anisotropic lateral strains seen in the CP data.  
Both models pass the litmus test for the characteristic kink-like 
behavior of $E_{xx}$ in L and T compression, which arises from twinning.
Between them, the 2S model generally shows a better quantitative comparison to the CP data. 
A similar kink is also seen in the S-tension, which is captured well by the 3S model; 
the 2S model tends to smear out this subtle feature. 

For off-axis directions (LT, SL, TS) as well, 
both models predict the lateral strains trends. 
Both models predict the kink in LT-compression, 
which highlights their ability to capture transverse isotropy. 
The 3S model shows better quantitative agreement
with the CP data. Notably, it reproduces
the nonlinearity of $E_{xx}$ in the TS direction and 
of $E_{zz}$ in the SL direction. 
On the other hand, the 2S model fails to capture
one of the lateral strains when loading along
TS or SL.

Overall, reasonable agreement across six orientations, 
without any direct calibration to lateral strain data, 
demonstrates that both models encode physically meaningful anisotropy.
The predictive ability of the 2S model may be improved
if one set of lateral strain data were included in calibration.

\subsubsection{Relative activities}
The ability of the 2S and 3S models to predict lateral deformation anisotropy (cf.~\cref{sss:defaniso})
motivates the assessment of the underlying mechanisms that govern these macroscopic observations. 
Because these quantities were also excluded from calibration, 
they serve as another stringent test of the internal structure of the models. 
We present two illustrative assessments.
First, for the relative cumulative glide activity, 
$\xi^{\rm g} = p^{\rm g}/p$,
where $p = \sum_{\alpha} p^\alpha$ with $\alpha \in \{{\rm g, t}\}$ (2S)
and $\alpha \in \{{\rm s, h, t} \}$ (3S). 
And, $p^{\rm g}=p^{\rm s}+p^{\rm h}$ for 3S.
The relative activities of the twinning mechanisms
can be inferred from its complementary value $1-p^{\rm g}/p$. 
In other words, if $\xi^{\rm g}$ is small, the twinning mechanisms dominate.
This definition is consistent with its counterpart in CP except that the CP framework 
incorporates a richer set of mechanisms.
In addition, an assessment of the 3S model in predicting relative soft glide activity, 
$\xi^{\rm s} = p^{\rm s}/p$, is also presented.
\begin{figure}[t!]
	\centering
	\includegraphics[width=0.95\textwidth,page=11]{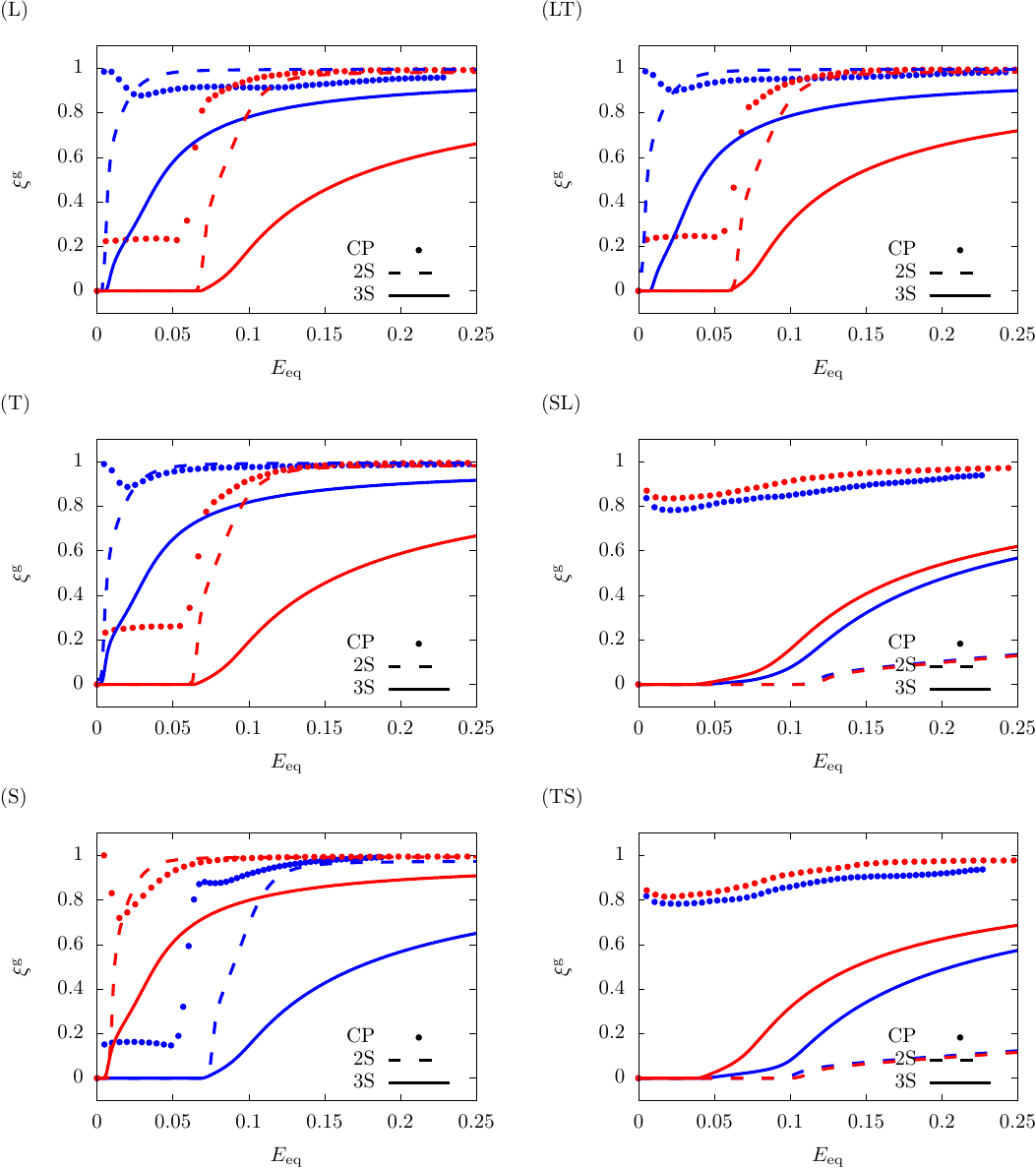}
	\caption{Predicted relative cumulative activity of glide, $\xi^{\rm g} = p^{\rm g}/p$, for material B with $\bar d = 10^4 \mu$m 
    under uniaxial loading along principal material (left column) and off-axis (right column) directions. 
    {\bf Symbols}: CP data \citep{Baweja23},
	{\bf Dashed lines}: 2S model, {\bf Solid lines}: 3S model. {\red Red}: Compressive responses, {\blue Blue}: Tensile responses}
	\label{fig:B_pg}
\end{figure}

\cref{fig:B_pg} compares the evolution of the relative glide activity $\xi^{\rm g}$ for material~B.
The CP simulations reveal distinct deformation sequences across six orientations. 
Under tension along L and T, the deformation is dominated by glide, 
while under compression, twinning activates early, 
saturates near $E_{\mathrm{eq}}\!\approx\!0.06$,
and is gradually replaced by glide.  
The trend is reversed for loading along the S-direction. 
Off‐axis loadings (SL, TS) show mild twinning and are glide-dominated under both, tension and compression. 
By contrast, the LT orientation shows a twinning-dominated behavior in compression
and a glide-dominated behavior in tension. 

Both models capture overall trends, but differ in details. 
Under twinning-dominated loading states (compression along L, T and LT, and S-tension),
they do not predict glide in the early stages. 
In the absence of the abundance of mechanisms that are encoded in CP,
models must choose twinning to be the sole mechanism driving the response in these conditions. 
Post-twinning, the rapid increase in $\xi^{\rm g}$ is captured well by 2S,
while the 3S model tends to exhibit a continued contribution, 
which implies that the soft and hard glide mechanisms contribute 
less in this regime compared to the 2S model. 
In glide-dominated loading states (L-tension, T-tension and LT-tension, and S-compression),
the 3S model completely suppresses twinning activity; on the other hand, 
the 2S model exhibits an initial surge in twinning that rapidly dissipates
giving way to glide. 
For this reason, the post-twinning stress-strain responses
of the 2S model tend to show a better corroboration with the CP data
for these loading orientations.

On the other hand, the predictive abilities of the 3S model are far better
when considering off-axes (SL and TS) responses.
Unlike the 2S model, which shows isotropic twinning-governed behaviors in both
tension and compression along these directions, the 3S model
captures the nonlinear trends and predicts them quantitatively as well. 
At least for these loading orientations, while the 2S model reproduces 
the macroscopic stress-strain responses quite well (cf.~\cref{fig:B_stress}, SL and TS)
it obfuscates the underlying mechanistic basis. Recall that the macroscopic lateral
deformation anisotropy predicted by the 2S model 
for these orientations is insufficient (cf.~\cref{fig:B_Exx}, \ref{fig:B_Ezz}).

\begin{figure}[h!]
	\centering
	\includegraphics[width=0.95\textwidth,page=12]{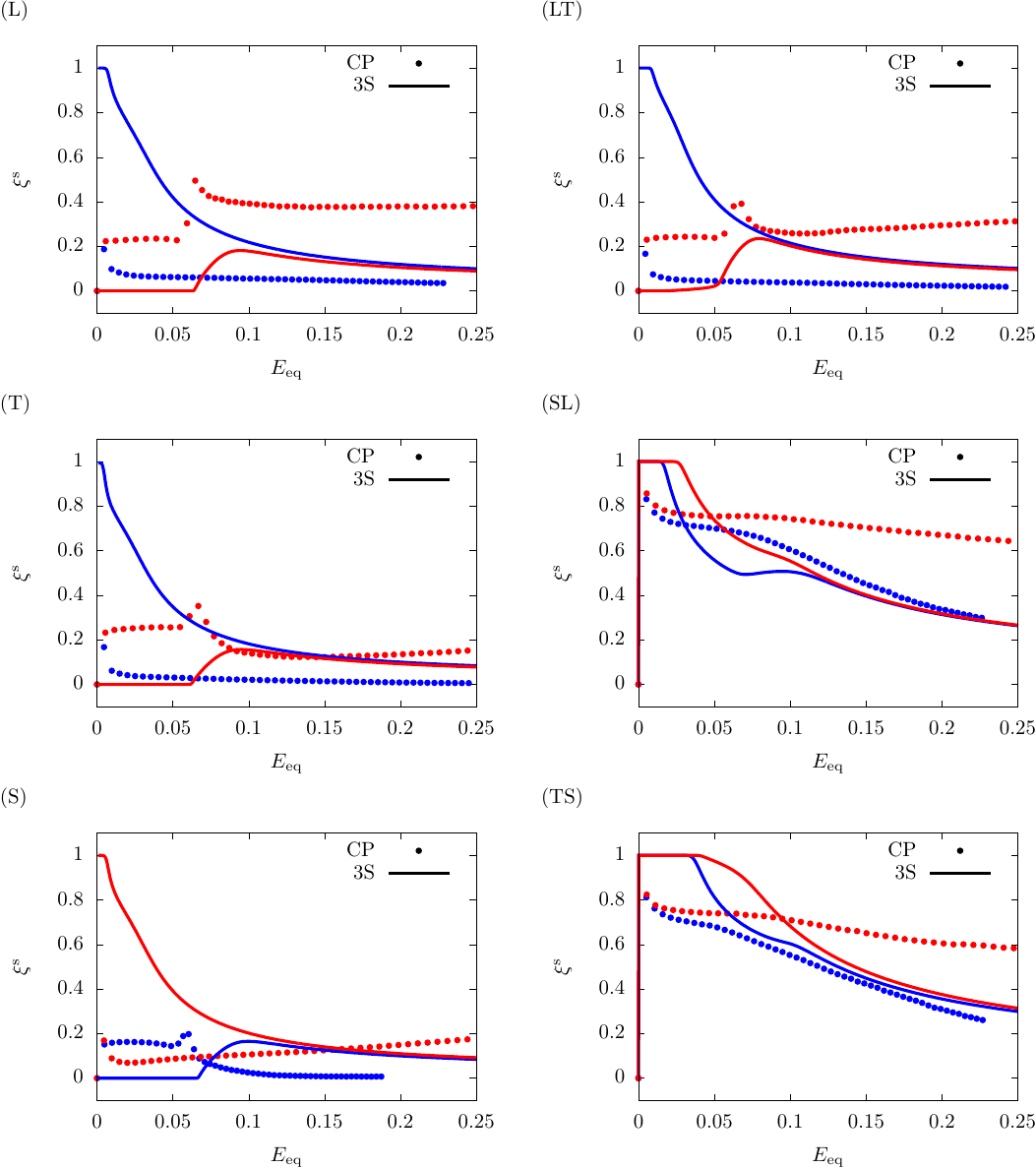}
	\caption{Predicted relative activity of soft glide ($\xi^{\rm s} = p^{\rm s}/p$) by the 3S model
    for material B with $\bar d = 10^4 \mu$m 
    under uniaxial loading along principal material (left column) and off-axis (right column) directions. 
    {\bf Symbols}: CP data \citep{Baweja23},
	{\red Red}: Compressive responses, {\blue Blue}: Tensile responses}
	\label{fig:B_ps}
\end{figure}
Next, \cref{fig:B_ps} compares the evolution of the relative soft glide activity $\xi^{\rm s}$ for material~B.
The CP simulations show that under tension along L, T, S, and LT, the deformation is not dominated by basal slip, 
while under compression, except for L, SL, and TS, 
basal slip contributes about 20\% of the total activity.  
Off‐axis loadings (SL, TS) are dominated by soft glide mode under both tension and compression, but that decreases with increasing strain.

The 3S model captures the trends quite well, except in the earlier stages of deformation.
Under tension along L, T and LT, and S-compression,
the model predicts soft-glide from the outset,
while under compression along L, T and LT, and S-tension,
it does not predict soft-glide, contrary to the CP data.
Nevertheless, after $E_{\mathrm{eq}}\!\approx\!0.06$, an increase in $\xi^{\rm s}$ is captured by the 3S model.
With deformation, a saturating soft-glide activity of about 10\% is predicted by the model.

The predictive power of the 3S model is shown for loading along the SL and TS directions.
As in CP, a soft-glide dominated yielding mode is predicted from the outset for tension and compression loadings.
This helped in accurately capturing the evolution of lateral strains by the 3S model (cf.~\cref{fig:B_Exx}, \ref{fig:B_Ezz}). 

\subsection{A weak anisotropic material}
Next, we assess the predictive capabilities of the two coarse-grained models for a weakly anisotropic 
material K with $\bar d = 10^4 \mu$m. 
\begin{figure}[h!]
	\centering
	\includegraphics[width=0.95\textwidth,page=61]{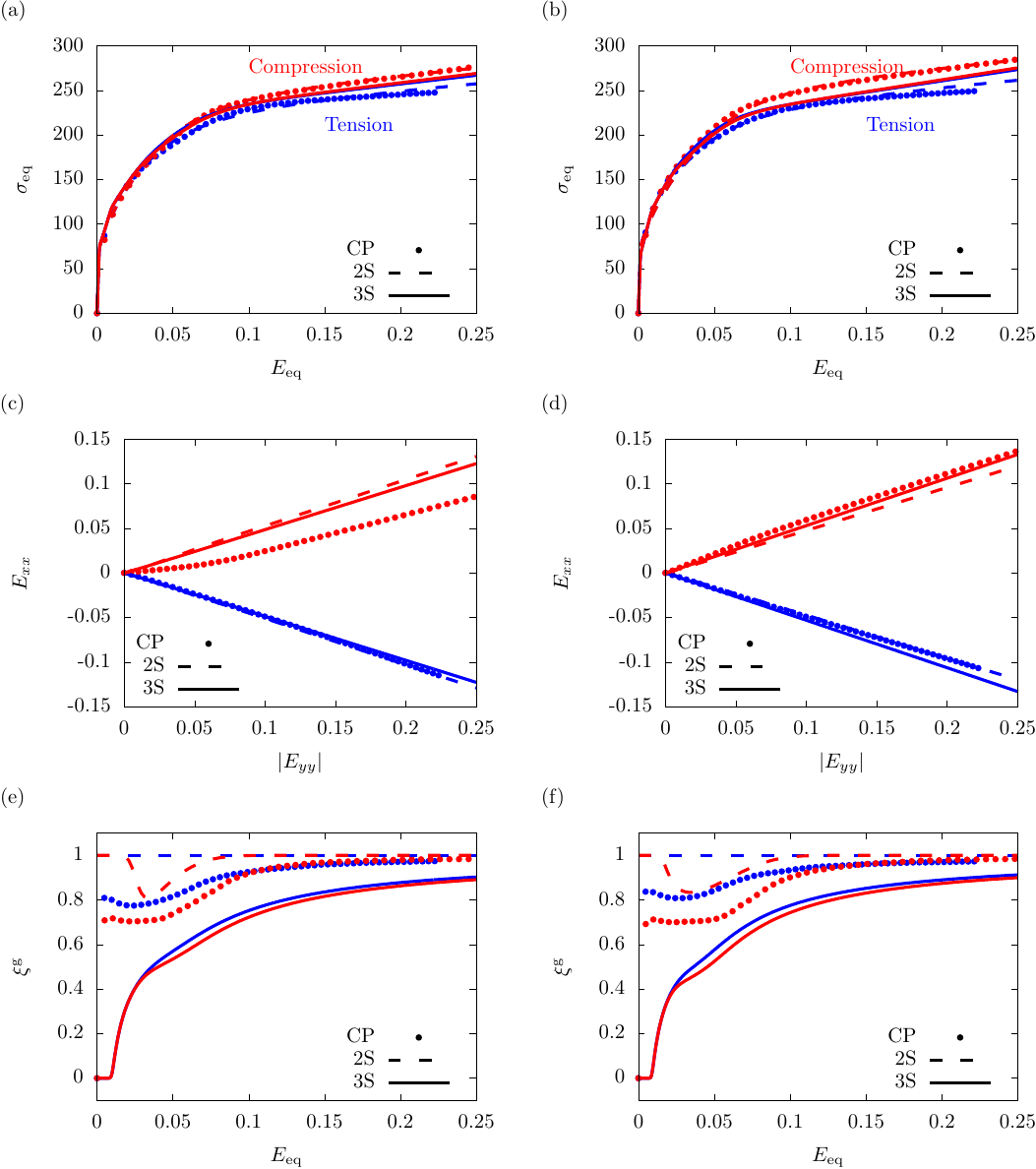}
	\caption{(a,b) Calibrated stress-strain response for material K ($\bar d = 10^4 \mu$m) in L and SL directions, respectively. 
    The corresponding predicted lateral strains, $E_{\rm xx}$ (Panels (c) and (d)) 
    and predicted glide relative activities (Panels (e) and (f)) are shown.
    {\bf Symbols}: CP data \citep{Baweja23},
	{\bf Dashed lines}: 2S model, {\bf Solid lines}: 3S model. {\red Red}: Compressive responses, {\blue Blue}: Tensile responses
	}
	\label{fig:K}
\end{figure}
Given the near-isotropic behaviors seen in the CP data \citep{Baweja23},
\cref{fig:K} collates the responses for only two representative directions, L (left column) and SL (right column).  
The CP simulations exhibit modest tension--compression asymmetry in the 
stress-strain responses (\cref{fig:K}{\blue a,b}) with power-law hardening, 
reflecting limited twinning propensity.  
Both coarse‐grained models capture these overall features.  
The 2S model reproduces the stress magnitudes slightly more closely
in both tension and compression, 
while the 3S model yields a marginally softer response 
beyond $E_{\mathrm{eq}}\!\approx\!0.05$, consistent with the load sharing among its soft and hard glide modes.  
The small discrepancy between tension and compression is thus retained, but slightly muted, in the 3S results.

The lateral strains (\cref{fig:K}{\blue c,d}) vary almost linearly with $|E_{yy}|$.
Both models predict these trends with comparable fidelity,
and no directional inflections such as those seen in the material B case appear.  
Minor overprediction of transverse expansion under $L$‐compression
is observed in both models, likely reflecting 
the weak coupling between the glide and twinning modes once twinning activity subsides.

\begin{figure}[h!]
	\centering
	\includegraphics[width=0.95\textwidth,page=66]{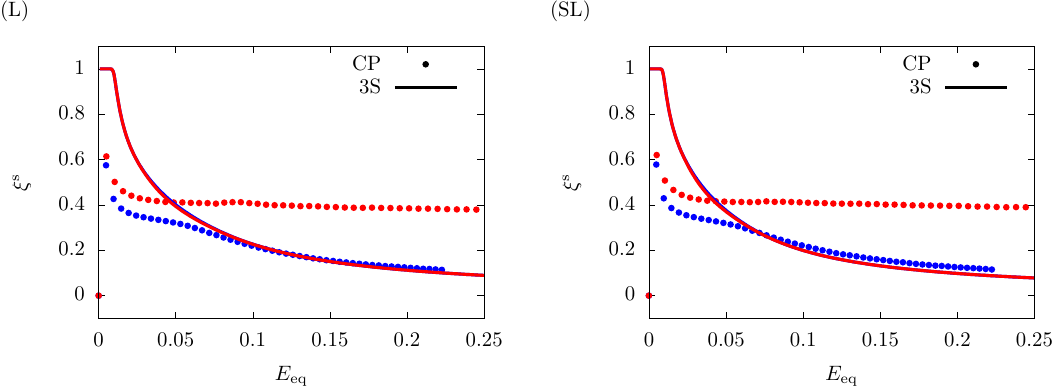}
	\caption{Predicted relative activity of soft glide, $\xi^{\rm s} = p^{\rm s}/p$, for material K with $\bar d = 10^4 \mu$m 
    under uniaxial loading along L (a) and SL (b) directions. 
    {\bf Symbols}: CP data \citep{Baweja23},
	{\bf Solid lines}: 3S model. {\red Red}: Compressive responses, {\blue Blue}: Tensile responses
	}
	\label{fig:K_ps}
\end{figure}

The predicted evolution of relative glide activity (\cref{fig:K}{\blue e,f}) confirms
that the deformation in material~K is glide‐controlled.
In the CP data, glide activity remains above 70\%,
and tends to 100\% by $E_{\mathrm{eq}}\!\approx\!0.1$. 
As seen in \cref{fig:K_ps}, soft-glide contributes approximately 40\% under tension
and approximately 15\% under compression.
Both, 2S and 3S models reproduce this glide dominated yielding; 
the 3S model appears to show a slightly better corroboration with the CP trends
with a reasonable corroboration of the soft-glide activity.

\subsection{Error Estimation}
\begin{figure}[h!]
	\centering
	\includegraphics[width=0.95\textwidth,page=67]{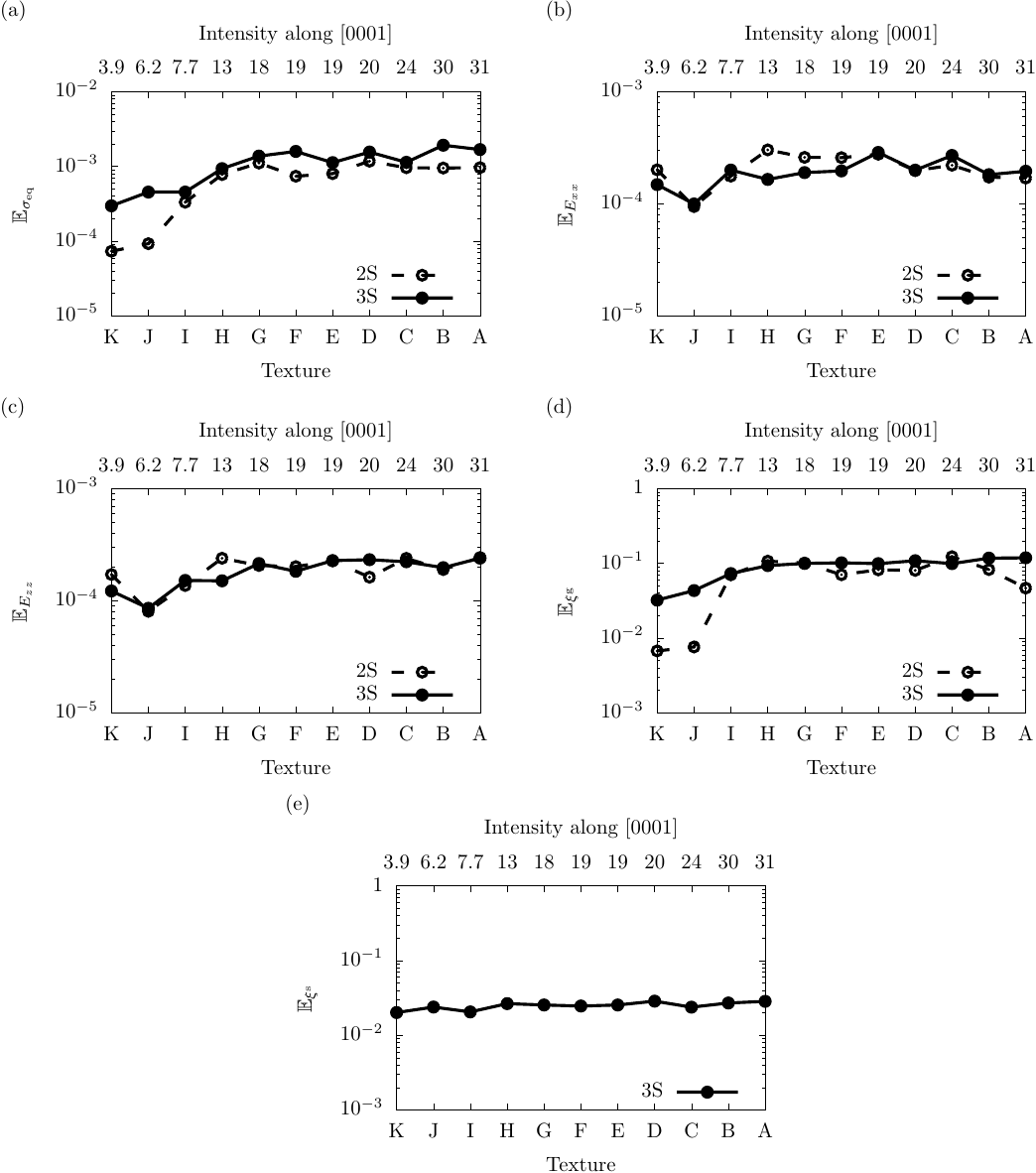}
	\caption{Total error in the difference between CP versus the 2S and 3S models for all textures with a grain size of $10^4 \mu$m}
	\label{fig:err}
\end{figure}

The results presented in the preceding sections
highlight the capabilities of the two models in reproducing the trends of the
macroscopic and microscopic field variables for extreme cases of material anisotropy 
and tension-compression asymmetry. 
In this study, we conducted a similar exhaustive analysis of the remaining nine materials, cf.~\cref{fig:all_init_txt}. 
\cref{fig:err} collates the performance of the two models relative to the CP 
by quantifying a {\it global error} for calibrated and predicted 
variables averaged over the entire regime of deformation. For brevity, the results are 
shown for $\bar d = 10^4 \mu$m. 

The global error in the variable $\beta$ is defined here as:  

\begin{equation}
\mathbb{E}_\beta \ =\frac{1}{N} \sum_{i=1}^{N} \frac{1}{\Eeq^i} \int_0^{\Eeq^i} 
\left( \beta_{\rm{CP}}^i - \beta_{\rm model}^i \right)^2 {\rm d} \Eeq^i
\qquad \forall \ {\rm \beta} \ \in
\left\lbrace E_{xx},E_{zz},\xi^{\rm g}, \xi^{\rm s} \right\rbrace
\label{eq:err}
\end{equation}
where $N=12$ is the total number of simulations per material. 
In addition, the error in the volume-averaged equivalent stress is calculated from \cref{eq:cost}
as $\mathbb{E}_{\Sigma_{\rm eq}} = \mathcal{E}/N$. 
Furthermore, we also include the error in the relative activity of the soft glide $\xi^{\rm s}$
mechanism, which is only tracked in the 3S model. 
Note that there is no normalization in \cref{eq:err} because the quantities therein are on the order of unity.
On the other hand, the error in the stress is normalized
by the maximum value of equivalent stress in the CP data (cf.~\cref{eq:cost}).
Each point in \cref{fig:err} reflects a measure of the
average error in the quantity of interest per material per simulation.  
Error values are plotted against materials with increasing $[0001]$ textural intensity. 

Details aside, the following salient observations are made. 
First, both models show relatively low errors across all the 
variables shown in \cref{fig:err}. 
The 2S model tends to show a better performance in reproducing the
macroscopic stress and its performance improves relative to 
the 3S model with decreasing texture strength (i.e., decreasing plastic anisotropy and tension-compression asymmetry). 
On the whole, the lateral deformation anisotropy ($E_{xx}$ and $E_{zz}$) is 
predicted with level of error that is remarkably low for both, 2S and 3S models.
The biggest difference between the 2S and 3S models is seen in the glide relative activity. 
Across the spectrum of materials (A-K), the 2S model performs better than the 3S 
model even for weakly textured materials, e.g., H, I, and J.

\subsection{Model parameters}
\begin{figure}[t!]
	\centering
	\includegraphics[width=0.95\textwidth,page=113]{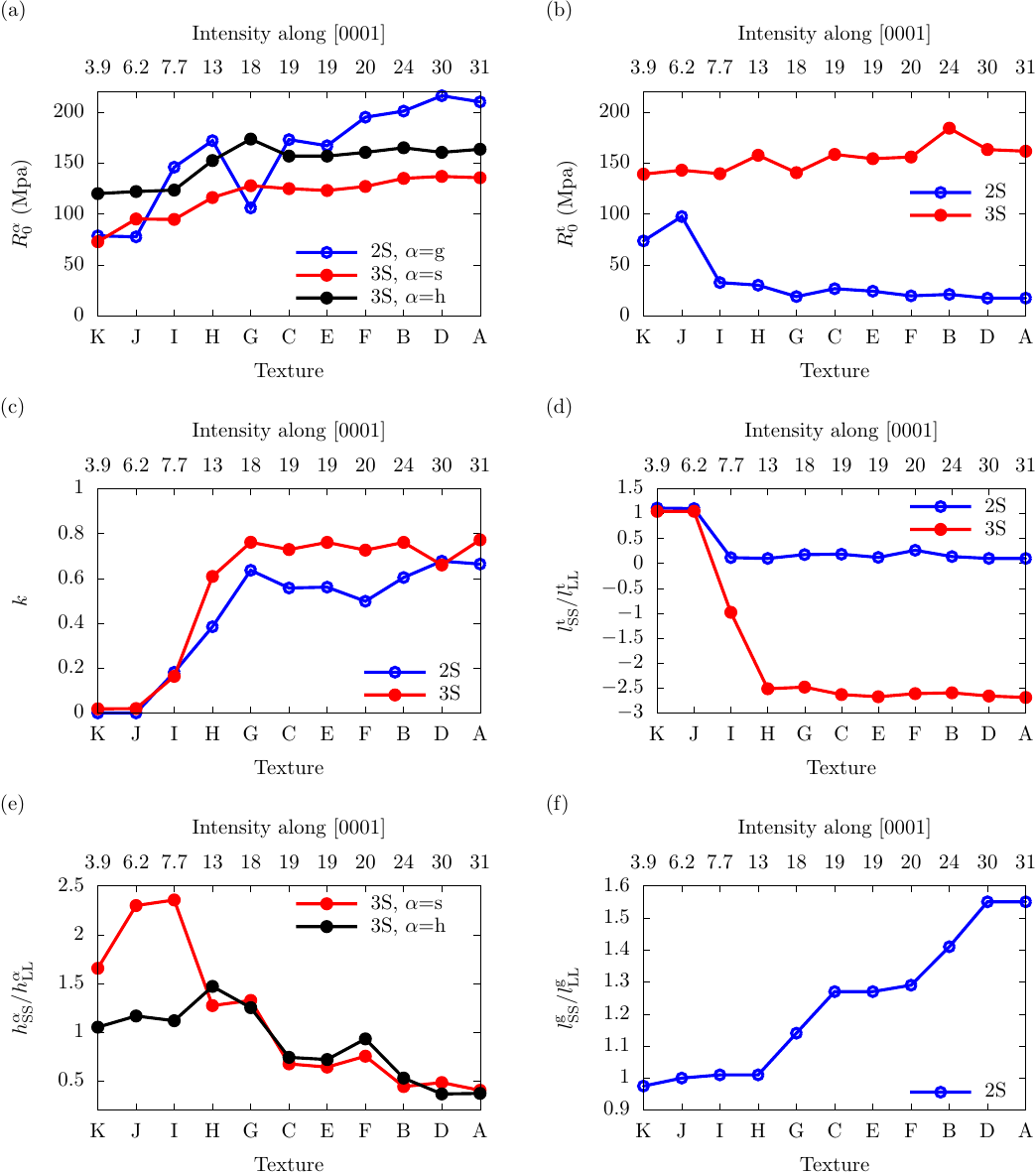}
	\caption{Calibrated parameters for the 2S and 3S models as a function of 
    material texture for $\bar d = 10^4 \mu$m. (a) initial glide strengths, (b) initial twinning strength, (c) tension-compression asymmetry parameter
	(d) ratio of normalized twinning anisotropy coefficient ($l_{\rm SS}^{\rm t}/l_{\rm LL}^{\rm t}$), 
    ratio of glide anisotropy coefficients for (e) 3S model ($h_{\rm SS}^{\alpha}/l_{\rm LL}^{\alpha}, \ (\alpha = {\rm s, h})$)
    and (f) 2S model ($l_{\rm SS}^{\rm g}/l_{\rm LL}^{\rm g}$)}
	\label{fig:params}
\end{figure}

\cref{fig:params} collates and compares the calibrated parameters obtained from the 2S and 3S models across the 11 materials (for $\bar d = 10^4 \mu$m).  
Both sets follow consistent physical trends, though they differ in how the anisotropy and hardening are internally distributed.

For glide strengths (\cref{fig:params}{\blue a}), both models show an
increase in $R_0^\alpha$ with a stronger $[0001]$ textural intensity,
indicating a texture-induced material strengthening.  
In the 3S formulation, the consistently lower yield strength for 
the soft glide compared to the hard glide is evident. 
By comparison, the glide strength in the 2S model fluctuates 
between these limiting values of the 3S model
for intermediate and weakly textured materials, but trends somewhat higher
for the more strongly textured materials,
which has implications on its role in accommodating both basal and non-basal slip through one yielding mode.  

The twinning yield strength, $R_0^{\mathrm{t}}$ (\cref{fig:params}{\blue b}), 
shows a different trend with textural intensity in the 2S and 3S models. 
While the 3S model shows a weak increase in $R_0^{\mathrm{t}}$ with increasing $[0001]$ textural intensity, 
the 2S model shows the opposite trend. Moreover, in the 3S model, 
the $R_0^{\mathrm{t}}$ values are consistently higher than 
their glide counterparts, while they are generally lower in the 2S model. 
Recall that the yield potential and the hardening function adopted for twinning are the same for the 2S and 3S models. 
The significantly different dependence of $R_0^{\mathrm{t}}$ on the material textures
suggests the role played by distinguishing the soft glide from the hard glide in the 3S model.  
 
\cref{fig:params}{\blue c} and {\blue d} respectively illustrate 
how twinning asymmetry and anisotropy are reflected in the two models. 
The asymmetry parameter $k$ (\cref{fig:params}{\blue c}) in the twinning yield potential
rises systematically with texture strength in both models, 
confirming that the degree of tension–compression asymmetry is captured consistently.  
\cref{fig:params}{\blue d} illustrates the variation of the normalized twinning anisotropy coefficient,
$l_{\rm SS}^{\rm t}/l_{\rm LL}^{\rm t}$ (\cref{eq:Lt})
along a principal material direction (S). 
This ratio is one of the key parameters for the interplay between glide-dominated 
and twinning-dominated yielding during the initial deformation phase.
For example, under tension along the S-direction, a lower value of $l_{\rm SS}^{\rm t}/l_{\rm LL}^{\rm t}$
activates twinning before glide. 
For weak textures, the ratio is closer to 1 and hence twinning is not a primary mode; see \cref{fig:K}{\blue e,f}. 
For intermediate and strong textures, the ratio decreases rapidly. 
For the 2S model, it becomes negligible, whereas for the 3S model it achieves a nearly constant negative value. 
In both cases, twinning dominates under twin-friendly loading orientations, particularly for intermediate and strong textures. 

Similarly, \cref{fig:params}{\blue e} shows the ratio of anisotropy coefficients of the
soft and hard glide modes for the 3S model, cf.~\cref{eq:H}. 
The normalized values for the soft ($h_{\rm SS}^{\rm s}/h_{\rm LL}^{\rm s}$) 
and hard ($h_{\rm SS}^{\rm h}/h_{\rm LL}^{\rm h}$) glide modes are presented\footnote{Given the transverse isotropy,
$h_{\rm LL}^{\alpha} \approx h_{\rm TT}^{\alpha} \ \forall \alpha \in \{{\rm s, h}\}$}.
These ratios along with $l_{\rm SS}^{\rm t}/l_{\rm LL}^{\rm t}$ in \cref{fig:params}{\blue d} play an important role in driving the
competition between the three mechanisms for a given material. 
Interestingly, in the space of moderate to strong textures, both ratios are nearly identical 
for a particular texture; furthermore, they decrease with increasing textural intensity. 
The corresponding ratio that sets up the glide anisotropy in the 2S model is $l_{\rm SS}^{\rm g}/l_{\rm LL}^{\rm g}$  (\cref{eq:Lg}); 
it increases with increasing intensity, \cref{fig:params}{\blue f}. 
Although it is not possible to infer from the trends of the anisotropy coefficient ratios from the 
two models, it is nevertheless interesting that they are inversely correlated.

\begin{figure}[t!]
    \centering	\includegraphics[width=0.95\textwidth,page=114]{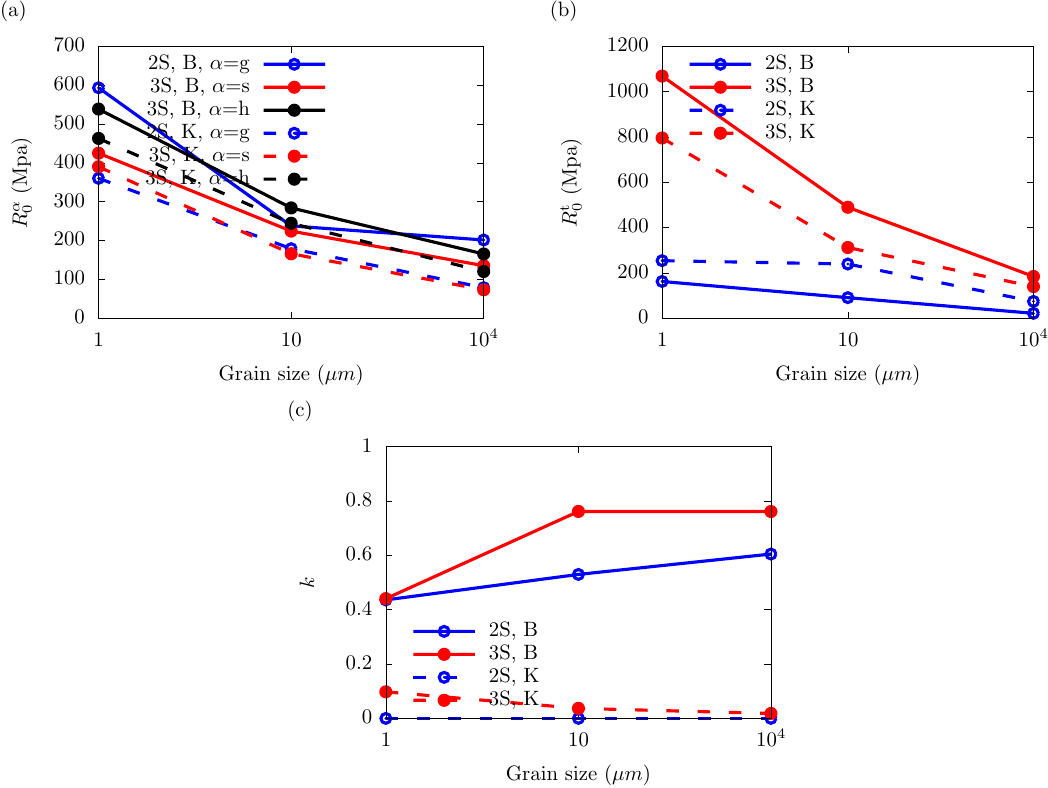}
    \caption{Calibrated parameters for the 2S and 3S models as a function of 
    grain size for two material textures: strong (B) and weak (K).
    (a) initial glide strengths, (b) initial twinning strength, and (c) tension-compression asymmetry parameter}
	\label{fig:params_gs}
\end{figure}

The role of grain size on these parameters is also assessed. 
For brevity, we show how the initial yield strengths in
glide (\cref{fig:params_gs}{\blue a}) and twinning (\cref{fig:params_gs}{\blue b}), and
the tension-compression asymmetry parameter (\cref{fig:params_gs}{\blue c})
depend on the grain size for two textures, B (strong) and K (random). 
As expected, the glide yield strengths decrease with increasing grain size ({\it a l{\' a}} Hall-Petch effect). 
For both models, the weaker texture (K) has a
lower glide strength than the strong texture (B) for all grain sizes.
A similar grain size dependence is also evident in the twinning yield strength ($R_0^{\rm t}$).
Notably, the 2S model shows a weaker dependence of $R_0^{\rm t}$ compared to
the 3S model. For both models, the grain size yield strengthening for 
the weaker texture (K) is less pronounced compared to the strong texture (B). 

The tension-compression asymmetry parameter ($k$, cf.~\cref{eq:sbart}) shows similar dependence on grain size 
for both models. For the weaker texture (K), it is small ($\ll 1$) and shows a
negligible dependence on grain size.
By contrast, for the strong texture (B), $k$ shows a strong inverse dependence on grain size, 
indicating a reduction in the tension-compression asymmetry with decreasing grain size.

Thus, while textural strengthening tends to increase tension-compression asymmetry
(for a fixed grain size),  grain size refinement tends to decrease it (for a fixed texture). 

\section{Discussion} \label{sec:disc}

The coarse-graining of the constitutive descriptions in both models is 
rather extreme: each material point represents a polycrystalline ensemble, 
and the homogenized response emerging from the grain-scale crystal plasticity 
is modeled using either two or three yield potentials. 
The double-blind calibration of the two models is performed 
to an extensive CP dataset representing synthetic surrogates of real magnesium alloys. 
The foregoing results sample a subset of the vast number of calculations performed in this work
to illustrate the main features of both models and to compare their performances. 

The success of both models lies in the fact that they ably predict
complex features of macroscopic deformation anisotropy and the underlying mechanisms
across the range of loading orientations, material textures, and grain sizes. 

It is not the aim to offer a verdict on the quality of the two models. Yet, if 
one seeks a simple takeaway from the results, it may be the following:
A quantitative estimate of the global errors suggests that,
the 2S model does a better job of fitting the macroscopic stress-strain characteristics,
while the 3S model is more capable at predicting the 
deformation anisotropy and undelying micromechanics.
That said, neither model exhibits any major departures 
from the ground-truth (i.e., CP) data in terms of the global errors
in stresses, strains, and relative mechanism activities. 

As both models turn out to be quite capable, we close with a discussion
on two points:
(i) when and why there is a potential need for a third surface and
(ii) the prospective of coarse-grained models in developing damage models.

The three-surface model \citep{Indurkar23} was developed to account for the
disparate nature of slip in the basal and non-basal systems.
In pure magnesium single crystals, the yield strength of the basal slip is typically
two orders of magnitude lower than that of the pyramidal slip \citep{Zhang12}.
In addition, the basal and nonbasal yielding modes exhibit distinct self- and latent hardening behaviors,
which have implications on dictating ductility.
For example, the basal slip may trigger macroscopic shear instability while the nonbasal slip modes may not \citep{Indurkar23,indurkar2022micromechanics}.
Therefore, in situations where instability occurs, one may need a third surface that captures slip in the basal planes.

One of the lessons drawn from the systematic comparisons between the 2S and 3S models
is that macroscopic quantities, such as the flow stress or the plastic strain,
may not discriminate the two models. However, microscopic quantities, such as relative
activities, do reveal differences that can have implications where local stress buildup matters.
An important consideration is to employ the coarse-grained models in developing void nucleation criteria
applicable to Mg alloys, where twin cracks have been reported in the literature
\citep{barnett2007twinning,RodriguezAM}.
Work along these lines that employs a crystal plasticity formulation, \citep[e.g.,][]{Cheng15},
or atomistics, \citep[e.g.,][]{Beyerlein10b}, has been reported
by several investigators.
However, in the spirit of computational efficiency, the type of polycrystal
calculations carried out by \citet{Baweja23} can be repeated using the MSM formulations
per grain. Such calculations would deliver quantitative measures of stress concentration
where twin activity is maximum with the objective of developing a practical,
physics-informed microcrack nucleation criterion in Mg alloys.


Ultimately, the ductility of a polycrystalline material 
is connected to the physical mechanisms of nucleation, growth, and coalescence of voids.
Therefore, micromechanical simulations of voided unit cells 
are needed to discover superior microstructures (textures and grain sizes) that have improved ductility.
As mentioned earlier, highly resolved damage-free polycrystal plasticity simulations
\citep[e.g.,][]{Baweja23,indurkar2020predicting} are expensive,
let alone when voids are explicitly modeled \citep{baweja_phd2023}.
As coarse-grained MSMs can offer substantive gains in
CPU times over their CP counterparts \citep{Indurkar23},
they enable failure investigations from a damage-tolerant materials design 
standpoint that were hitherto impractical.
In the era of data-driven machine learning, MSMs offer a rigorous pathway to
generate meaningful data affordably.

On the other hand, it is analytically cumbersome to develop homogenized damage models
while the matrix is represented by a crystal plasticity framework.
The coarse-grained plasticity models enable the development of physics-based coupled plasticity-damage models;
a field in infancy \citep{Vignesh22}.
Thereby, structural simulations can be accelerated by several orders of magnitude.

\section{Conclusion}

This work uses a large suite of crystal-plasticity dataset,
spanning eleven textures, four grain sizes, and 
six loading orientations in tension and compression, 
to test how far plasticity in magnesium polycrystals can be coarse-grained
while still retaining the features that matter for design:
evolving anisotropy, tension–compression asymmetry,
and the partitioning of slip and twinning.
Within that setting, both the two-surface (2S) 
and three-surface (3S) multisurface models perform well
given the level of homogenization where 
a single material point represents an entire polycrystalline ensemble. 

Both models reproduce key mechanistic fingerprints 
that guide the macroscopic responses across the vast
spectrum of material microstructures tested in this work.
The small global errors in stress, lateral strain, 
and relative mechanism activities establishes them 
as credible surrogates for computationally prohibitive crystal plasticity,
particularly to explore the design space of 
microstructural degrees of freedom more economically 
than fully resolved polycrystal simulations would allow.
The comparative assessment clarifies the distinct roles of the two formulations.
The 2S model is effective at reproducing orientation-dependent stress–strain curves.
When the primary need is to capture global load-deformation behavior, for example,
in structural analyses or forming simulations that do not explicitly track individual
slip modes the 2S model offers an attractive reduced-order description.
By constrast, the 3S model allows for a better reproducibility 
of the deformation anisotropy and capturing the competition between 
soft and hard glide in orientations where multiple mechanisms are active.
In that sense, it mimics crystal plasticity more faithfully, 
particularly for strongly textured materials and for off-axis loading states.
This added structure becomes relevant in settings where instability, 
shear banding, or mechanism-dependent ductility limits are controlled 
by how basal slip, non-basal slip, and twinning interact.

The calibrated parameters for both representations show robust trends
with respect to the underlying microstructure in terms of their dependence 
of the textural strengths and the grain size. 
Beyond this, the way these trends differ between the 2S and 3S models,
especially in the evolution of twinning strength and asymmetry,
underlines the value of distinguishing basal and non-basal glide
when a closer link to the fine-scale deformation modes is desired.

In summary, the outcomes here suggest a pragmatic view of model choice.
For problems in which macroscopic stress-strain behavior over large portions 
of the texture/grain-size space is the main target, 
the 2S model appears sufficient and computationally attractive.
For problems aimed at predicting damage and failure by void mechanisms
or by instability in which the details of twinning and basal versus non-basal activities are central,
the 3S framework may offer a more suitable foundation for 
embedding micromechanically informed criteria. 

As a next step, the capabilities of these models under heterogeneous loading should be assessed, 
e.g.~under multiaxial conditions. 
One prospect is to develop homogenization-based porous 
multisurface plasticity formulations to enable mechanism-aware failure modeling 
in magnesium alloys at a cost that is tenable with large-scale structural simulations.





\section*{Acknowledgements}
RV and AAB acknowledge financial support from the National Science Foundation
(grant CMMI-1932975) and are grateful for the high performance research computing
(HPRC) resources provided by Texas A\&M University.
SD and SPJ acknowledge financial support from the National Science Foundation (grant CMMI-1932976). 
SPJ and SD acknowledge the use of the Carya cluster and the advanced support from the 
Research Computing Data Core (RCDC) at the University of Houston, United States to carry out the research presented here.

\begin{appendices}
	\renewcommand{\theequation}{\thesection-\arabic{equation}}
	\renewcommand\thefigure{\thesection-\arabic{figure}}
	\renewcommand\thetable{\thesection-\arabic{table}}
	\crefalias{section}{appendix}
	
	\setcounter{table}{0}
	\setcounter{figure}{0}
	\setcounter{equation}{0} 
\section{Two surface model parameters} \label{app:2S}

The optimization strategy for the 2S model follows \citep{Vignesh23}.
The Levernberg-Marquardt algorithm of the Z-set package \citep{Zset} was used.
The 24 model parameters were calibrated together in one stage.
And, bounds for the parameters were set to avoid convergence or oscillation to unphysical values.
The initial yield strengths $R^{\rm g}_0$, $R^{\rm t}_0$ and the hardening parameters
$Q_{1}^{\rm g}$, $b_1^{\rm g}$, $Q_{1}^{\rm g}$, $b_2^{\rm g}$, $Q^{\rm t}$, $b^{\rm t}$,
$\mathcal{H}^{\rm tg}$, $\mathcal{H}^{\rm gt}$ were set to a minimum value of zero.
The anisotropic coefficients $l_{\rm XY}^{\rm g}$, $l_{\rm XY}^{\rm t}$, $\mathcal{L}_{\rm XY}^{\rm t}$ were
set between 0.1 and 10.
Also, the asymmetry parameter $k$ must have a value in the range (-1,1).

The calibration was first performed for texture A.
The initial value for the 24 parameters of texture A was taken as in \cite{Kondori19}.
Then, twelve material point calculations (tension and compression along principal and off-axes directions)
were carried out in parallel.
Each simulation took $\sim$30 seconds.
The cost is then calculated using \cref{eq:cost}.
Based on the increase or decrease of the cost, the optimizer suggests a new parameter set for the next iteration.
This process was repeated until the cost reaches a minimum value or further minimization is impossible.
A total of $\sim$500 iterations were required to reach convergence.
For texture B, the initial values of parameters were taken the same as the final converged values of A,
and similarly, the initial values of C were taken from the final values of texture B, and so on.
This choice prevents the optimizer from getting stuck at a local minimum.
The converged model parameters are listed in \cref{tab:2sparam}.
All the values are rounded to atmost 2 significant digits to save space.

{\footnotesize
\tabcolsep=3pt  
\begin{longtable}{ | p{0.6in} c c c c c c c c c c c | c c c | c c c |}
    \hline
Texture&	K&	J&	I&	H&	G&	C&	E&	F&	B&	D&	A&	K&	E&	B&	K&	E & B\\[1.5ex]
Grain size& \multicolumn{11}{c|}{$10^4\mu$m} & \multicolumn{3}{c|}{$10\mu$m} & \multicolumn{3}{c|}{$1\mu$m} \\[1.5ex]
Intensity&	3.9&	6.2&	7.7&	13&	18&	19&	19&	20&	24&	30&	31&	3.9&	19&	24&	3.9&	19&	24\\[2ex]
\hline
\hline                                   
\endfirsthead  
\endhead
\hline
Texture&	K&	J&	I&	H&	G&	C&	E&	F&	B&	D&	A&	K&	E&	B&	K&	E & B\\[1.5ex]
Grain size& \multicolumn{11}{c|}{$10^4\mu$m} & \multicolumn{3}{c|}{$10\mu$m} & \multicolumn{3}{c|}{$1\mu$m} \\[1.5ex]
Intensity&	3.9&	6.2&	7.7&	13&	18&	19&	19&	20&	24&	30&	31&	3.9&	19&	24&	3.9&	19&	24\\[1.5ex]  
\hline
\hline  
\endhead
\hline
\multicolumn{18}{r}{(continued)}
\endfoot
\endlastfoot
$l_{\rm TT}^{\rm g}$&	1&	0.99&	1&	0.99&	0.98&	0.94&	0.97&	0.96&	0.91&	1&	0.98&	0.98&	1&	1.1&	1&	1&	1.2\\[1.5ex]
$l_{\rm SS}^{\rm g}$&	0.98&	1&	1&	1&	1.1&	1.3&	1.3&	1.3&	1.4&	1.6&	1.6&	0.94&	1.3&	1.5&	0.97&	1.3&	1.6\\[1.5ex]
$l_{\rm TS}^{\rm g}$&	0.96&	1&	0.99&	1&	1.1&	1.1&	1.2&	1.2&	1.6&	0.67&	0.53&	0.95&	1.2&	1.5&	0.95&	1.3&	1.9\\[1.5ex]
$l_{\rm SL}^{\rm g}$&	0.95&	0.96&	1&	1&	1&	1.2&	1.2&	1.2&	2.3&	1.6&	2.1&	0.94&	1.2&	1.5&	0.98&	1.2&	2\\[1.5ex]
$l_{\rm LT}^{\rm g}$&	1&	1&	1&	1&	1.1&	1.2&	1.2&	1.2&	1.3&	1.4&	1.4&	0.98&	1.2&	1.4&	1&	1.2&	1.4\\[1.5ex]
$R_0^{\rm g}$&	79&	78&	150&	170&	110&	170&	170&	200&	200&	220&	210&	180&	220&	240&	360&	470&	590\\[1.5ex]
$Q_1^{\rm g}$&	280&	370&	230&	210&	89&	350&	390&	170&	620&	340&	440&	420&	220&	500&	430&	460&	6800\\[1.5ex]
$b_1^{\rm g}$&	0.77&	0.56&	1.3&	1.4&	5.9&	0.95&	0.87&	2.3&	0.46&	0.97&	0.68&	0.79&	79&	0.79&	3.1&	44&	0.17\\[1.5ex]
$Q_2^{\rm g}$&	140&	150&	74&	54&	130&	88&	95&	75&	88&	98&	110&	180&	1200&	270&	330&	940&	580\\[1.5ex]
$b_2^{\rm g}$&	30&	32&	84&	110&	160&	140&	140&	110&	150&	130&	130&	25&	0.29&	100&	24&	1.1&	57\\[1.5ex]
$\mathcal{H}^{\rm tg}$&	0&	0&	0&	0&	0&	0&	0&	0&	0&	0&	0&	0&	0&	0&	0&	0&	0\\[1.5ex]
$l_{\rm TT}^{\rm t}$&	0.74&	0.68&	1.2&	1.2&	1.2&	1.2&	1.1&	1.2&	0.89&	1.1&	1.1&	0.94&	0.94&	0.87&	0.57&	0.99&	1\\[1.5ex]
$l_{\rm SS}^{\rm t}$&	1.1&	1.1&	0.12&	0.1&	0.18&	0.19&	0.12&	0.26&	0.14&	0.1&	0.1&	1.6&	0.33&	0.21&	1.1&	0.32&	0.23\\[1.5ex]
$l_{\rm TS}^{\rm t}$&	0.36&	0.43&	0.65&	0.87&	1.1&	1.4&	1.2&	0.95&	3.5&	2.3&	2.9&	0.57&	0.95&	2.3&	0.45&	0.78&	1.1\\[1.5ex]
$l_{\rm SL}^{\rm t}$&	0.1&	0.1&	0.71&	1&	2.4&	2.4&	2.2&	1.9&	3&	6.1&	5.9&	0.1&	1.1&	1.5&	0.1&	0.97&	1.2\\[1.5ex]
$l_{\rm LT}^{\rm t}$&	0.1&	0.1&	0.48&	0.47&	0.41&	0.5&	0.49&	0.42&	0.39&	0.38&	0.38&	0.1&	0.27&	0.34&	0.1&	0.3&	0.4\\[1.5ex]
$\mathcal{L}_{\rm LT}^{\rm t}$&	1.2&	1.2&	0.67&	0.67&	0.69&	0.59&	0.58&	0.68&	0.54&	0.68&	0.67&	1.4&	0.7&	0.63&	1.1&	0.69&	0.63\\[1.5ex]
$\mathcal{L}_{\rm SL}^{\rm t}$&	1.1&	0.98&	0.65&	0.54&	0.42&	0.33&	0.32&	0.46&	0.2&	0.49&	0.46&	1.5&	0.34&	0.21&	1.1&	0.31&	0.65\\[1.5ex]
$\mathcal{L}_{\rm TS}^{\rm t}$&	0.71&	0.64&	0.19&	0.13&	0.12&	0.1&	0.11&	0.17&	0.33&	0.35&	0.34&	1&	0.42&	0.61&	0.6&	0.32&	0.74\\[1.5ex]
$k$&	0&	0&	0.18&	0.38&	0.64&	0.56&	0.56&	0.5&	0.6&	0.68&	0.66&	0.93&	0.58&	0.53&	0.35&	0.99&	0.44\\[1.5ex]
$R^{\rm t}_0$&	74&	98&	33&	30&	19&	27&	24&	20&	21&	18&	18&	240&	82&	90&	250&	280&	160\\[1.5ex]
$Q^{\rm t}$&	64&	110&	160&	240&	860&	22&	32&	730&	9.5&	14&	12&	390&	1.1&	0.016&	30&	7&	2.4\\[1.5ex]
$b^{\rm t}$&	0.59&	0.93&	2.8&	2.4&	1.3&	19&	16&	1&	21&	18&	20&	1.4&	39&	81&	47&	47&	27\\[1.5ex]
$\mathcal{H}^{\rm gt}$&	33&	35&	0.02&	0.02&	0.01&	0.02&	0.022&	0&	0&	0.02&	0.02&	0&	0&	0&	0&	0.02&	0\\[1.5ex]
\hline
\caption{2S model parameters for various textures and grain sizes}
\label{tab:2sparam}
\end{longtable}
}

\section{Three surface model parameters} \label{app:3S}
The optimization strategy for the 3S model follows \citep{Indurkar23}.
The calibration of 35 model parameters is carried out in two stages. 
In the first stage, the soft and hard glide yield strength and their hardening parameters, 
$R^{\rm s}_0$, $\mathcal{H}^{\rm s}$, $n$, $\mathcal{H}^{\rm hs}$,
$R^{\rm h}_0$, $\mathcal{H}^{\rm h}$, $Q^{\rm h}$, $b^{\rm h}$, $\mathcal{H}^{\rm sh}$
entering \cref{eq:Rs,eq:Rh} are identified.
And, in the next stage, the anisotropy coefficients of the two glide modes
$h_{\rm XY}^{\rm s}$, $h_{\rm XY}^{\rm h}$ entering \cref{eq:H},
the anisotropy coefficients of the twinning mode
$l_{\rm XY}^{\rm t}$, $\mathcal{L}_{\rm XY}^{\rm t}$, the
asymmetry parameter $k$ entering \cref{eq:Lt} and
the twinning yield strength and its parameters
$R^{\rm t}_0$, $\mathcal{H}^{\rm t},Q^{\rm t},b^{\rm t}$, $p^{\rm c}$, $\mathcal{H}^{\rm st}$,
$\mathcal{H}^{\rm ht}$ entering \cref{eq:Rt1} are identified.

For the first stage, only the stress-strain response for the tensile loading along L is considered.
Since the response to tensile loading in the L direction is known to have negligible influence from extension twinning, 
the nine parameters of the soft and hard glide modes can be calibrated independently of the twinning mode using \cref{eq:cost} with $N=1$.
The optimization problem is solved using the \texttt{fmincon} function of the MATLAB\textsuperscript{\textregistered} Optimization Toolbox.
The \texttt{fmincon} function is an interior point algorithm combined with a central difference scheme for the gradient approximation.
In this problem, each material parameter is perturbed twice, once by a small positive increment and once by a small negative increment,
while all other parameters are kept constant. Each of these perturbations is referred to as a function evaluation or \textit{f-count}.

For every \textit{f-count}, a 3S model simulation is executed using a single element model in ABAQUS/standard\textsuperscript{\textregistered},
utilizing the current set of parameters generated by the optimizer.
During optimization of the parameters, the remaining parameters are set to fixed values (if specified previously) or
assigned random values and held constant throughout the process.
After each simulation using the 3S model, the resulting stress-strain response is extracted, and the cost is evaluated.
A complete cycle of perturbing all nine variables once is termed an \textit{iteration}.
After each iteration, the optimizer updates the parameter set based on how the cost function responded to each individual \textit{f-count}.
In essence, whether a parameter change led to an increase or decrease in the cost function determines how that parameter is adjusted in the subsequent iteration. 

Also, we impose the condition that soft glide must always be active and should dominate over hard glide, particularly during the early stages of deformation.
This constraint is based on the well-established understanding in crystal plasticity that basal slip activates more easily than nonbasal slip systems.
Consequently, the L-direction response should initially yield through the soft glide mechanism. 
To enforce this constraint, we require that $R_0^{\rm s} < R_0^{\rm h}$.
If the optimizer generates parameter sets that violate this inequality,
the associated cost function is penalized by a factor of 10 to discourage the selection of such values in future iterations.
The initial guess for $R_0^{\rm s}$ is taken as the stress at 0.2\% strain from the CP data.
Once the L-tension response is successfully optimized, the resulting parameter values
enter the second stage of the optimization process. 

In the second stage, we focus on identifying the remaining 25 model parameters.
Throughout this stage, all but $\mathcal{H}^{\rm h}$ from the first stage remain fixed.
The parameter $\mathcal{H}^{\rm h}$ is not fixed in order to capture different hardening trends under tension and compression loadings.
Now, all the stress-strain responses are considered using $N=12$ in the cost function \cref{eq:cost}.
The optimization is again carried out using the same \texttt{fmincon} function.
The perturbation and parameter updation processes are the same as in the first stage.
Due to the high dimensionality of the 25+1-parameter space, the optimization process is
susceptible to becoming trapped in a local minimum.
To address this, the calibration is terminated if the total cost
does not decrease by more than 5\% over three consecutive \textit{iterations}.
The final calibrated parameters are listed in \cref{tab:3sparam}.
As before, all the values are rounded to atmost 2 significant digits to save space.
Also, the $h_{\rm SS}^{\alpha}$ parameters are
calculated using the relation $h_{\rm SS}^\alpha = 6 - 4h_{\rm LL}^\alpha - h_{\rm TT}^\alpha$
for all $\alpha \in \{\rm s, h\}$.

{\footnotesize
\tabcolsep=1pt  
\begin{longtable}{ | p{0.6in} c c c c c c c c c c c | c c c | c c c |}
    \hline
Texture&	K&	J&	I&	H&	G&	C&	E&	F&	B&	D&	A&	K&	E&	B&	K&	E & B\\[1.5ex]
Grain size& \multicolumn{11}{c|}{$10^4\mu$m} & \multicolumn{3}{c|}{$10\mu$m} & \multicolumn{3}{c|}{$1\mu$m} \\[1.5ex]
Intensity&	3.9&	6.2&	7.7&	13&	18&	19&	19&	20&	24&	30&	31&	3.9&	19&	24&	3.9&	19&	24\\[2ex]
\hline
\hline                                   
\endfirsthead  
\hline
Texture&	K&	J&	I&	H&	G&	C&	E&	F&	B&	D&	A&	K&	E&	B&	K&	E & B\\[1.5ex]
Grain size& \multicolumn{11}{c|}{$10^4\mu$m} & \multicolumn{3}{c|}{$10\mu$m} & \multicolumn{3}{c|}{$1\mu$m} \\[1.5ex]
Intensity&	3.9&	6.2&	7.7&	13&	18&	19&	19&	20&	24&	30&	31&	3.9&	19&	24&	3.9&	19&	24\\[1.5ex]  
\hline
\hline  
\endhead
\hline
\multicolumn{18}{r}{(continued)}
\endfoot
\endlastfoot
$h_{\rm LL}^{\rm s}$&	0.85&	0.76&	0.77&	1&	1&	1.1&	1.1&	1.1&	1.1&	1.1&	1.1&	0.88&	1.1&	1.1&	0.94&	1.1&	1.1\\[1.5ex]
$h_{\rm TT}^{\rm s}$&	1.2&	1.2&	1.1&	0.73&	0.53&	0.82&	0.93&	0.75&	1&	0.86&	0.96&	1.14&	0.66&	0.92&	1.1&	0.72&	0.96\\[1.5ex]
$h_{\rm LT}^{\rm s}$&	1.4&	2&	1.7&	1.8&	1.8&	2.3&	2.4&	2.2&	2.3&	2.2&	2.4&	1.64&	1.6&	2.2&	1.2&	1.4&	2.1\\[1.5ex]
$h_{\rm SL}^{\rm s}$&	1.3&	1.8&	2.4&	11&	10&	14&	12&	10&	34&	23&	32&	1.49&	4.4&	9.2&	1.3&	2.5&	6.2\\[1.5ex]
$h_{\rm TS}^{\rm s}$&	1.4&	2.2&	2.9&	21&	19&	29&	25&	23&	48&	89&	78&	1.59&	7&	13&	1.5&	3.5&	7.8\\[1.5ex]
$h_{\rm LL}^{\rm h}$&	0.99&	0.97&	0.97&	0.9&	0.94&	1&	1&	0.99&	1&	1.1&	1.1&	0.96&	1.1&	1.1&	0.97&	1.1&	1.1\\[1.5ex]
$h_{\rm TT}^{\rm h}$&	1&	0.98&	1&	1.1&	1.1&	1.2&	1.2&	1.1&	1.4&	1.2&	1.2&	1.04&	1.1&	0.96&	0.99&	1.1&	1\\[1.5ex]
$h_{\rm LT}^{\rm h}$&	1.2&	1.1&	1.1&	1.2&	1.2&	1.4&	1.3&	1.3&	1.5&	1.2&	1.3&	1.19&	1.2&	1.2&	1.3&	1.3&	1.4\\[1.5ex]
$h_{\rm SL}^{\rm h}$&	1.1&	1.2&	1.2&	1.7&	1.8&	2.2&	1.8&	2&	3.1&	1.6&	2.2&	1.18&	1.4&	2.3&	1.2&	1.6&	3.6\\[1.5ex]
$h_{\rm TS}^{\rm h}$&	0.99&	1.3&	1.2&	2&	2.6&	2.4&	2.1&	2.5&	3.3&	3&	3.7&	1.10&	1.7&	2.2&	1.1&	2&	3.5\\[1.5ex]
$k$&	0.018&	0.019&	0.16&	0.61&	0.76&	0.73&	0.76&	0.73&	0.76&	0.66&	0.77&	0.04&	0.65&	0.76&	0.098&	0.38&	0.44\\[1.5ex]
$\mathcal{L}_{\rm LT}^{\rm t}$&	0.05&	0.05&	0.058&	7.3&	8&	6.1&	6.6&	6.3&	5.2&	5.6&	5.5&	0.05&	6.1&	6.1&	0.015&	5.5&	3.4\\[1.5ex]
$\mathcal{L}_{\rm SL}^{\rm t}$&	0.05&	0.05&	0.049&	0.23&	0.19&	0.23&	0.23&	0.22&	0.24&	0.24&	0.24&	0.05&	0.23&	0.23&	0.025&	0.22&	0.21\\[1.5ex]
$l_{\rm TT}^{\rm t}$&	1&	1&	0.99&	2.4&	3.4&	2.2&	1.8&	2.2&	1.8&	1.9&	1.8&	0.86&	3&	2.1&	0.85&	1.4&	1.4\\[1.5ex]
$\mathcal{L}_{\rm TS}^{\rm t}$&	0.045&	0.045&	0.045&	-2.3&	-3.1&	-2.8&	-2.9&	-3&	-3&	-3&	-2.9&	0.04&	-2.4&	-2.7&	0.043&	-2.3&	-0.81\\[1.5ex]
$l_{\rm SS}^{\rm t}$&	1&	1&	-0.98&	-2.5&	-2.5&	-2.6&	-2.7&	-2.6&	-2.6&	-2.7&	-2.7&	0.88&	-2.6&	-2.2&	0.98&	-2.5&	-1.9\\[1.5ex]
$l_{\rm LT}^{\rm t}$&	1.1&	1.1&	1.1&	5.6&	6.4&	4.5&	4.2&	4.7&	4&	4.1&	4&	0.89&	4.2&	4.1&	0.82&	4.1&	2.3\\[1.5ex]
$l_{\rm SL}^{\rm t}$&	1&	1.1&	-1.2&	-4.1&	-4.6&	-5.1&	-3.9&	-4&	7.5&	-3.6&	-4.2&	0.86&	-4.6&	-8.6&	0.83&	-4.1&	-4.1\\[1.5ex]
$l_{\rm TS}^{\rm t}$&	1&	1.1&	-1.2&	-3.3&	-4.2&	-3.5&	-3.9&	-3.8&	-4&	-5.6&	-5.4&	0.85&	-4.2&	-5.6&	0.78&	-3.8&	-2.3\\[1.5ex]
$R^{\rm s}_0$ &	73&	95&	95&	120&	130&	120&	120&	130&	140&	140&	140&	165.68&	220&	220&	390&	470&	420\\[1.5ex]
$\mathcal{H}^{\rm s}$&	5000&	3300&	3800&	3300&	3400&	3400&	3300&	3300&	3600&	3600&	3600&	4852.20&	5100&	5600&	8000&	6000&	11000\\[1.5ex]
$\mathcal{H}^{\rm hs}$&	500&	1300&	1300&	80&	80&	80&	80&	80&	80&	80&	80&	549.74&	30&	80&	1600&	380&	80\\[1.5ex]
$R^{\rm h}_0$ &	120&	120&	120&	150&	170&	160&	160&	160&	170&	160&	160&	244.27&	310&	280&	460&	700&	540\\[1.5ex]
$Q^{\rm h}$ &	100&	100&	100&	80&	64&	88&	88&	79&	88&	93&	90&	127.82&	75&	110&	380&	140&	270\\[1.5ex]
$b^{\rm h}$ &	40&	40&	51&	90&	110&	120&	110&	110&	200&	130&	160&	47.69&	99&	98&	22&	45&	39\\[1.5ex]
$\mathcal{H}^{\rm sh}$ &	90&	280&	230&	480&	580&	330&	280&	550&	200&	180&	250&	162.07&	210&	210&	360&	430&	1400\\[1.5ex]
$R^{\rm t}_0$ &	140&	140&	140&	160&	140&	160&	150&	160&	180&	160&	160&	311.49&	460&	490&	790&	800&	1100\\[1.5ex]
$Q^{\rm t}$ &	0.16&	0.16&	0.17&	0.15&	0.15&	0.15&	0.15&	0.15&	0.15&	0.22&	0.22&	0.32&	0.15&	0.15&	0.12&	0.16&	0.15\\[1.5ex]
$b^{\rm t}$ &	72&	71&	71&	260&	310&	260&	230&	260&	260&	250&	230&	92.04&	290&	310&	71&	370&	410\\[1.5ex]
$\mathcal{H}^{\rm st}$ &	1200&	1300&	1400&	1400&	1100&	1500&	1600&	1500&	1700&	1700&	1500&	847.30&	1400&	1600&	1200&	1400&	1600\\[1.5ex]
$\mathcal{H}^{\rm ht}$ &	1300&	1100&	1500&	3900&	3600&	5600&	6000&	5300&	6200&	6100&	6100&	887.60&	5800&	5600&	1300&	5500&	5900\\[1.5ex]
$N$&	1&	1.1&	1.1&	1.5&	1.5&	1.5&	1.5&	1.5&	1.5&	1.5&	1.5&	1.15&	1.1&	1.7&	1&	1&	1.5\\[1.5ex]
$\mathcal{H}^{\rm t}$ &	970&	940&	1400&	2100&	1900&	1400&	1200&	1500&	1200&	1400&	1300&	1130&	1600&	1700&	1400&	1800&	1800\\[1.5ex]
$\mathcal{H}^{\rm h}$ &	170&	150&	130&	150&	130&	170&	140&	130&	230&	83&	130&	257.37&	260&	150&	330&	610&	740\\[1.5ex]
$p^{\rm c}$&	0.062&	0.065&	0.05&	0.05&	0.059&	0.048&	0.046&	0.051&	0.048&	0.047&	0.044&	0.03&	0.046&	0.048&	0.063&	0.048&	0.04\\[1.5ex]
\hline
\caption{3S model parameters for various textures and grain sizes}
\label{tab:3sparam}
\end{longtable}
}

\end{appendices}

\bibliographystyle{unsrt-short-author}
\bibliography{vignesh.bib}

@article{hosford_msea98,
  title={Reflections on the dependence of plastic anisotropy on texture},
  author={Hosford, William F},
  journal={Materials Science and Engineering: A},
  volume={257},
  number={1},
  pages={1--8},
  year={1998},
  publisher={Elsevier}
}

@article{wenk_rev_min_geo02,
  title={Texture and anisotropy},
  author={Wenk, Hans-Rudolph},
  journal={Reviews in mineralogy and geochemistry},
  volume={51},
  number={1},
  pages={291--329},
  year={2002},
  publisher={Mineralogical Society of America}
}

@phdthesis{baweja_phd2023,
  title={Understanding Structure-Property Linkages in Magnesium Alloys via Size-Dependent Crystal Plasticity Modeling},
  author={Baweja, Shahmeer},
  year={2023},
  school = {University of Houston}
}

@article{hufnagel_mecmat22,
  title={Magnesium alloy design: Examples from the materials in extreme dynamic environments metals collaborative research group},
  author={Hufnagel, Todd C and Lloyd, Jeffrey T and Weihs, Timothy P and Kecskes, Laszlo J and Sano, Tomoko},
  journal={Mechanics of Materials},
  volume={165},
  pages={104136},
  year={2022},
  publisher={Elsevier}
}

@Article{Yu_JMST18,
	author = {Yu, Huihui and Xin, Yunchang and Wang, Maoyin and Liu, Qing},
	journal = {Journal of Materials Science and Technology},
	title = {{Hall-Petch relationship in Mg alloys: A review}},
	year = {2018},
	issn = {1005-0302},
	month = feb,
	number = {2},
	pages = {248--256},
	volume = {34},
	doi = {10.1016/j.jmst.2017.07.022},
	publisher = {Chinese Society of Metals}
}

@article{wei_AM21,
  title={Grain size effect on tensile properties and slip systems of pure magnesium},
  author={Wei, Kang and Hu, Rong and Yin, Dongdi and Xiao, Lirong and Pang, Song and Cao, Yang and Zhou, Hao and Zhao, Yonghao and Zhu, Yuntian},
  journal={Acta Materialia},
  volume={206},
  pages={116604},
  year={2021},
  publisher={Elsevier}
}

@article{masood_CritRev22,
  title={Designing highly ductile magnesium alloys: Current status and future challenges},
  author={Masood Chaudry, Umer and Tekumalla, Sravya and Gupta, Manoj and Jun, Tea-Sung and Hamad, Kotiba},
  journal={Critical Reviews in Solid State and Materials Sciences},
  volume={47},
  number={2},
  pages={194--281},
  year={2022},
  publisher={Taylor \& Francis}
}

@article{selvarajou_AM17,
  title={Three dimensional simulations of texture and triaxiality effects on the plasticity of magnesium alloys},
  author={Selvarajou, Balaji and Joshi, Shailendra P and Benzerga, A Amine},
  journal={Acta Materialia},
  volume={127},
  pages={54--72},
  year={2017},
  publisher={Elsevier}
}

@article{li2010,
  title={An efficient constitutive model for room-temperature, low-rate plasticity of annealed {M}g {AZ31B} sheet},
  author={Li, M and Lou, XY and Kim, JH and Wagoner, RH},
  journal={International Journal of Plasticity},
  volume={26},
  number={6},
  pages={820--858},
  year={2010},
  publisher={Elsevier}
}

@article{kim2017,
  title={Distortional hardening concept for modeling anisotropic/asymmetric plastic behavior of {AZ31B} magnesium alloy sheets},
  author={Lee, J and Kim, S-J and Lee, Y-S and Lee, J-Y and Kim, D and Lee, M-G},
  journal={International Journal of Plasticity},
  volume={94},
  pages={74--97},
  year={2017},
  publisher={Elsevier}
}

@string{ IJSS = "Int. J. Solids Struct."}

@article{Chang15,
	title        = {A variational constitutive model for slip-twinning interactions in hcp metals: application to single-and polycrystalline magnesium},
	author       = {Chang, Yingrui and Kochmann, Dennis M},
	year         = 2015,
	journal      = {International Journal of Plasticity},
	publisher    = {Elsevier},
	volume       = 73,
	pages        = {39--61},
}

@article{indurkar2022micromechanics,
	title        = {On the micromechanics of void mediated failure in HCP crystals},
	author       = {Indurkar, Padmeya P and Joshi, Shailendra P and Benzerga, A Amine},
	year         = 2022,
	journal      = {Journal of the Mechanics and Physics of Solids},
	publisher    = {Elsevier},
	pages        = 104923,
}

@article{Hill48,
	title        = {A theory of the yielding and plastic flow of anisotropic metals},
	author       = {Hill, Rodney},
	year         = 1948,
	journal      = {Proceedings of the Royal Society of London. Series A. Mathematical and Physical Sciences},
	publisher    = {The Royal Society London},
	volume       = 193,
	number       = 1033,
	pages        = {281--297},
}

@article{Steglich16,
	title        = {Mechanism-based modelling of plastic deformation in magnesium alloys},
	author       = {Steglich, Dirk and Tian, Xiaowei and Besson, Jacques},
	year         = 2016,
	journal      = {European Journal of Mechanics-A/Solids},
	publisher    = {Elsevier},
	volume       = 55,
	pages        = {289--303},
}

@article{steglich2018,
  title={Fracture prediction based on a two-surface plasticity law for the anisotropic magnesium alloys {AZ31} and {ZE10}},
  author={Lee, Jeong-Yeon and Steglich, Dirk and Lee, Myoung-Gyu},
  journal={International Journal of Plasticity},
  volume={105},
  pages={1--23},
  year={2018},
  publisher={Elsevier}
}

@article{Baweja23,
  title={Three-dimensional computational characterization of grain size and texture effects in magnesium alloys},
  author={Baweja, Shahmeer and Joshi, Shailendra P},
  journal={Journal of Magnesium and Alloys},
  volume={11},
  number={10},
  pages={3657--3672},
  year={2023},
  publisher={Elsevier}
}

@article{indurkar2020predicting,
	title        = {Predicting textural variability effects in the anisotropic plasticity and stability of hexagonal metals: Application to magnesium and its alloys},
	author       = {Indurkar, Padmeya P and Baweja, Shahmeer and Perez, Robert and Joshi, Shailendra P},
	year         = 2020,
	journal      = {International Journal of Plasticity},
	publisher    = {Elsevier},
	volume       = 132,
	pages        = 102762,
}

@article{barnett2007twinning,
	title        = {Twinning and the ductility of magnesium alloys: Part I:“Tension” twins},
	author       = {Barnett, MR},
	year         = 2007,
	journal      = {Materials Science and Engineering: A},
	publisher    = {Elsevier},
	volume       = 464,
	number       = {1-2},
	pages        = {1--7},
}

@article{Barlat91,
	title        = {A six-component yield function for anisotropic materials},
	author       = {Barlat, Fr{\'e}d{\'e}ric and Lege, Daniel J and Brem, John C},
	year         = 1991,
	journal      = {International journal of plasticity},
	publisher    = {Elsevier},
	volume       = 7,
	number       = 7,
	pages        = {693--712},
}

@article{Kalidindi98,
	title        = {Incorporation of deformation twinning in crystal plasticity models},
	author       = {Kalidindi, Surya R},
	year         = 1998,
	journal      = {Journal of the Mechanics and Physics of Solids},
	publisher    = {Elsevier},
	volume       = 46,
	number       = 2,
	pages        = {267--290},
}

@article{Zhang12,
	title        = {Phenomenological crystal plasticity modeling and detailed micromechanical investigations of pure magnesium},
	author       = {Zhang, Jing and Joshi, Shailendra P},
	year         = 2012,
	journal      = {Journal of the Mechanics and Physics of Solids},
	publisher    = {Elsevier},
	volume       = 60,
	number       = 5,
	pages        = {945--972},
}

@article{Agnew06,
	title        = {Validating a polycrystal model for the elastoplastic response of magnesium alloy AZ31 using in situ neutron diffraction},
	author       = {Agnew, SR and Brown, DW and Tom{\'e}, CN},
	year         = 2006,
	journal      = {Acta materialia},
	publisher    = {Elsevier},
	volume       = 54,
	number       = 18,
	pages        = {4841--4852},
}

@article{Lebensohn93,
	title        = {A self-consistent anisotropic approach for the simulation of plastic deformation and texture development of polycrystals: application to zirconium alloys},
	author       = {Lebensohn, Ricardo A and Tom{\'e}, CN},
	year         = 1993,
	journal      = {Acta metallurgica et materialia},
	publisher    = {Elsevier},
	volume       = 41,
	number       = 9,
	pages        = {2611--2624},
}

@article{Cazacu06,
	title        = {Orthotropic yield criterion for hexagonal closed packed metals},
	author       = {Cazacu, Oana and Plunkett, Brian and Barlat, Fr{\'e}d{\'e}ric},
	year         = 2006,
	journal      = {International Journal of Plasticity},
	publisher    = {Elsevier},
	volume       = 22,
	number       = 7,
	pages        = {1171--1194},
}

@article{Becker16,
	title        = {A reduced-order crystal model for HCP metals: application to Mg},
	author       = {Becker, R and Lloyd, JT},
	year         = 2016,
	journal      = {Mechanics of Materials},
	publisher    = {Elsevier},
	volume       = 98,
	pages        = {98--110},
}

@article{Kondori19,
	title        = {Evolution of the 3D plastic anisotropy of HCP metals: experiments and modeling},
	author       = {Kondori, B and Madi, Yazid and Besson, Jacques and Benzerga, AA},
	year         = 2019,
	journal      = {International Journal of Plasticity},
	publisher    = {Elsevier},
	volume       = 117,
	pages        = {71--92},
}

@article{Benzerga01,
	title        = {Plastic potentials for anisotropic porous solids},
	author       = {Benzerga, Ahmed Amine and Besson, Jacques},
	year         = 2001,
	journal      = {European Journal of Mechanics-A/Solids},
	publisher    = {Elsevier},
	volume       = 20,
	number       = 3,
	pages        = {397--434},
}

@article{Indurkar23,
	title        = {A mechanism-based multisurface plasticity model for hexagonal close-packed materials with detailed validation and assessment},
	author       = {Indurkar, Padmeya P and Joshi, Shailendra P},
	year         = 2023,
	journal      = {Journal of the Mechanics and Physics of Solids},
	publisher    = {Elsevier},
	volume       = 176,
	pages        = 105302,
}

@article{Vignesh23,
  title={Assessment of a two-surface plasticity model for hexagonal materials},
  author={Vigneshwaran, R and Benzerga, AA},
  journal={Journal of Magnesium and Alloys},
    volume={11},
  number={12},
  pages={4431--4444},
  year={2023},
  publisher={Elsevier}
}

@misc{Zset,
	title        = {software package by Ecole des Mines ParisTech (France) and Onera-the French Aerospace Lab},
	author       = {Z-set/ZeBuLoN},
	year         = {9.1.3},
	howpublished = {http://www.zset-software.com},
}

@incollection{Vignesh22,
	title        = {A Predictive Multisurface Approach to Damage Modeling in Mg Alloys},
	author       = {Vigneshwaran, R and Benzerga, A A},
	year         = 2022,
	booktitle    = {Magnesium Technology 2022},
	publisher    = {Springer},
	pages        = {293--297},
}

@article{Ravaji21,
  title={A crystal plasticity investigation of grain size-texture interaction in magnesium alloys},
  author={Ravaji, Babak and Joshi, Shailendra P},
  journal={Acta Materialia},
  volume={208},
  pages={116743},
  year={2021},
  publisher={Elsevier}
}

@incollection{Herrington19,
  title={Modeling the 3D plastic anisotropy of a magnesium alloy processed using severe plastic deformation},
  author={Herrington, JS and Madi, Yazid and Besson, Jacques and Benzerga, AA},
  booktitle={Magnesium technology 2019},
  pages={283--287},
  year={2019},
  publisher={Springer}
}

@article{Basu17,
	title        = {Towards designing anisotropy for ductility enhancement: A theory-driven investigation in Mg-alloys},
	author       = {Basu, Shamik and Dogan, E and Kondori, B and Karaman, I and Benzerga, AA},
	year         = 2017,
	journal      = {Acta Materialia},
	publisher    = {Elsevier},
	volume       = 131,
	pages        = {349--362},
}

@string{ ActaMat = "Acta Mater."}

@article{bohlen2007,
  title={The texture and anisotropy of magnesium--zinc--rare earth alloy sheets},
  author={Bohlen, Jan and N{\"u}rnberg, Marcus R and Senn, Jeremy W and Letzig, Dietmar and Agnew, Sean R},
  journal={Acta Materialia},
  volume={55},
  number={6},
  pages={2101--2112},
  year={2007},
  publisher={Elsevier}
}

@article{agnew2005plastic,
  title={Plastic anisotropy and the role of non-basal slip in magnesium alloy {AZ31B}},
  author={Agnew, Sean R and Duygulu, {\"O}zg{\"u}r},
  journal={International Journal of plasticity},
  volume={21},
  number={6},
  pages={1161--1193},
  year={2005},
  publisher={Elsevier}
}

@article{jedidi20,
  title={Prediction of necking in HCP sheet metals using a two-surface plasticity model},
  author={Jedidi, Mohamed Yassine and Bettaieb, M Ben and Abed-Meraim, Farid and Khabou, Mohamed Taoufik and Bouguecha, Anas and Haddar, Mohamed},
  journal={International Journal of Plasticity},
  volume={128},
  pages={102641},
  year={2020},
  publisher={Elsevier}
}

@article{lei23,
  title={Multi-mechanism constitutive model for uniaxial ratchetting of extruded AZ31 magnesium alloy at room temperature},
  author={Lei, Yu and Yu, Chao and Wang, Ziyi and Xu, Xiang and Li, Hang and Kang, Guozheng},
  journal={Mechanics of Materials},
  volume={179},
  pages={104607},
  year={2023},
  publisher={Elsevier}
}

@article{SoareIJSS,
    author="Soare, S. and Benzerga, A. A.",
    title="{On the modeling of asymmetric yield functions}",
    journal=IJSS,
    volume=80, pages="486-500", year=2016}

@ARTICLE{Cheng15,
author={Cheng, J. and Ghosh, S.},
title={{A crystal plasticity FE model for deformation with twin nucleation in magnesium alloys}},
journal= IJP,
year={2015},
volume={67},
pages={148--170}
}

@article{Beyerlein10b,
Author = {Wang, J. and Beyerlein, I. J. and Tome, C. N.},
Title = {{An atomic and probabilistic perspective on twin nucleation in Mg}}, 
Journal = SM,
Year = "2010", Volume = "63", Pages = "741-746"}

@article{RodriguezAM,
    author="Rodriguez, A. K. and Ayoub, G. and Mansoor, B. and Benzerga, A. A.",
    title="Effect of strain rate and temperature on fracture of {AZ31B} magnesium alloy",
    journal=ActaMat,
    volume=112, pages="194--208", year=2016}

\newpage

\setcounter{table}{0}
\setcounter{figure}{0}
\setcounter{equation}{0}
\setcounter{section}{0}
\setcounter{subsection}{0}
\setcounter{page}{1}

\newcounter{pno}
\setcounter{pno}{1}
\startcontents[part2toc]
\renewcommand\thefigure{S-\arabic{figure}}
\renewcommand{\thepage}{S\arabic{page}}

\begin{center}
\Large
    Supplementary material for ``An assessment of mechanism-based plasticity models for polycrystalline magnesium alloys"
\end{center}

The model parameters are calibrated for all eleven textures (A--K) having the base grain size, $\bar d = 10^4 \mu$m.
Furthermore, the models are calibrated for two more grain sizes $\bar d = 10\mu$m and 1 $\mu$m
of three selected textures, B, E, and K, which are representative
of the strong, intermediate, and weak group, respectively.
In all, the model parameters are calibrated for 17 different sets of materials.
For brevity, comparisons between the models and the CP data are presented
only for materials B and K with the base grain size in the main text,
the additional results are presented here.

\begin{singlespace}
    \printcontents[part2toc]{l}{1}{{\section*{Contents}}}
\end{singlespace}
\section{Texture A}

\begin{figure}[h!]
	\centering
	\includegraphics[width=0.95\textwidth,page=\thepno]{Recursive_plot.pdf}
	\caption{Calibrated stress-strain responses for \underline{material A} with $\bar d = 10^4 \mu$m 
    under uniaxial loading along principal material (left column) and off-axis (right column) directions. 
    {\bf Symbols}: CP data \citep{Baweja23},
	{\bf Dashed lines}: 2S model, {\bf Solid lines}: 3S model. 
    {\red Red}: Compressive responses, {\blue Blue}: Tensile responses.
    }
	\label{fig:A_stress}
\end{figure}
\stepcounter{pno}





\begin{figure}[h!]
	\centering
	\includegraphics[width=0.95\textwidth,page=\thepno]{Recursive_plot.pdf}
	\caption{Predicted lateral strain, $E_{xx}$, for \underline{material A} with $\bar d = 10^4 \mu$m 
    under uniaxial loading along principal material (left column) and off-axis (right column) directions. 
    {\bf Symbols}: CP data \citep{Baweja23},
	{\bf Dashed lines}: 2S model, {\bf Solid lines}: 3S model. 
    {\red Red}: Compressive responses, {\blue Blue}: Tensile responses.
    }
	\label{fig:A_Exx}
\end{figure}
\stepcounter{pno}

\begin{figure}[h!]
	\centering
	\includegraphics[width=0.95\textwidth,page=\thepno]{Recursive_plot.pdf}
	\caption{Predicted lateral strain, $E_{zz}$, for \underline{material A} with $\bar d = 10^4 \mu$m 
    under uniaxial loading along principal material (left column) and off-axis (right column) directions. 
    {\bf Symbols}: CP data \citep{Baweja23},
	{\bf Dashed lines}: 2S model, {\bf Solid lines}: 3S model. 
    {\red Red}: Compressive responses, {\blue Blue}: Tensile responses.
    }
	\label{fig:A_Ezz}
\end{figure}
\stepcounter{pno}
\stepcounter{pno}

\begin{figure}[h!]
	\centering
	\includegraphics[width=0.95\textwidth,page=\thepno]{Recursive_plot.pdf}
	\caption{Predicted relative cumulative activity of glide, $\xi^{\rm g} = p^{\rm g}/p$, for \underline{material A} with $\bar d = 10^4 \mu$m 
    under uniaxial loading along principal material (left column) and off-axis (right column) directions. 
    {\bf Symbols}: CP data \citep{Baweja23},
	{\bf Dashed lines}: 2S model, {\bf Solid lines}: 3S model. 
    {\red Red}: Compressive responses, {\blue Blue}: Tensile responses.
    }
	\label{fig:A_pg}
\end{figure}
\stepcounter{pno}




\begin{figure}[h!]
	\centering
	\includegraphics[width=0.95\textwidth,page=\thepno]{Recursive_plot.pdf}
	\caption{Predicted relative activity of soft glide ($\xi^{\rm s} = p^{\rm s}/p$) by the 3S model for \underline{material A} with $\bar d = 10^4 \mu$m 
    under uniaxial loading along principal material (left column) and off-axis (right column) directions. 
    {\bf Symbols}: CP data \citep{Baweja23}. 
    {\red Red}: Compressive responses, {\blue Blue}: Tensile responses.
    }
	\label{fig:A_ps}
\end{figure}
\stepcounter{pno}

\clearpage
\section{Texture B}
\stepcounter{pno}
\stepcounter{pno}
\stepcounter{pno}
\stepcounter{pno}
\stepcounter{pno}
\stepcounter{pno}

\setcounter{pno}{74}
\subsection{Grain size, $\bar d = 10 \mu$m}

\begin{figure}[h!]
	\centering
	\includegraphics[width=0.95\textwidth,page=\thepno]{Recursive_plot.pdf}
	\caption{Calibrated stress-strain responses for \underline{material B} with $\bar d = 10 \mu$m 
    under uniaxial loading along principal material (left column) and off-axis (right column) directions. 
    {\bf Symbols}: CP data \citep{Baweja23},
	{\bf Dashed lines}: 2S model, {\bf Solid lines}: 3S model. 
    {\red Red}: Compressive responses, {\blue Blue}: Tensile responses.
    }
	\label{fig:B_stress_10}
\end{figure}
\stepcounter{pno}

\begin{figure}[h!]
	\centering
	\includegraphics[width=0.95\textwidth,page=\thepno]{Recursive_plot.pdf}
	\caption{Predicted lateral strain, $E_{xx}$, for \underline{material B} with $\bar d = 10 \mu$m 
    under uniaxial loading along principal material (left column) and off-axis (right column) directions. 
    {\bf Symbols}: CP data \citep{Baweja23},
	{\bf Dashed lines}: 2S model, {\bf Solid lines}: 3S model. 
    {\red Red}: Compressive responses, {\blue Blue}: Tensile responses.
    }
	\label{fig:B_Exx_10}
\end{figure}
\stepcounter{pno}

\begin{figure}[h!]
	\centering
	\includegraphics[width=0.95\textwidth,page=\thepno]{Recursive_plot.pdf}
	\caption{Predicted lateral strain, $E_{zz}$, for \underline{material B} with $\bar d = 10 \mu$m 
    under uniaxial loading along principal material (left column) and off-axis (right column) directions. 
    {\bf Symbols}: CP data \citep{Baweja23},
	{\bf Dashed lines}: 2S model, {\bf Solid lines}: 3S model. 
    {\red Red}: Compressive responses, {\blue Blue}: Tensile responses.
    }
	\label{fig:B_Ezz_10}
\end{figure}
\stepcounter{pno}
\stepcounter{pno}

\begin{figure}[h!]
	\centering
	\includegraphics[width=0.95\textwidth,page=\thepno]{Recursive_plot.pdf}
	\caption{Predicted relative cumulative activity of glide, $\xi^{\rm g} = p^{\rm g}/p$, for \underline{material B} with $\bar d = 10 \mu$m 
    under uniaxial loading along principal material (left column) and off-axis (right column) directions. 
    {\bf Symbols}: CP data \citep{Baweja23},
	{\bf Dashed lines}: 2S model, {\bf Solid lines}: 3S model. 
    {\red Red}: Compressive responses, {\blue Blue}: Tensile responses.
    }
	\label{fig:B_pg_10}
\end{figure}
\stepcounter{pno}

\begin{figure}[h!]
	\centering
	\includegraphics[width=0.95\textwidth,page=\thepno]{Recursive_plot.pdf}
	\caption{Predicted relative activity of soft glide ($\xi^{\rm s} = p^{\rm s}/p$) by the 3S model for \underline{material B} with $\bar d = 10 \mu$m 
    under uniaxial loading along principal material (left column) and off-axis (right column) directions. 
    {\bf Symbols}: CP data \citep{Baweja23}. 
    {\red Red}: Compressive responses, {\blue Blue}: Tensile responses.
    }
	\label{fig:B_ps_10}
\end{figure}
\stepcounter{pno}
\clearpage

\setcounter{pno}{92}
\subsection{Grain size, $\bar d = 1 \mu$m}

\begin{figure}[h!]
	\centering
	\includegraphics[width=0.95\textwidth,page=\thepno]{Recursive_plot.pdf}
	\caption{Calibrated stress-strain responses for \underline{material B} with $\bar d = 1 \mu$m 
    under uniaxial loading along principal material (left column) and off-axis (right column) directions. 
    {\bf Symbols}: CP data \citep{Baweja23},
	{\bf Dashed lines}: 2S model, {\bf Solid lines}: 3S model. 
    {\red Red}: Compressive responses, {\blue Blue}: Tensile responses.
    }
	\label{fig:B_stress_1}
\end{figure}
\stepcounter{pno}

\begin{figure}[h!]
	\centering
	\includegraphics[width=0.95\textwidth,page=\thepno]{Recursive_plot.pdf}
	\caption{Predicted lateral strain, $E_{xx}$, for \underline{material B} with $\bar d = 1 \mu$m 
    under uniaxial loading along principal material (left column) and off-axis (right column) directions. 
    {\bf Symbols}: CP data \citep{Baweja23},
	{\bf Dashed lines}: 2S model, {\bf Solid lines}: 3S model. 
    {\red Red}: Compressive responses, {\blue Blue}: Tensile responses.
    }
	\label{fig:B_Exx_1}
\end{figure}
\stepcounter{pno}

\begin{figure}[h!]
	\centering
	\includegraphics[width=0.95\textwidth,page=\thepno]{Recursive_plot.pdf}
	\caption{Predicted lateral strain, $E_{zz}$, for \underline{material B} with $\bar d = 1 \mu$m 
    under uniaxial loading along principal material (left column) and off-axis (right column) directions. 
    {\bf Symbols}: CP data \citep{Baweja23},
	{\bf Dashed lines}: 2S model, {\bf Solid lines}: 3S model. 
    {\red Red}: Compressive responses, {\blue Blue}: Tensile responses.
    }
	\label{fig:B_Ezz_1}
\end{figure}
\stepcounter{pno}
\stepcounter{pno}

\begin{figure}[h!]
	\centering
	\includegraphics[width=0.95\textwidth,page=\thepno]{Recursive_plot.pdf}
	\caption{Predicted relative cumulative activity of glide, $\xi^{\rm g} = p^{\rm g}/p$, for \underline{material B} with $\bar d = 1 \mu$m 
    under uniaxial loading along principal material (left column) and off-axis (right column) directions. 
    {\bf Symbols}: CP data \citep{Baweja23},
	{\bf Dashed lines}: 2S model, {\bf Solid lines}: 3S model. 
    {\red Red}: Compressive responses, {\blue Blue}: Tensile responses.
    }
	\label{fig:B_pg_1}
\end{figure}
\stepcounter{pno}

\begin{figure}[h!]
	\centering
	\includegraphics[width=0.95\textwidth,page=\thepno]{Recursive_plot.pdf}
	\caption{Predicted relative activity of soft glide ($\xi^{\rm s} = p^{\rm s}/p$) by the 3S model for \underline{material B} with $\bar d = 1 \mu$m 
    under uniaxial loading along principal material (left column) and off-axis (right column) directions. 
    {\bf Symbols}: CP data \citep{Baweja23}. 
    {\red Red}: Compressive responses, {\blue Blue}: Tensile responses.
    }
	\label{fig:B_ps_1}
\end{figure}
\stepcounter{pno}

\setcounter{pno}{13}

\clearpage
\section{Texture C}

\begin{figure}[h!]
	\centering
	\includegraphics[width=0.95\textwidth,page=\thepno]{Recursive_plot.pdf}
	\caption{Calibrated stress-strain responses for \underline{material C} with $\bar d = 10^4 \mu$m 
    under uniaxial loading along principal material (left column) and off-axis (right column) directions. 
    {\bf Symbols}: CP data \citep{Baweja23},
	{\bf Dashed lines}: 2S model, {\bf Solid lines}: 3S model. 
    {\red Red}: Compressive responses, {\blue Blue}: Tensile responses.
    }
	\label{fig:C_stress}
\end{figure}
\stepcounter{pno}

\begin{figure}[h!]
	\centering
	\includegraphics[width=0.95\textwidth,page=\thepno]{Recursive_plot.pdf}
	\caption{Predicted lateral strain, $E_{xx}$, for \underline{material C} with $\bar d = 10^4 \mu$m 
    under uniaxial loading along principal material (left column) and off-axis (right column) directions. 
    {\bf Symbols}: CP data \citep{Baweja23},
	{\bf Dashed lines}: 2S model, {\bf Solid lines}: 3S model. 
    {\red Red}: Compressive responses, {\blue Blue}: Tensile responses.
    }
	\label{fig:C_Exx}
\end{figure}
\stepcounter{pno}

\begin{figure}[h!]
	\centering
	\includegraphics[width=0.95\textwidth,page=\thepno]{Recursive_plot.pdf}
	\caption{Predicted lateral strain, $E_{zz}$, for \underline{material C} with $\bar d = 10^4 \mu$m 
    under uniaxial loading along principal material (left column) and off-axis (right column) directions. 
    {\bf Symbols}: CP data \citep{Baweja23},
	{\bf Dashed lines}: 2S model, {\bf Solid lines}: 3S model. 
    {\red Red}: Compressive responses, {\blue Blue}: Tensile responses.
    }
	\label{fig:C_Ezz}
\end{figure}
\stepcounter{pno}

\stepcounter{pno}

\begin{figure}[h!]
	\centering
	\includegraphics[width=0.95\textwidth,page=\thepno]{Recursive_plot.pdf}
	\caption{Predicted relative cumulative activity of glide, $\xi^{\rm g} = p^{\rm g}/p$, for \underline{material C} with $\bar d = 10^4 \mu$m 
    under uniaxial loading along principal material (left column) and off-axis (right column) directions. 
    {\bf Symbols}: CP data \citep{Baweja23},
	{\bf Dashed lines}: 2S model, {\bf Solid lines}: 3S model. 
    {\red Red}: Compressive responses, {\blue Blue}: Tensile responses.
    }
	\label{fig:C_pg}
\end{figure}
\stepcounter{pno}

\begin{figure}[h!]
	\centering
	\includegraphics[width=0.95\textwidth,page=\thepno]{Recursive_plot.pdf}
	\caption{Predicted relative activity of soft glide ($\xi^{\rm s} = p^{\rm s}/p$) by the 3S model for \underline{material C} with $\bar d = 10^4 \mu$m 
    under uniaxial loading along principal material (left column) and off-axis (right column) directions. 
    {\bf Symbols}: CP data \citep{Baweja23}. 
    {\red Red}: Compressive responses, {\blue Blue}: Tensile responses.
    }
	\label{fig:C_ps}
\end{figure}
\stepcounter{pno}

\clearpage
\section{Texture D}

\begin{figure}[h!]
	\centering
	\includegraphics[width=0.95\textwidth,page=\thepno]{Recursive_plot.pdf}
	\caption{Calibrated stress-strain responses for \underline{material D} with $\bar d = 10^4 \mu$m 
    under uniaxial loading along principal material (left column) and off-axis (right column) directions. 
    {\bf Symbols}: CP data \citep{Baweja23},
	{\bf Dashed lines}: 2S model, {\bf Solid lines}: 3S model. 
    {\red Red}: Compressive responses, {\blue Blue}: Tensile responses.
    }
	\label{fig:D_stress}
\end{figure}
\stepcounter{pno}

\begin{figure}[h!]
	\centering
	\includegraphics[width=0.95\textwidth,page=\thepno]{Recursive_plot.pdf}
	\caption{Predicted lateral strain, $E_{xx}$, for \underline{material D} with $\bar d = 10^4 \mu$m 
    under uniaxial loading along principal material (left column) and off-axis (right column) directions. 
    {\bf Symbols}: CP data \citep{Baweja23},
	{\bf Dashed lines}: 2S model, {\bf Solid lines}: 3S model. 
    {\red Red}: Compressive responses, {\blue Blue}: Tensile responses.
    }
	\label{fig:D_Exx}
\end{figure}
\stepcounter{pno}

\begin{figure}[h!]
	\centering
	\includegraphics[width=0.95\textwidth,page=\thepno]{Recursive_plot.pdf}
	\caption{Predicted lateral strain, $E_{zz}$, for \underline{material D} with $\bar d = 10^4 \mu$m 
    under uniaxial loading along principal material (left column) and off-axis (right column) directions. 
    {\bf Symbols}: CP data \citep{Baweja23},
	{\bf Dashed lines}: 2S model, {\bf Solid lines}: 3S model. 
    {\red Red}: Compressive responses, {\blue Blue}: Tensile responses.
    }
	\label{fig:D_Ezz}
\end{figure}
\stepcounter{pno}
\stepcounter{pno}

\begin{figure}[h!]
	\centering
	\includegraphics[width=0.95\textwidth,page=\thepno]{Recursive_plot.pdf}
	\caption{Predicted relative cumulative activity of glide, $\xi^{\rm g} = p^{\rm g}/p$, for \underline{material D} with $\bar d = 10^4 \mu$m 
    under uniaxial loading along principal material (left column) and off-axis (right column) directions. 
    {\bf Symbols}: CP data \citep{Baweja23},
	{\bf Dashed lines}: 2S model, {\bf Solid lines}: 3S model. 
    {\red Red}: Compressive responses, {\blue Blue}: Tensile responses.
    }
	\label{fig:D_pg}
\end{figure}
\stepcounter{pno}

\begin{figure}[h!]
	\centering
	\includegraphics[width=0.95\textwidth,page=\thepno]{Recursive_plot.pdf}
	\caption{Predicted relative activity of soft glide ($\xi^{\rm s} = p^{\rm s}/p$) by the 3S model for \underline{material D} with $\bar d = 10^4 \mu$m 
    under uniaxial loading along principal material (left column) and off-axis (right column) directions. 
    {\bf Symbols}: CP data \citep{Baweja23}. 
    {\red Red}: Compressive responses, {\blue Blue}: Tensile responses.
    }
	\label{fig:D_ps}
\end{figure}
\stepcounter{pno}

\clearpage
\section{Texture E}
\subsection{Grain size, $\bar d = 10^4 \mu$m}

\begin{figure}[h!]
	\centering
	\includegraphics[width=0.95\textwidth,page=\thepno]{Recursive_plot.pdf}
	\caption{Calibrated stress-strain responses for \underline{material E} with $\bar d = 10^4 \mu$m 
    under uniaxial loading along principal material (left column) and off-axis (right column) directions. 
    {\bf Symbols}: CP data \citep{Baweja23},
	{\bf Dashed lines}: 2S model, {\bf Solid lines}: 3S model. 
    {\red Red}: Compressive responses, {\blue Blue}: Tensile responses.
    }
	\label{fig:E_stress}
\end{figure}
\stepcounter{pno}

\begin{figure}[h!]
	\centering
	\includegraphics[width=0.95\textwidth,page=\thepno]{Recursive_plot.pdf}
	\caption{Predicted lateral strain, $E_{xx}$, for \underline{material E} with $\bar d = 10^4 \mu$m 
    under uniaxial loading along principal material (left column) and off-axis (right column) directions. 
    {\bf Symbols}: CP data \citep{Baweja23},
	{\bf Dashed lines}: 2S model, {\bf Solid lines}: 3S model. 
    {\red Red}: Compressive responses, {\blue Blue}: Tensile responses.
    }
	\label{fig:E_Exx}
\end{figure}
\stepcounter{pno}

\begin{figure}[h!]
	\centering
	\includegraphics[width=0.95\textwidth,page=\thepno]{Recursive_plot.pdf}
	\caption{Predicted lateral strain, $E_{zz}$, for \underline{material E} with $\bar d = 10^4 \mu$m 
    under uniaxial loading along principal material (left column) and off-axis (right column) directions. 
    {\bf Symbols}: CP data \citep{Baweja23},
	{\bf Dashed lines}: 2S model, {\bf Solid lines}: 3S model. 
    {\red Red}: Compressive responses, {\blue Blue}: Tensile responses.
    }
	\label{fig:E_Ezz}
\end{figure}
\stepcounter{pno}
\stepcounter{pno}

\begin{figure}[h!]
	\centering
	\includegraphics[width=0.95\textwidth,page=\thepno]{Recursive_plot.pdf}
	\caption{Predicted relative cumulative activity of glide, $\xi^{\rm g} = p^{\rm g}/p$, for \underline{material E} with $\bar d = 10^4 \mu$m 
    under uniaxial loading along principal material (left column) and off-axis (right column) directions. 
    {\bf Symbols}: CP data \citep{Baweja23},
	{\bf Dashed lines}: 2S model, {\bf Solid lines}: 3S model. 
    {\red Red}: Compressive responses, {\blue Blue}: Tensile responses.
    }
	\label{fig:E_pg}
\end{figure}
\stepcounter{pno}

\begin{figure}[h!]
	\centering
	\includegraphics[width=0.95\textwidth,page=\thepno]{Recursive_plot.pdf}
	\caption{Predicted relative activity of soft glide ($\xi^{\rm s} = p^{\rm s}/p$) by the 3S model for \underline{material E} with $\bar d = 10^4 \mu$m 
    under uniaxial loading along principal material (left column) and off-axis (right column) directions. 
    {\bf Symbols}: CP data \citep{Baweja23}. 
    {\red Red}: Compressive responses, {\blue Blue}: Tensile responses.
    }
	\label{fig:E_ps}
\end{figure}
\stepcounter{pno}

\setcounter{pno}{80}
\clearpage
\subsection{Grain size, $\bar d = 10 \mu$m}

\begin{figure}[h!]
	\centering
	\includegraphics[width=0.95\textwidth,page=\thepno]{Recursive_plot.pdf}
	\caption{Calibrated stress-strain responses for \underline{material E} with $\bar d = 10 \mu$m 
    under uniaxial loading along principal material (left column) and off-axis (right column) directions. 
    {\bf Symbols}: CP data \citep{Baweja23},
	{\bf Dashed lines}: 2S model, {\bf Solid lines}: 3S model. 
    {\red Red}: Compressive responses, {\blue Blue}: Tensile responses.
    }
	\label{fig:E_stress_10}
\end{figure}
\stepcounter{pno}

\begin{figure}[h!]
	\centering
	\includegraphics[width=0.95\textwidth,page=\thepno]{Recursive_plot.pdf}
	\caption{Predicted lateral strain, $E_{xx}$, for \underline{material E} with $\bar d = 10 \mu$m 
    under uniaxial loading along principal material (left column) and off-axis (right column) directions. 
    {\bf Symbols}: CP data \citep{Baweja23},
	{\bf Dashed lines}: 2S model, {\bf Solid lines}: 3S model. 
    {\red Red}: Compressive responses, {\blue Blue}: Tensile responses.
    }
	\label{fig:E_Exx_10}
\end{figure}
\stepcounter{pno}

\begin{figure}[h!]
	\centering
	\includegraphics[width=0.95\textwidth,page=\thepno]{Recursive_plot.pdf}
	\caption{Predicted lateral strain, $E_{zz}$, for \underline{material E} with $\bar d = 10 \mu$m 
    under uniaxial loading along principal material (left column) and off-axis (right column) directions. 
    {\bf Symbols}: CP data \citep{Baweja23},
	{\bf Dashed lines}: 2S model, {\bf Solid lines}: 3S model. 
    {\red Red}: Compressive responses, {\blue Blue}: Tensile responses.
    }
	\label{fig:E_Ezz_10}
\end{figure}
\stepcounter{pno}
\stepcounter{pno}

\begin{figure}[h!]
	\centering
	\includegraphics[width=0.95\textwidth,page=\thepno]{Recursive_plot.pdf}
	\caption{Predicted relative cumulative activity of glide, $\xi^{\rm g} = p^{\rm g}/p$, for \underline{material E} with $\bar d = 10 \mu$m 
    under uniaxial loading along principal material (left column) and off-axis (right column) directions. 
    {\bf Symbols}: CP data \citep{Baweja23},
	{\bf Dashed lines}: 2S model, {\bf Solid lines}: 3S model. 
    {\red Red}: Compressive responses, {\blue Blue}: Tensile responses.
    }
	\label{fig:E_pg_10}
\end{figure}
\stepcounter{pno}

\begin{figure}[h!]
	\centering
	\includegraphics[width=0.95\textwidth,page=\thepno]{Recursive_plot.pdf}
	\caption{Predicted relative activity of soft glide ($\xi^{\rm s} = p^{\rm s}/p$) by the 3S model for \underline{material E} with $\bar d = 10 \mu$m 
    under uniaxial loading along principal material (left column) and off-axis (right column) directions. 
    {\bf Symbols}: CP data \citep{Baweja23}. 
    {\red Red}: Compressive responses, {\blue Blue}: Tensile responses.
    }
	\label{fig:E_ps_10}
\end{figure}
\stepcounter{pno}

\setcounter{pno}{98}
\clearpage
\subsection{Grain size, $\bar d = 1 \mu$m}

\begin{figure}[h!]
	\centering
	\includegraphics[width=0.95\textwidth,page=\thepno]{Recursive_plot.pdf}
	\caption{Calibrated stress-strain responses for \underline{material E} with $\bar d = 1 \mu$m 
    under uniaxial loading along principal material (left column) and off-axis (right column) directions. 
    {\bf Symbols}: CP data \citep{Baweja23},
	{\bf Dashed lines}: 2S model, {\bf Solid lines}: 3S model. 
    {\red Red}: Compressive responses, {\blue Blue}: Tensile responses.
    }
	\label{fig:E_stress_1}
\end{figure}
\stepcounter{pno}

\begin{figure}[h!]
	\centering
	\includegraphics[width=0.95\textwidth,page=\thepno]{Recursive_plot.pdf}
	\caption{Predicted lateral strain, $E_{xx}$, for \underline{material E} with $\bar d = 1 \mu$m 
    under uniaxial loading along principal material (left column) and off-axis (right column) directions. 
    {\bf Symbols}: CP data \citep{Baweja23},
	{\bf Dashed lines}: 2S model, {\bf Solid lines}: 3S model. 
    {\red Red}: Compressive responses, {\blue Blue}: Tensile responses.
    }
	\label{fig:E_Exx_1}
\end{figure}
\stepcounter{pno}

\begin{figure}[h!]
	\centering
	\includegraphics[width=0.95\textwidth,page=\thepno]{Recursive_plot.pdf}
	\caption{Predicted lateral strain, $E_{zz}$, for \underline{material E} with $\bar d = 1 \mu$m 
    under uniaxial loading along principal material (left column) and off-axis (right column) directions. 
    {\bf Symbols}: CP data \citep{Baweja23},
	{\bf Dashed lines}: 2S model, {\bf Solid lines}: 3S model. 
    {\red Red}: Compressive responses, {\blue Blue}: Tensile responses.
    }
	\label{fig:E_Ezz_1}
\end{figure}
\stepcounter{pno}
\stepcounter{pno}

\begin{figure}[h!]
	\centering
	\includegraphics[width=0.95\textwidth,page=\thepno]{Recursive_plot.pdf}
	\caption{Predicted relative cumulative activity of glide, $\xi^{\rm g} = p^{\rm g}/p$, for \underline{material E} with $\bar d = 1 \mu$m 
    under uniaxial loading along principal material (left column) and off-axis (right column) directions. 
    {\bf Symbols}: CP data \citep{Baweja23},
	{\bf Dashed lines}: 2S model, {\bf Solid lines}: 3S model. 
    {\red Red}: Compressive responses, {\blue Blue}: Tensile responses.
    }
	\label{fig:E_pg_1}
\end{figure}
\stepcounter{pno}

\begin{figure}[h!]
	\centering
	\includegraphics[width=0.95\textwidth,page=\thepno]{Recursive_plot.pdf}
	\caption{Predicted relative activity of soft glide ($\xi^{\rm s} = p^{\rm s}/p$) by the 3S model for \underline{material E} with $\bar d = 1 \mu$m 
    under uniaxial loading along principal material (left column) and off-axis (right column) directions. 
    {\bf Symbols}: CP data \citep{Baweja23}. 
    {\red Red}: Compressive responses, {\blue Blue}: Tensile responses.
    }
	\label{fig:E_ps_1}
\end{figure}
\stepcounter{pno}

\setcounter{pno}{31}
\clearpage
\section{Texture F}

\begin{figure}[h!]
	\centering
	\includegraphics[width=0.95\textwidth,page=\thepno]{Recursive_plot.pdf}
	\caption{Calibrated stress-strain responses for \underline{material F} with $\bar d = 10^4 \mu$m 
    under uniaxial loading along principal material (left column) and off-axis (right column) directions. 
    {\bf Symbols}: CP data \citep{Baweja23},
	{\bf Dashed lines}: 2S model, {\bf Solid lines}: 3S model. 
    {\red Red}: Compressive responses, {\blue Blue}: Tensile responses.
    }
	\label{fig:F_stress}
\end{figure}
\stepcounter{pno}

\begin{figure}[h!]
	\centering
	\includegraphics[width=0.95\textwidth,page=\thepno]{Recursive_plot.pdf}
	\caption{Predicted lateral strain, $E_{xx}$, for \underline{material F} with $\bar d = 10^4 \mu$m 
    under uniaxial loading along principal material (left column) and off-axis (right column) directions. 
    {\bf Symbols}: CP data \citep{Baweja23},
	{\bf Dashed lines}: 2S model, {\bf Solid lines}: 3S model. 
    {\red Red}: Compressive responses, {\blue Blue}: Tensile responses.
    }
	\label{fig:F_Exx}
\end{figure}
\stepcounter{pno}

\begin{figure}[h!]
	\centering
	\includegraphics[width=0.95\textwidth,page=\thepno]{Recursive_plot.pdf}
	\caption{Predicted lateral strain, $E_{zz}$, for \underline{material F} with $\bar d = 10^4 \mu$m 
    under uniaxial loading along principal material (left column) and off-axis (right column) directions. 
    {\bf Symbols}: CP data \citep{Baweja23},
	{\bf Dashed lines}: 2S model, {\bf Solid lines}: 3S model. 
    {\red Red}: Compressive responses, {\blue Blue}: Tensile responses.
    }
	\label{fig:F_Ezz}
\end{figure}
\stepcounter{pno}
\stepcounter{pno}

\begin{figure}[h!]
	\centering
	\includegraphics[width=0.95\textwidth,page=\thepno]{Recursive_plot.pdf}
	\caption{Predicted relative cumulative activity of glide, $\xi^{\rm g} = p^{\rm g}/p$, for \underline{material F} with $\bar d = 10^4 \mu$m 
    under uniaxial loading along principal material (left column) and off-axis (right column) directions. 
    {\bf Symbols}: CP data \citep{Baweja23},
	{\bf Dashed lines}: 2S model, {\bf Solid lines}: 3S model. 
    {\red Red}: Compressive responses, {\blue Blue}: Tensile responses.
    }
	\label{fig:F_pg}
\end{figure}
\stepcounter{pno}

\begin{figure}[h!]
	\centering
	\includegraphics[width=0.95\textwidth,page=\thepno]{Recursive_plot.pdf}
	\caption{Predicted relative activity of soft glide ($\xi^{\rm s} = p^{\rm s}/p$) by the 3S model for \underline{material F} with $\bar d = 10^4 \mu$m 
    under uniaxial loading along principal material (left column) and off-axis (right column) directions. 
    {\bf Symbols}: CP data \citep{Baweja23}. 
    {\red Red}: Compressive responses, {\blue Blue}: Tensile responses.
    }
	\label{fig:F_ps}
\end{figure}
\stepcounter{pno}
\clearpage
\clearpage
\section{Texture G}

\begin{figure}[h!]
	\centering
	\includegraphics[width=0.95\textwidth,page=\thepno]{Recursive_plot.pdf}
	\caption{Calibrated stress-strain responses for \underline{material G} with $\bar d = 10^4 \mu$m 
    under uniaxial loading along principal material (left column) and off-axis (right column) directions. 
    {\bf Symbols}: CP data \citep{Baweja23},
	{\bf Dashed lines}: 2S model, {\bf Solid lines}: 3S model. 
    {\red Red}: Compressive responses, {\blue Blue}: Tensile responses.
    }
	\label{fig:G_stress}
\end{figure}
\stepcounter{pno}

\begin{figure}[h!]
	\centering
	\includegraphics[width=0.95\textwidth,page=\thepno]{Recursive_plot.pdf}
	\caption{Predicted lateral strain, $E_{xx}$, for \underline{material G} with $\bar d = 10^4 \mu$m 
    under uniaxial loading along principal material (left column) and off-axis (right column) directions. 
    {\bf Symbols}: CP data \citep{Baweja23},
	{\bf Dashed lines}: 2S model, {\bf Solid lines}: 3S model. 
    {\red Red}: Compressive responses, {\blue Blue}: Tensile responses.
    }
	\label{fig:G_Exx}
\end{figure}
\stepcounter{pno}

\begin{figure}[h!]
	\centering
	\includegraphics[width=0.95\textwidth,page=\thepno]{Recursive_plot.pdf}
	\caption{Predicted lateral strain, $E_{zz}$, for \underline{material G} with $\bar d = 10^4 \mu$m 
    under uniaxial loading along principal material (left column) and off-axis (right column) directions. 
    {\bf Symbols}: CP data \citep{Baweja23},
	{\bf Dashed lines}: 2S model, {\bf Solid lines}: 3S model. 
    {\red Red}: Compressive responses, {\blue Blue}: Tensile responses.
    }
	\label{fig:G_Ezz}
\end{figure}
\stepcounter{pno}
\stepcounter{pno}

\begin{figure}[h!]
	\centering
	\includegraphics[width=0.95\textwidth,page=\thepno]{Recursive_plot.pdf}
	\caption{Predicted relative cumulative activity of glide, $\xi^{\rm g} = p^{\rm g}/p$, for \underline{material G} with $\bar d = 10^4 \mu$m 
    under uniaxial loading along principal material (left column) and off-axis (right column) directions. 
    {\bf Symbols}: CP data \citep{Baweja23},
	{\bf Dashed lines}: 2S model, {\bf Solid lines}: 3S model. 
    {\red Red}: Compressive responses, {\blue Blue}: Tensile responses.
    }
	\label{fig:G_pg}
\end{figure}
\stepcounter{pno}

\begin{figure}[h!]
	\centering
	\includegraphics[width=0.95\textwidth,page=\thepno]{Recursive_plot.pdf}
	\caption{Predicted relative activity of soft glide ($\xi^{\rm s} = p^{\rm s}/p$) by the 3S model for \underline{material G} with $\bar d = 10^4 \mu$m 
    under uniaxial loading along principal material (left column) and off-axis (right column) directions. 
    {\bf Symbols}: CP data \citep{Baweja23}. 
    {\red Red}: Compressive responses, {\blue Blue}: Tensile responses.
    }
	\label{fig:G_ps}
\end{figure}
\stepcounter{pno}

\clearpage
\section{Texture H}

\begin{figure}[h!]
	\centering
	\includegraphics[width=0.95\textwidth,page=\thepno]{Recursive_plot.pdf}
	\caption{Calibrated stress-strain responses for \underline{material H} with $\bar d = 10^4 \mu$m 
    under uniaxial loading along principal material (left column) and off-axis (right column) directions. 
    {\bf Symbols}: CP data \citep{Baweja23},
	{\bf Dashed lines}: 2S model, {\bf Solid lines}: 3S model. 
    {\red Red}: Compressive responses, {\blue Blue}: Tensile responses.
    }
	\label{fig:H_stress}
\end{figure}
\stepcounter{pno}

\begin{figure}[h!]
	\centering
	\includegraphics[width=0.95\textwidth,page=\thepno]{Recursive_plot.pdf}
	\caption{Predicted lateral strain, $E_{xx}$, for \underline{material H} with $\bar d = 10^4 \mu$m 
    under uniaxial loading along principal material (left column) and off-axis (right column) directions. 
    {\bf Symbols}: CP data \citep{Baweja23},
	{\bf Dashed lines}: 2S model, {\bf Solid lines}: 3S model. 
    {\red Red}: Compressive responses, {\blue Blue}: Tensile responses.
    }
	\label{fig:H_Exx}
\end{figure}
\stepcounter{pno}

\begin{figure}[h!]
	\centering
	\includegraphics[width=0.95\textwidth,page=\thepno]{Recursive_plot.pdf}
	\caption{Predicted lateral strain, $E_{zz}$, for \underline{material H} with $\bar d = 10^4 \mu$m 
    under uniaxial loading along principal material (left column) and off-axis (right column) directions. 
    {\bf Symbols}: CP data \citep{Baweja23},
	{\bf Dashed lines}: 2S model, {\bf Solid lines}: 3S model. 
    {\red Red}: Compressive responses, {\blue Blue}: Tensile responses.
    }
	\label{fig:H_Ezz}
\end{figure}
\stepcounter{pno}
\stepcounter{pno}

\begin{figure}[h!]
	\centering
	\includegraphics[width=0.95\textwidth,page=\thepno]{Recursive_plot.pdf}
	\caption{Predicted relative cumulative activity of glide, $\xi^{\rm g} = p^{\rm g}/p$, for \underline{material H} with $\bar d = 10^4 \mu$m 
    under uniaxial loading along principal material (left column) and off-axis (right column) directions. 
    {\bf Symbols}: CP data \citep{Baweja23},
	{\bf Dashed lines}: 2S model, {\bf Solid lines}: 3S model. 
    {\red Red}: Compressive responses, {\blue Blue}: Tensile responses.
    }
	\label{fig:H_pg}
\end{figure}
\stepcounter{pno}

\begin{figure}[h!]
	\centering
	\includegraphics[width=0.95\textwidth,page=\thepno]{Recursive_plot.pdf}
	\caption{Predicted relative activity of soft glide ($\xi^{\rm s} = p^{\rm s}/p$) by the 3S model for \underline{material H} with $\bar d = 10^4 \mu$m 
    under uniaxial loading along principal material (left column) and off-axis (right column) directions. 
    {\bf Symbols}: CP data \citep{Baweja23}. 
    {\red Red}: Compressive responses, {\blue Blue}: Tensile responses.
    }
	\label{fig:H_ps}
\end{figure}
\stepcounter{pno}

\clearpage
\section{Texture I}

\begin{figure}[h!]
	\centering
	\includegraphics[width=0.95\textwidth,page=\thepno]{Recursive_plot.pdf}
	\caption{Calibrated stress-strain responses for \underline{material I} with $\bar d = 10^4 \mu$m 
    under uniaxial loading along principal material (left column) and off-axis (right column) directions. 
    {\bf Symbols}: CP data \citep{Baweja23},
	{\bf Dashed lines}: 2S model, {\bf Solid lines}: 3S model. 
    {\red Red}: Compressive responses, {\blue Blue}: Tensile responses.
    }
	\label{fig:I_stress}
\end{figure}
\stepcounter{pno}

\begin{figure}[h!]
	\centering
	\includegraphics[width=0.95\textwidth,page=\thepno]{Recursive_plot.pdf}
	\caption{Predicted lateral strain, $E_{xx}$, for \underline{material I} with $\bar d = 10^4 \mu$m 
    under uniaxial loading along principal material (left column) and off-axis (right column) directions. 
    {\bf Symbols}: CP data \citep{Baweja23},
	{\bf Dashed lines}: 2S model, {\bf Solid lines}: 3S model. 
    {\red Red}: Compressive responses, {\blue Blue}: Tensile responses.
    }
	\label{fig:I_Exx}
\end{figure}
\stepcounter{pno}

\begin{figure}[h!]
	\centering
	\includegraphics[width=0.95\textwidth,page=\thepno]{Recursive_plot.pdf}
	\caption{Predicted lateral strain, $E_{zz}$, for \underline{material I} with $\bar d = 10^4 \mu$m 
    under uniaxial loading along principal material (left column) and off-axis (right column) directions. 
    {\bf Symbols}: CP data \citep{Baweja23},
	{\bf Dashed lines}: 2S model, {\bf Solid lines}: 3S model. 
    {\red Red}: Compressive responses, {\blue Blue}: Tensile responses.
    }
	\label{fig:I_Ezz}
\end{figure}
\stepcounter{pno}
\stepcounter{pno}

\begin{figure}[h!]
	\centering
	\includegraphics[width=0.95\textwidth,page=\thepno]{Recursive_plot.pdf}
	\caption{Predicted relative cumulative activity of glide, $\xi^{\rm g} = p^{\rm g}/p$, for \underline{material I} with $\bar d = 10^4 \mu$m 
    under uniaxial loading along principal material (left column) and off-axis (right column) directions. 
    {\bf Symbols}: CP data \citep{Baweja23},
	{\bf Dashed lines}: 2S model, {\bf Solid lines}: 3S model. 
    {\red Red}: Compressive responses, {\blue Blue}: Tensile responses.
    }
	\label{fig:I_pg}
\end{figure}
\stepcounter{pno}

\begin{figure}[h!]
	\centering
	\includegraphics[width=0.95\textwidth,page=\thepno]{Recursive_plot.pdf}
	\caption{Predicted relative activity of soft glide ($\xi^{\rm s} = p^{\rm s}/p$) by the 3S model for \underline{material I} with $\bar d = 10^4 \mu$m 
    under uniaxial loading along principal material (left column) and off-axis (right column) directions. 
    {\bf Symbols}: CP data \citep{Baweja23}. 
    {\red Red}: Compressive responses, {\blue Blue}: Tensile responses.
    }
	\label{fig:I_ps}
\end{figure}
\stepcounter{pno}

\clearpage
\section{Texture J}

\begin{figure}[h!]
	\centering
	\includegraphics[width=0.95\textwidth,page=\thepno]{Recursive_plot.pdf}
	\caption{Calibrated stress-strain responses for \underline{material J} with $\bar d = 10^4 \mu$m 
    under uniaxial loading along principal material (left column) and off-axis (right column) directions. 
    {\bf Symbols}: CP data \citep{Baweja23},
	{\bf Dashed lines}: 2S model, {\bf Solid lines}: 3S model. 
    {\red Red}: Compressive responses, {\blue Blue}: Tensile responses.
    }
	\label{fig:J_stress}
\end{figure}
\stepcounter{pno}

\begin{figure}[h!]
	\centering
	\includegraphics[width=0.95\textwidth,page=\thepno]{Recursive_plot.pdf}
	\caption{Predicted lateral strain, $E_{xx}$, for \underline{material J} with $\bar d = 10^4 \mu$m 
    under uniaxial loading along principal material (left column) and off-axis (right column) directions. 
    {\bf Symbols}: CP data \citep{Baweja23},
	{\bf Dashed lines}: 2S model, {\bf Solid lines}: 3S model. 
    {\red Red}: Compressive responses, {\blue Blue}: Tensile responses.
    }
	\label{fig:J_Exx}
\end{figure}
\stepcounter{pno}

\begin{figure}[h!]
	\centering
	\includegraphics[width=0.95\textwidth,page=\thepno]{Recursive_plot.pdf}
	\caption{Predicted lateral strain, $E_{zz}$, for \underline{material J} with $\bar d = 10^4 \mu$m 
    under uniaxial loading along principal material (left column) and off-axis (right column) directions. 
    {\bf Symbols}: CP data \citep{Baweja23},
	{\bf Dashed lines}: 2S model, {\bf Solid lines}: 3S model. 
    {\red Red}: Compressive responses, {\blue Blue}: Tensile responses.
    }
	\label{fig:J_Ezz}
\end{figure}
\stepcounter{pno}
\stepcounter{pno}

\begin{figure}[h!]
	\centering
	\includegraphics[width=0.95\textwidth,page=\thepno]{Recursive_plot.pdf}
	\caption{Predicted relative cumulative activity of glide, $\xi^{\rm g} = p^{\rm g}/p$, for \underline{material J} with $\bar d = 10^4 \mu$m 
    under uniaxial loading along principal material (left column) and off-axis (right column) directions. 
    {\bf Symbols}: CP data \citep{Baweja23},
	{\bf Dashed lines}: 2S model, {\bf Solid lines}: 3S model. 
    {\red Red}: Compressive responses, {\blue Blue}: Tensile responses.
    }
	\label{fig:J_pg}
\end{figure}
\stepcounter{pno}

\begin{figure}[h!]
	\centering
	\includegraphics[width=0.95\textwidth,page=\thepno]{Recursive_plot.pdf}
	\caption{Predicted relative activity of soft glide ($\xi^{\rm s} = p^{\rm s}/p$) by the 3S model for \underline{material J} with $\bar d = 10^4 \mu$m 
    under uniaxial loading along principal material (left column) and off-axis (right column) directions. 
    {\bf Symbols}: CP data \citep{Baweja23}. 
    {\red Red}: Compressive responses, {\blue Blue}: Tensile responses.
    }
	\label{fig:J_ps}
\end{figure}
\stepcounter{pno}

\clearpage
\section{Texture K}
\stepcounter{pno}
\stepcounter{pno}
\stepcounter{pno}
\stepcounter{pno}
\stepcounter{pno}
\stepcounter{pno}

\setcounter{pno}{86}
\subsection{Grain size, $\bar d = 10 \mu$m}

\begin{figure}[h!]
	\centering
	\includegraphics[width=0.95\textwidth,page=\thepno]{Recursive_plot.pdf}
	\caption{Calibrated stress-strain responses for \underline{material K} with $\bar d = 10 \mu$m 
    under uniaxial loading along principal material (left column) and off-axis (right column) directions. 
    {\bf Symbols}: CP data \citep{Baweja23},
	{\bf Dashed lines}: 2S model, {\bf Solid lines}: 3S model. 
    {\red Red}: Compressive responses, {\blue Blue}: Tensile responses.
    }
	\label{fig:K_stress_10}
\end{figure}
\stepcounter{pno}

\begin{figure}[h!]
	\centering
	\includegraphics[width=0.95\textwidth,page=\thepno]{Recursive_plot.pdf}
	\caption{Predicted lateral strain, $E_{xx}$, for \underline{material K} with $\bar d = 10 \mu$m 
    under uniaxial loading along principal material (left column) and off-axis (right column) directions. 
    {\bf Symbols}: CP data \citep{Baweja23},
	{\bf Dashed lines}: 2S model, {\bf Solid lines}: 3S model. 
    {\red Red}: Compressive responses, {\blue Blue}: Tensile responses.
    }
	\label{fig:K_Exx_10}
\end{figure}
\stepcounter{pno}

\begin{figure}[h!]
	\centering
	\includegraphics[width=0.95\textwidth,page=\thepno]{Recursive_plot.pdf}
	\caption{Predicted lateral strain, $E_{zz}$, for \underline{material K} with $\bar d = 10 \mu$m 
    under uniaxial loading along principal material (left column) and off-axis (right column) directions. 
    {\bf Symbols}: CP data \citep{Baweja23},
	{\bf Dashed lines}: 2S model, {\bf Solid lines}: 3S model. 
    {\red Red}: Compressive responses, {\blue Blue}: Tensile responses.
    }
	\label{fig:K_Ezz_10}
\end{figure}
\stepcounter{pno}
\stepcounter{pno}

\begin{figure}[h!]
	\centering
	\includegraphics[width=0.95\textwidth,page=\thepno]{Recursive_plot.pdf}
	\caption{Predicted relative cumulative activity of glide, $\xi^{\rm g} = p^{\rm g}/p$, for \underline{material K} with $\bar d = 10 \mu$m 
    under uniaxial loading along principal material (left column) and off-axis (right column) directions. 
    {\bf Symbols}: CP data \citep{Baweja23},
	{\bf Dashed lines}: 2S model, {\bf Solid lines}: 3S model. 
    {\red Red}: Compressive responses, {\blue Blue}: Tensile responses.
    }
	\label{fig:K_pg_10}
\end{figure}
\stepcounter{pno}

\begin{figure}[h!]
	\centering
	\includegraphics[width=0.95\textwidth,page=\thepno]{Recursive_plot.pdf}
	\caption{Predicted relative activity of soft glide ($\xi^{\rm s} = p^{\rm s}/p$) by the 3S model for \underline{material K} with $\bar d = 10 \mu$m 
    under uniaxial loading along principal material (left column) and off-axis (right column) directions. 
    {\bf Symbols}: CP data \citep{Baweja23}. 
    {\red Red}: Compressive responses, {\blue Blue}: Tensile responses.
    }
	\label{fig:K_ps_10}
\end{figure}
\stepcounter{pno}

\setcounter{pno}{104}
\clearpage
\subsection{Grain size, $\bar d = 1 \mu$m}

\begin{figure}[h!]
	\centering
	\includegraphics[width=0.95\textwidth,page=\thepno]{Recursive_plot.pdf}
	\caption{Calibrated stress-strain responses for \underline{material K} with $\bar d = 1 \mu$m 
    under uniaxial loading along principal material (left column) and off-axis (right column) directions. 
    {\bf Symbols}: CP data \citep{Baweja23},
	{\bf Dashed lines}: 2S model, {\bf Solid lines}: 3S model. 
    {\red Red}: Compressive responses, {\blue Blue}: Tensile responses.
    }
	\label{fig:K_stress_1}
\end{figure}
\stepcounter{pno}

\begin{figure}[h!]
	\centering
	\includegraphics[width=0.95\textwidth,page=\thepno]{Recursive_plot.pdf}
	\caption{Predicted lateral strain, $E_{xx}$, for \underline{material K} with $\bar d = 1 \mu$m 
    under uniaxial loading along principal material (left column) and off-axis (right column) directions. 
    {\bf Symbols}: CP data \citep{Baweja23},
	{\bf Dashed lines}: 2S model, {\bf Solid lines}: 3S model. 
    {\red Red}: Compressive responses, {\blue Blue}: Tensile responses.
    }
	\label{fig:K_Exx_1}
\end{figure}
\stepcounter{pno}

\begin{figure}[h!]
	\centering
	\includegraphics[width=0.95\textwidth,page=\thepno]{Recursive_plot.pdf}
	\caption{Predicted lateral strain, $E_{zz}$, for \underline{material K} with $\bar d = 1 \mu$m 
    under uniaxial loading along principal material (left column) and off-axis (right column) directions. 
    {\bf Symbols}: CP data \citep{Baweja23},
	{\bf Dashed lines}: 2S model, {\bf Solid lines}: 3S model. 
    {\red Red}: Compressive responses, {\blue Blue}: Tensile responses.
    }
	\label{fig:K_Ezz_1}
\end{figure}
\stepcounter{pno}
\stepcounter{pno}

\begin{figure}[h!]
	\centering
	\includegraphics[width=0.95\textwidth,page=\thepno]{Recursive_plot.pdf}
	\caption{Predicted relative cumulative activity of glide, $\xi^{\rm g} = p^{\rm g}/p$, for \underline{material K} with $\bar d = 1 \mu$m 
    under uniaxial loading along principal material (left column) and off-axis (right column) directions. 
    {\bf Symbols}: CP data \citep{Baweja23},
	{\bf Dashed lines}: 2S model, {\bf Solid lines}: 3S model. 
    {\red Red}: Compressive responses, {\blue Blue}: Tensile responses.
    }
	\label{fig:K_pg_1}
\end{figure}
\stepcounter{pno}

\begin{figure}[h!]
	\centering
	\includegraphics[width=0.95\textwidth,page=\thepno]{Recursive_plot.pdf}
	\caption{Predicted relative activity of soft glide ($\xi^{\rm s} = p^{\rm s}/p$) by the 3S model for \underline{material K} with $\bar d = 1 \mu$m 
    under uniaxial loading along principal material (left column) and off-axis (right column) directions. 
    {\bf Symbols}: CP data \citep{Baweja23}. 
    {\red Red}: Compressive responses, {\blue Blue}: Tensile responses.
    }
	\label{fig:K_ps_1}
\end{figure}
\stepcounter{pno}

\stopcontents[part2toc]

\end{document}